\documentclass[runningheads,envcountsame]{llncs}
\usepackage[utf8]{inputenc}
\usepackage[T1]{fontenc}
\usepackage[letterpaper,margin=1in,asymmetric,left=1.005in,right=1.025in,bottom=1.05in]{geometry}
\usepackage{lmodern}
\usepackage{amsmath,amssymb}
\usepackage{dsfont}
\usepackage{graphicx,booktabs,microtype}
\usepackage{algorithm}
\usepackage{algpseudocode}
\usepackage{xcolor}
\usepackage{tikz}
\usetikzlibrary{arrows.meta,positioning,calc}
\DeclareMathOperator{\Var}{Var}

\newcommand{\E}{\mathbb{E}}
\newcommand{\Prob}{\mathbb{P}}
\newcommand{\ind}{\mathds{1}}
\newcommand{\pos}[1]{\left[#1\right]_{+}}
\newcommand{\qlo}{q^{\ominus}}
\newcommand{\qhi}{q^{\oplus}}
\newcommand{\pe}{P_{\eta}}
\newcommand{\qe}{q_{\eta}}
\newcommand{\ee}{E_{\eta}}
\newcommand{\qP}{q_{_{\mathrm P}}}
\newcommand{\qL}{q_{_{\mathrm L}}}
\newcommand{\qPL}{q_{_{\mathrm{PL}}}}
\newcommand{\eb}{\delta_y}
\newcommand{\ef}{\delta_\pi}
\newcommand{\lb}{\ell_y}
\newcommand{\ub}{u_y}
\newcommand{\pup}{\overline{\pi}}

\newcommand{\sidechi}[1]{\chi_{_{#1}}}
\newcommand{\stat}[1]{\csname paperstat@#1\endcsname}
\expandafter\def\csname paperstat@common1systems\endcsname{36}
\expandafter\def\csname paperstat@common1pairs\endcsname{873}
\expandafter\def\csname paperstat@common1spearman\endcsname{0.9955}
\expandafter\def\csname paperstat@common1spearmanshort\endcsname{0.996}
\expandafter\def\csname paperstat@common1pearson\endcsname{0.9993}
\expandafter\def\csname paperstat@common1difference\endcsname{0.1059}
\expandafter\def\csname paperstat@common2systems\endcsname{36}
\expandafter\def\csname paperstat@common2pairs\endcsname{873}
\expandafter\def\csname paperstat@common2spearman\endcsname{0.9917}
\expandafter\def\csname paperstat@common2spearmanshort\endcsname{0.992}
\expandafter\def\csname paperstat@common2pearson\endcsname{0.9980}
\expandafter\def\csname paperstat@common2difference\endcsname{0.1946}
\expandafter\def\csname paperstat@common4systems\endcsname{36}
\expandafter\def\csname paperstat@common4pairs\endcsname{873}
\expandafter\def\csname paperstat@common4spearman\endcsname{0.9814}
\expandafter\def\csname paperstat@common4spearmanshort\endcsname{0.981}
\expandafter\def\csname paperstat@common4pearson\endcsname{0.9960}
\expandafter\def\csname paperstat@common4difference\endcsname{0.2665}
\expandafter\def\csname paperstat@pair1systems\endcsname{36}
\expandafter\def\csname paperstat@pair1pairs\endcsname{903}
\expandafter\def\csname paperstat@pair1spearman\endcsname{0.9954}
\expandafter\def\csname paperstat@pair1spearmanshort\endcsname{0.995}
\expandafter\def\csname paperstat@pair1pearson\endcsname{0.9992}
\expandafter\def\csname paperstat@pair1difference\endcsname{0.1067}
\expandafter\def\csname paperstat@pair2systems\endcsname{36}
\expandafter\def\csname paperstat@pair2pairs\endcsname{899}
\expandafter\def\csname paperstat@pair2spearman\endcsname{0.9915}
\expandafter\def\csname paperstat@pair2spearmanshort\endcsname{0.992}
\expandafter\def\csname paperstat@pair2pearson\endcsname{0.9977}
\expandafter\def\csname paperstat@pair2difference\endcsname{0.1910}
\expandafter\def\csname paperstat@pair4systems\endcsname{36}
\expandafter\def\csname paperstat@pair4pairs\endcsname{901}
\expandafter\def\csname paperstat@pair4spearman\endcsname{0.9835}
\expandafter\def\csname paperstat@pair4spearmanshort\endcsname{0.983}
\expandafter\def\csname paperstat@pair4pearson\endcsname{0.9971}
\expandafter\def\csname paperstat@pair4difference\endcsname{0.2696}
\expandafter\def\csname paperstat@refsystems\endcsname{9}
\expandafter\def\csname paperstat@refruns\endcsname{900}
\expandafter\def\csname paperstat@refpositions\endcsname{5--8}
\expandafter\def\csname paperstat@refassignments\endcsname{1.005--10.080}
\expandafter\def\csname paperstat@refmaxassignments\endcsname{10.08}
\expandafter\def\csname paperstat@refwidthmin\endcsname{0.6111}
\expandafter\def\csname paperstat@refwidthmax\endcsname{0.6142}
\expandafter\def\csname paperstat@runtimedesigns\endcsname{136}
\expandafter\def\csname paperstat@runtimesystems\endcsname{38}
\expandafter\def\csname paperstat@runtimemark\endcsname{118}
\expandafter\def\csname paperstat@runtimepack\endcsname{136}
\expandafter\def\csname paperstat@runtimerunning\endcsname{11}
\expandafter\def\csname paperstat@runtimefailed\endcsname{0}
\expandafter\def\csname paperstat@runtimepairs\endcsname{118}
\expandafter\def\csname paperstat@runtimematched\endcsname{118}
\expandafter\def\csname paperstat@runtimefixed\endcsname{80}
\expandafter\def\csname paperstat@runtimepending\endcsname{0}
\expandafter\def\csname paperstat@runtimecleanoriginal\endcsname{81}
\expandafter\def\csname paperstat@runtimemissing\endcsname{2}
\expandafter\def\csname paperstat@runtimeratiomin1\endcsname{2.8}
\expandafter\def\csname paperstat@runtimeratiomax1\endcsname{1088.7}
\expandafter\def\csname paperstat@runtimeratiomin2\endcsname{2.80}
\expandafter\def\csname paperstat@runtimeratiomax2\endcsname{1088.69}
\expandafter\def\csname paperstat@tail1tested\endcsname{56}
\expandafter\def\csname paperstat@tail1rejected\endcsname{3}
\expandafter\def\csname paperstat@tail1remaining\endcsname{53}
\expandafter\def\csname paperstat@tail1above\endcsname{3}
\expandafter\def\csname paperstat@tail2tested\endcsname{58}
\expandafter\def\csname paperstat@tail2rejected\endcsname{1}
\expandafter\def\csname paperstat@tail2remaining\endcsname{57}
\expandafter\def\csname paperstat@tail2above\endcsname{5}
\expandafter\def\csname paperstat@tail4tested\endcsname{68}
\expandafter\def\csname paperstat@tail4rejected\endcsname{0}
\expandafter\def\csname paperstat@tail4remaining\endcsname{68}
\expandafter\def\csname paperstat@tail4above\endcsname{4}
\expandafter\def\csname paperstat@tailtested\endcsname{182}
\expandafter\def\csname paperstat@tailremaining\endcsname{178}
\expandafter\def\csname paperstat@tailbelow1\endcsname{84}
\expandafter\def\csname paperstat@tailbetween1and5\endcsname{61}
\expandafter\def\csname paperstat@tailbetween5and20\endcsname{21}
\expandafter\def\csname paperstat@tailatmost20\endcsname{166}
\expandafter\def\csname paperstat@tailabove\endcsname{12}
\expandafter\def\csname paperstat@tailsingle\endcsname{91}
\expandafter\def\csname paperstat@tailmedian\endcsname{1.432}
\expandafter\def\csname paperstat@tailp95\endcsname{25.250}
\expandafter\def\csname paperstat@tailmaximum\endcsname{61.830}
\expandafter\def\csname paperstat@tailkeptpercent\endcsname{93.3}
\expandafter\def\csname paperstat@tailallpercent\endcsname{91.2}
\expandafter\def\csname paperstat@runtimecancelled\endcsname{0}
\expandafter\def\csname paperstat@runtimecensored\endcsname{18}
\expandafter\def\csname paperstat@runtimeallpairs\endcsname{136}
\expandafter\def\csname paperstat@runtimetimeouts\endcsname{16}
\expandafter\def\csname paperstat@runtimeboundmax1\endcsname{2282.1}
\expandafter\def\csname paperstat@runtimeauthorbounds\endcsname{2}
\expandafter\def\csname paperstat@runtimemaxdesign\endcsname{5it3\_flex\_p10}
\expandafter\def\csname paperstat@runtimemaxmarkhours2\endcsname{148.70}
\expandafter\def\csname paperstat@runtimemaxpackminutes2\endcsname{8.20}
\expandafter\def\csname paperstat@runtimemaxmarkscores\endcsname{34}
\expandafter\def\csname paperstat@runtimemaxpackscores\endcsname{30}
\expandafter\def\csname paperstat@runtimemarkobserved\endcsname{136}
\expandafter\def\csname paperstat@runtimemarkonly\endcsname{0}
\expandafter\def\csname paperstat@runtimeboundmin1\endcsname{465.8}
\expandafter\def\csname paperstat@runtimeboundmin2\endcsname{465.86}
\expandafter\def\csname paperstat@runtimeboundmax2\endcsname{2282.09}
\expandafter\def\csname paperstat@tailtotal\endcsname{6065}
\expandafter\def\csname paperstat@runtimeplotdesigns\endcsname{136}
\expandafter\def\csname paperstat@runtimeplotmark\endcsname{136}

\let\small\normalsize
\let\footnotesize\normalsize
\let\scriptsize\normalsize
\microtypesetup{protrusion=false,expansion=true}
\usepackage{flafter}
\usepackage{refcount,longtable,array,chapterbib}
\usepackage{xurl}
\usepackage{hyperref}
\hypersetup{hidelinks,hypertexnames=false}

\definecolor{packblue}{HTML}{526B82}
\definecolor{packolive}{HTML}{68734A}
\definecolor{packbrick}{HTML}{A94F45}
\newlength{\introparskip}
\let\PackstarNormalSize\normalsize
\makeatletter
\g@addto@macro\normalsize{%
  \setlength{\abovedisplayskip}{4pt plus 2pt minus 1pt}%
  \setlength{\belowdisplayskip}{4pt plus 2pt minus 1pt}%
  \setlength{\abovedisplayshortskip}{0pt plus 2pt}%
  \setlength{\belowdisplayshortskip}{4pt plus 2pt minus 2pt}}
\makeatother

\newenvironment{auditedlemma}[1]{\renewcommand{\thelemma}{#1}\begin{lemma}}{\end{lemma}}
\newenvironment{auditedtheorem}[1]{\renewcommand{\thetheorem}{#1}\begin{theorem}}{\end{theorem}}
\newenvironment{auditedproposition}[1]{\renewcommand{\theproposition}{#1}\begin{proposition}}{\end{proposition}}
\newenvironment{auditedcorollary}[1]{\renewcommand{\thecorollary}{#1}\begin{corollary}}{\end{corollary}}
\newcounter{suppresult}
\newenvironment{supplemma}{\stepcounter{suppresult}\begin{auditedlemma}{S\arabic{suppresult}}}{\end{auditedlemma}}
\newenvironment{suppproposition}{\stepcounter{suppresult}\begin{auditedproposition}{S\arabic{suppresult}}}{\end{auditedproposition}}

\ExplSyntaxOn
\cs_new:Npn \packstar_si_key:n #1
 {
  \str_if_eq:eeTF {\str_range:nnn {#1}{1}{5}} {MAIN-}
   {\str_range:nnn {#1}{6}{-1}}
   {SI-#1}
 }
\NewExpandableDocumentCommand{\PackstarSIKey}{m}{\packstar_si_key:n{#1}}
\ExplSyntaxOff
\makeatletter
\newcommand{\PackstarBeginAppendix}{%
 \clearpage
 \let\normalsize\PackstarNormalSize\normalsize
 \setlength{\parskip}{0pt plus 1pt}%
 \renewcommand{\topfraction}{0.9}%
 \renewcommand{\bottomfraction}{0.85}%
 \renewcommand{\textfraction}{0.1}%
 \renewcommand{\floatpagefraction}{0.75}%
 \setlength{\textfloatsep}{10pt plus 2pt minus 2pt}%
 \setlength{\intextsep}{6pt plus 2pt minus 2pt}%
 \setlength{\abovecaptionskip}{3pt}%
 \setcounter{section}{0}\setcounter{subsection}{0}%
 \setcounter{equation}{0}\setcounter{figure}{0}\setcounter{table}{0}%
 \setcounter{theorem}{0}\setcounter{algorithm}{0}\setcounter{footnote}{0}%
 \pagenumbering{arabic}%
 \let\PackstarOriginalLabel\label
 \let\PackstarOriginalRef\ref
 \let\PackstarOriginalPageRef\pageref
 \let\PackstarOriginalRefNumber\getrefnumber
 \renewcommand{\label}[1]{\PackstarOriginalLabel{SI-##1}}%
 \let\ltx@label\label
 \renewcommand{\ref}[1]{\PackstarOriginalRef{\PackstarSIKey{##1}}}%
 \renewcommand{\pageref}[1]{\PackstarOriginalPageRef{\PackstarSIKey{##1}}}%
 \renewcommand{\getrefnumber}[1]{\PackstarOriginalRefNumber{\PackstarSIKey{##1}}}%
}
\makeatother
\begin{document}

\pagestyle{plain}
\pagenumbering{roman}
\title{PACK*: Probably Approximately Correct Partition Function Estimation for Ensemble-Based Protein Design}
\titlerunning{PACK*: Partition Function Estimation}
\author{Liyang Zhu\inst{1} \and Jonathan D. Jou\inst{2} \and
Allen C. McBride\inst{1} \and
Bruce R. Donald\inst{1,3,4,5}\thanks{Corresponding author: brd+recomb27@cs.duke.edu.
This work was supported by the National Institutes of Health under grant
R35-GM144042 to B.R.D. The PACK* implementation is free and open source and is
available at \url{https://github.com/liyang868/OSPREY3}.}}
\authorrunning{L. Zhu, J. D. Jou, A. C. McBride, and B. R. Donald}
\institute{Department of Computer Science, Duke University, Durham, NC
\and Computational Chemistry, Schr\"odinger, Inc., New York, NY
\and Department of Chemistry, Duke University, Durham, NC
\and Department of Biochemistry, Duke University, Durham, NC
\and Department of Mathematics, Duke University, Durham, NC}

\maketitle
\begin{abstract}
Ensemble-based protein design algorithms rank sequences by binding scores computed from partition functions of the bound complex, receptor, and ligand. Each residue conformation (RC) specifies an amino acid type and bounds on its side-chain dihedral angles. The partition functions sum Boltzmann weights over an RC assignment space that grows exponentially with the number of flexible residue positions. Furthermore, evaluating each assignment's energy requires expensive constrained local minimization to model the continuous side-chain flexibility for quantitative accuracy. We introduce PACK*, a randomized algorithm that formulates partition function evaluation as a statistical estimation problem to drastically reduce the number of expensive local minimizations. To estimate partition functions via importance sampling, we construct a cheaper distribution called the \emph{L-distribution}, over RC assignments. Dynamic programming on the branch decomposition exactly computes its normalization constant and enables independent sampling for importance reweighting. PACK* learns the L-distribution and yields finite-sample confidence intervals for the partition functions. The probability of reporting a confidence interval that meets the user-specified accuracy $\varepsilon$ but fails to contain the true partition function is at most $\delta$. We compared PACK* with MARK*, the previous state-of-the-art algorithm for provable partition function approximation in ensemble-based protein design. Across 873 sequence evaluations for 36 protein--ligand complexes, Spearman correlations between PACK* and MARK* binding scores had a median of 0.996. In a comprehensive scaling benchmark of 136 protein design problems derived from 38 protein--ligand complexes, with increasing numbers of mutable and flexible residue positions, PACK* achieved speedups of up to three orders of magnitude over MARK*.
Using PACK*, we generated and analyzed importance-weighted conformational
ensembles of the complex between SHV-1
$\beta$-lactamase---an enzyme that hydrolyzes $\beta$-lactam antibiotics---and
$\beta$-lactamase inhibitor protein (BLIP). PACK* revealed mutation-induced
reorganization of the side-chain dihedral-angle distributions of SHV-1 D104
following the BLIP E73M substitution. The population of conformations satisfying
a geometric proxy for the salt bridge increased from nearly zero in wild type to
approximately 52\% in E73M, providing an ensemble-based prediction consistent
with the experimentally observed interactions.
\end{abstract}
\thispagestyle{empty}
\clearpage
\pagenumbering{arabic}
\setlength{\introparskip}{\parskip}
\setlength{\parskip}{0pt}
\vspace{-0.1in}
\section{Introduction}
\label{sec:intro}
\vspace{-0.1in}
Ensembles of molecular conformations have long played a central role in structural biology and computational
protein design~\cite{allen2010msdfaster,karimi2018icfn,sauer2020multistate,jing2024alphaflow,wang2024confdiff,lewis2025bioemu,sengar2025ldfpg,sengar2026beyond}.
Ensemble partition functions provide a basis for estimating binding free
energies, which account for both conformational energies and conformational
entropy~\cite{gilson1997}.
Here, we study partition functions over residue conformation (RC) assignments.
An RC specifies an amino acid type and allowed ranges for the residue's
conformational degrees of freedom, such as side-chain dihedral angles.
These ranges define a \emph{voxel} in the residue's conformational space.
A \emph{residue position} is an index in the polymer chain.
An RC assignment selects one RC at every flexible residue position in the polymer chain.
Constrained local minimization within the conformational voxels maps each
RC assignment to a minimized conformation.
The partition function sums the Boltzmann weights associated with the
minimized energies of all RC assignments in the conformational space, with lower energies
receiving larger weights. We call this sum the \emph{target partition function} $q_t$;
normalizing these Boltzmann weights defines the \emph{target distribution} $P_t$.

Within the {\sc Osprey} framework~\cite{hallen2018osprey,long2025path}, the K* family~\cite{lilien2005kstar,ojewole2018bbkstar} pursues
provable ensemble-based protein design by combining conformational flexibility,
ensembles, and guarantees on partition function estimates with respect to the
input model. Outside the {\sc Osprey} framework, tools such as
FoldX~\cite{schymkowitz2005foldx}, Rosetta~\cite{Sora_2022}, and MM-GBSA~\cite{miller2012mmpbsa}
predict mutation effects and estimate binding energetics, but do not provide
these partition function guarantees.
{\sc Osprey}'s ensemble-based design
framework~\cite{gainza2013osprey,hallen2018osprey,Hallen_2015,Hallen_2017} has been used to design
experimentally validated inhibitors that restore CFTR activity in cystic fibrosis~\cite{roberts2012kstar},
design enzymes~\cite{chen2009enzyme},
predict mutations conferring resistance to antibacterial and anticancer
therapies~\cite{reeve2015resistance,guerin2022resistor}, design anti-HIV
antibodies~\cite{rudicell2014vrc07,holt2023v2apex}, and design
binders to KRas~\cite{lowegard2020kras}. K*-based computational inhibitor design
found allosteric inhibitors of the leukemia-associated Runx1--CBF$\beta$
interaction~\cite{gorczynski2007allosteric}. Therapeutics and vaccines designed
using {\sc Osprey} are currently in 34 clinical trials~\cite{dukecancer2025translation}.
The K* family provides \emph{$\varepsilon$-approximations} to partition functions.
For an interval $[\qlo,\qhi]$ guaranteed to contain the partition function
value, define $\varepsilon'=1-\qlo/\qhi$.
An $\varepsilon$-approximation requires
$\varepsilon'\le\varepsilon$, where $\varepsilon$ is the
user-specified accuracy.
Methods such as iMinDEE--A*--K*~\cite{gainza2012continuous} and
MARK*~\cite{jou2020markstar} compute $\varepsilon$-approximations to partition functions by
maintaining deterministic lower and upper bounds on the partition function
and progressively tightening them through costly constrained local
minimizations until the user-specified accuracy is attained.

Partition function intervals are useful not only for quantifying approximation accuracy but also for guiding sequence selection. In ensemble-based protein design, binding scores depend on the partition functions of the bound complex and its unbound states. Bounds on these partition functions can be used to derive bounds on binding scores, allowing sequences to be compared before their partition functions have been computed to high precision. For example, a candidate can be eliminated when its upper binding score bound falls below the lower bound of another candidate. This motivates algorithms that obtain informative partition function bounds to avoid unnecessary computation for less promising sequences.
Tightening partition function bounds entails two coupled computational costs:
the number of RC assignments grows exponentially with the number of flexible
residue positions, and constrained local minimization of each RC assignment is
computationally expensive. Consequently, tightening these bounds becomes
increasingly costly as conformational flexibility increases.
In our benchmarks, MARK*, the fastest existing algorithm with
$\varepsilon$-approximation guarantees for partition function estimation,
required hours to days for designs
involving only about a dozen flexible residue positions. At this cost,
evaluating large sequence spaces can become computationally impractical, which motivates faster evaluation as design spaces grow.

Instead of evaluating enough RC assignments to obtain a deterministic
$\varepsilon$-approximation, we formulate target partition function
evaluation as a statistical estimation problem. Much of the classical Monte
Carlo theory provides asymptotic guarantees~\cite{geweke1989montecarlo,RobertC04},
but for provable protein design, asymptotic convergence alone is insufficient
since it provides no error guarantee at any finite sample size.
We seek finite-sample bounds
that give an interval containing $q_t$
with high probability. We aim to estimate $q_t$ using far fewer
local minimizations of RC assignments. Let $\delta\in(0,1)$ denote the user-specified upper bound on the probability that the constructed confidence interval fails to contain $q_t$.
By relaxing deterministic $\varepsilon$-approximation to a high-probability guarantee,
PACK* retains user-specified accuracy $\varepsilon$ while allowing a small failure
probability $\delta$, drastically reducing computational cost. One general
approach to obtaining finite-sample confidence bounds on $q_t$
is to (i) construct a computable estimator of $q_t$,
(ii) control the sampling distribution and dependence
among samples, and (iii) bound the uncertainty arising from finite sampling.
RC assignments are sampled from the \emph{L-distribution}, also called the
\emph{proposal distribution} $\pe$. This Boltzmann distribution is defined by a
\emph{proposal energy function} $\ee$.
We use importance reweighting~\cite{RobertC04} to estimate the target partition function $q_t$ from
these samples and their normalized probabilities under $\pe$.
PACK* uses independent samples with computable sampling probabilities, as in
previous finite-sample importance sampling
analyses~\cite{block2026partition,lou2017dynamic,chatterjee2018samplesize,agapiou2017importance}.
For a Boltzmann proposal distribution, the proposal partition function $\qe$ is the sum
of its Boltzmann weights and provides the normalizing constant needed to calculate
sampling probabilities. The challenge is to compute $\qe$ efficiently despite
the exponentially large RC assignment space.

We compute the proposal partition function,
because approaches that avoid this computation, such as sequential
Monte Carlo (SMC)~\cite{delmoral2006smc,cerou2011nonasymptotic} and autoregressive
models~\cite{uria2016nade,germain2015made}, have their own limitations.
We address this problem by extending the role of graph decomposition
in protein design. Residue interaction graphs have long been
exploited for efficient side-chain packing in SCWRL~\cite{canutescu2003graph}, TreePack~\cite{xu2006packing}, and
SCWRL4~\cite{krivov2009scwrl4}, while BWM*~\cite{jou2016bwm} uses
branch decomposition to compute the sparse global minimum energy conformation
(GMEC) and enumerate subsequent conformations efficiently in gap-free order of increasing
sparse energy. In the present paper, we extend branch decomposition from energy
minimization and ordered conformation enumeration to exact summation of the
proposal Boltzmann weights over RC assignments.

PACK* uses branch decomposition to compute $\qe$
exactly and draw independent samples from $\pe$.
Exact proposal normalization makes the importance estimator computable, while
independent sampling controls dependence among samples, addressing requirements
(i) and (ii) above. Additive terms in $\ee$ become multiplicative
Boltzmann weight factors after exponentiation. Each factor depends
on the RC at one residue position or on the RCs at the two ends of an
interaction edge. A branch decomposition partitions the graph into
subproblems~\cite{robertson1991branch,jou2016bwm}. These subproblems interact
only through shared residue positions. Once the RCs at shared residue positions
have been determined, the Boltzmann weight sum for each subproblem can be computed
independently. First,
bottom-up sum-product dynamic programming (DP) combines the Boltzmann weight
contributions of child subproblems and sums over RC choices, computing
$\qe$ exactly~\cite{wainwright2008graphical}.
The factors used in the DP recurrence are defined in Section~\ref{sec:normalization-sampling}.
Second, top-down ancestral sampling uses the same DP tables to sample
RC choices according to their conditional probabilities (Eq.~\ref{eq:dp-conditional}), yielding complete
assignments distributed exactly according to $\pe$.
Independent traversals produce i.i.d. RC assignments.
For bounded RC domain sizes, the BWM* bag-size bound~\cite{jou2016bwm}
implies that our sum-product DP assignment count is bounded exponentially
in the width of the chosen branch decomposition and polynomially in the
number of residue positions. The counting and storage formulas are given
in Appendix Section~\ref{SI-sec:structural-proof}.
Efficient exact summation and independent sampling enable PACK* to iteratively
refine the proposal distribution $\pe$ and thereby approximate the target distribution $P_t$.
At each iteration, PACK* draws samples from the current $\pe$ and performs
constrained local minimization to obtain their target energies.
PACK* learns a better $\ee$ from these target energies.
DP then recomputes $\qe$ exactly, yielding the updated, normalized $\pe$.

Finite-sample guarantees for importance reweighting require assumptions that control the contribution of
RC assignments with small sampling probabilities and large importance weights,
which might not be observed~\cite[Section~10.1]{wang2023catoni}.
Apparently well-behaved sampled weights do not rule out rare, unobserved
contributions that substantially affect the mean importance weight under
$\pe$. A proposal distribution may place insufficient
probability mass on RC assignments that actually carry substantial probability
mass under the target distribution, resulting in rare but large importance
weights that can dominate the estimate.
Clipping caps importance weights at a chosen threshold.
To bound the uncertainty from finite sampling, as required by (iii) above,
we assume that, among importance weights exceeding the clipping threshold,
the average amount by which they exceed the threshold, divided by the threshold,
is bounded by a specified constant (Eq.~\ref{eq:severity}).
We call this the \emph{conditional mean excess
assumption}, a mild tail assumption that is weaker than requiring a known uniform upper bound on
individual importance weights. Combined with finite-sample bounds on the mean
clipped importance weight under $\pe$ and on the probability of exceeding the
clipping threshold, this yields a confidence interval
$[\qlo,\qhi]$ for $q_t$. Under the assumptions above,
Theorem~\ref{thm:fs} guarantees that $[\qlo,\qhi]$ contains
$q_t$ with probability at least $1-\delta$.
PACK* reports the interval only if $\varepsilon'\leq\varepsilon$.
We refer to this as a \emph{probably approximately correct} (PAC)-style
guarantee, motivating the name PACK*: Probably Approximately Correct K*.

Our main contributions are as follows:
\begin{list}{\labelenumi}{\usecounter{enumi}%
  \setlength{\topsep}{0pt}\setlength{\partopsep}{0pt}}
\item \textbf{A provable statistical framework for partition function estimation.}
We introduce the first use of branch decomposition in computational protein design
to compute $\qe$ exactly for a tractable proposal distribution
and to generate independent samples from it. These capabilities enable importance
reweighting with finite-sample confidence intervals for $q_t$
under a mild tail assumption on the importance weights.
Complete proofs are available in the appendix.

\item \textbf{Empirical reliability and computational acceleration.}
We validate PACK* against exact partition functions obtained by exhaustive
enumeration, with all 900 confidence intervals containing their targets and
meeting the specified accuracy. PACK* also achieves strong agreement with MARK*
in sequence ranking, with a median Spearman correlation of
\stat{common1spearmanshort}. While maintaining this empirical reliability, PACK*
achieves speedups of up to \stat{runtimeboundmax1}-fold over MARK* across a
large-scale protein design benchmark.

\item \textbf{An open-source implementation.}
We implement PACK* in the OSPREY software suite and make it freely available,
providing a practical tool for ensemble-based protein design.

\item \textbf{Generation and analysis of sequence-dependent conformational ensembles.}
Beyond partition function estimation, we use PACK* to generate
importance-weighted conformational ensembles and infer sequence-dependent
changes in side-chain conformations. In the complex of SHV-1
$\beta$-lactamase and $\beta$-lactamase inhibitor protein (BLIP), we compare
wild type (WT) and BLIP E73M and identify mutation-induced redistribution of D104
side-chain conformations toward a contact-forming rotamer basin.
The population of conformations satisfying a geometric proxy for the salt
bridge increased from nearly zero in WT to approximately 52\% in the E73M mutant,
providing an ensemble-based prediction consistent with the experimentally
observed interactions (Section~\ref{sec:mi-analysis}).

\end{list}

\setlength{\parskip}{\introparskip}

\setlength{\textfloatsep}{10pt plus 2pt minus 2pt}
\setlength{\intextsep}{6pt plus 2pt minus 2pt}
\setlength{\abovecaptionskip}{3pt}
\begin{figure}[!htbp]
\centering
\includegraphics[width=0.996\linewidth]{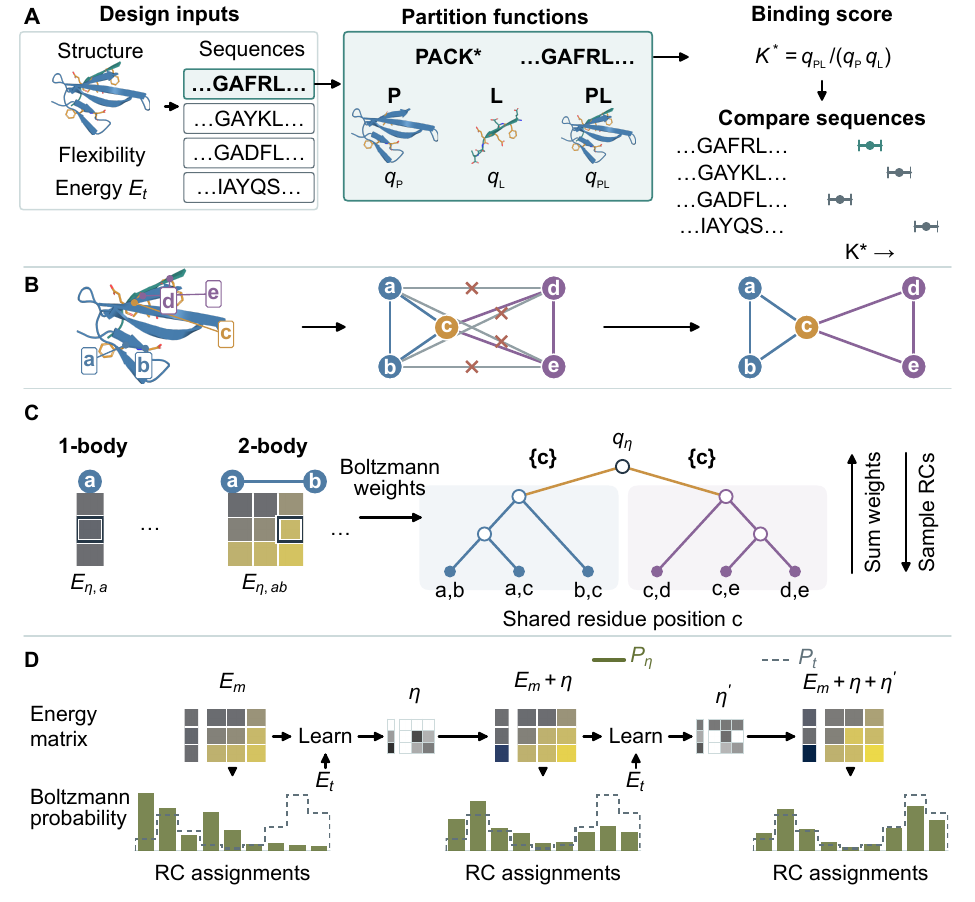}
\input{overview_caption_restored_20261011.tex}
\end{figure}
\vspace{-0.10in}
\section{PACK*: Algorithm and Methodology}
\label{sec:background}
\label{sec:algorithm}
\vspace{-0.05in}
\subsection{Problem formulation}
\vspace{-0.05in}
For a given sequence and molecular state, specify the structure,
 temperature $T>0$, interaction graph, force field, and
cyclic coordinate descent (CCD) protocol.
Let $V=\{1,\ldots,d\}$ index residue positions and
$R_i$ be the allowed RC set at residue position $i$.
An RC $r\in R_i$ at residue position $i$ specifies an amino acid type and
bounds on each conformational degree of freedom~\cite{hallen2016comets,bouchiba2021flexibility}.
These bounds define a voxel $\Omega(r)=I_1(r)\times\cdots\times I_k(r)$,
where $I_j(r)$ is the allowed interval for the $j$th degree of freedom.
In this work, these degrees of freedom are side-chain dihedral angles,
each allowed to vary within $\pm9^{\circ}$ of its RC angle center.
Let $C\subseteq\prod_{i\in V}R_i$ be the finite set of allowed RC assignments.
For the force field used here, the energy of a conformation is a sum of constant,
1-body, and 2-body terms. To evaluate the target energy $E_t(c)$, we jointly
minimize the total energy over the flexible side-chain dihedral angles within
the RC voxels specified by the complete assignment $c$.
The individual energy terms can then be evaluated at the
resulting jointly minimized coordinates, and their sum gives $E_t(c)$.
The target partition function and binding score are
\begin{equation}
 q_t=\sum_{c\in C}e^{-E_t(c)/RT},\qquad
 K^*(s)=\frac{\qPL(s)}{\qP(s)\qL(s)}.
 \label{eq:target}
\end{equation}
Here $\mathrm P$, $\mathrm L$, and $\mathrm{PL}$ denote the unbound protein,
unbound ligand, and complex, each with its own RC assignment space, energy
function, and positive finite partition function.
The target distribution has probability mass function $P_t(c)=e^{-E_t(c)/RT}/q_t$.

The residue interaction graph has residue positions in $V$ as vertices and
2-body interactions as edges. Let $\mathcal S_{t,2}$ and $\mathcal S_2$ denote
the edge sets used for $E_t$ and $E_m$, respectively. Both energies can be
decomposed into constant, 1-body, and 2-body contributions:
\begin{equation}
 E_t(c)=E_{t,0}+\sum_{i\in V}E_{t,i}(c)
 +\sum_{\{i,j\}\in\mathcal S_{t,2}}E_{t,ij}(c),
 \qquad
 E_m(c)=E_{m,0}+\sum_{i\in V}E_{m,i}(c_i)
 +\sum_{\{i,j\}\in\mathcal S_2}E_{m,ij}(c_i,c_j).
 \label{eq:pairwise-minimized}
\end{equation}
Here $E_{t,0}$ and $E_{m,0}$ are constant terms. The contributions
$E_{t,i}(c)$ and $E_{t,ij}(c)$ are evaluated at the jointly minimized
coordinates of the complete assignment $c$. The matrix entries
$E_{m,i}(c_i)$ and $E_{m,ij}(c_i,c_j)$ are instead obtained by minimizing
each corresponding energy term separately within its RC voxel or pair of voxels.
The pairwise-minimized energy matrix contains the constant $E_{m,0}$,
1-body vectors $E_{m,i}$ with $|R_i|$ entries, and 2-body matrices
$E_{m,ij}$ of size $|R_i|\times|R_j|$ on retained edges.
Complete RC assignments using the same 1-body or 2-body matrix entry can
yield different minimized side-chain dihedral angles and different values
of the corresponding physical energy term.
Removing physical interactions from the energy model changes both $E_t$ and
the corresponding $E_m$~\cite{jain2017critical}.
Alternatively, PACK* can use a sparse residue interaction graph for $E_m$ and
the learned corrections while retaining all physical interactions during the
minimization and evaluation of $E_t$.
\newpage
\vspace{-0.1in}
\subsection{Exact normalization and independent sampling}

\label{sec:normalization-sampling}
The proposal energy function is
\begin{equation}
 \ee(c)=E_m(c)+\eta(c),\qquad
 \eta(c)=\eta_0+\sum_{i\in V}\eta_i(c_i)
 +\sum_{\{i,j\}\in\mathcal S_2}\eta_{ij}(c_i,c_j).
 \label{eq:proposal-energy}
\end{equation}
The finite corrections comprise a constant $\eta_0$, 1-body correction vectors $\eta_i$
with $|R_i|$ entries, and 2-body correction matrices $\eta_{ij}$ of size
$|R_i|\times|R_j|$ on retained edges.
These corrections are fitted as described in Section~\ref{sec:proposal-learning},
with initialization and regression details in Appendix Section~\ref{SI-sec:proposal-fitting}.
A factor's \emph{scope} is the set of residue positions on which it depends.
Fitting changes the correction values while preserving their scopes and the
retained interaction graph.
Let $\xi(c)=E_t(c)-\ee(c)$. Define the proposal partition function
(normalizing constant) $\qe$, proposal distribution $\pe$, and importance
weight $w$ by
\begin{equation}
 \qe=\sum_{c\in C}e^{-\ee(c)/RT},\qquad
 \pe(c)=e^{-\ee(c)/RT}/\qe,\qquad
 w(c)=e^{-\xi(c)/RT}.
 \label{eq:proposal}
\end{equation}
The initial proposal distribution $P_m$ uses $\eta=0$.
For importance reweighting, assume $0<\qe<\infty$ and $\pe(c)>0$ for every
$c\in C$ with $P_t(c)>0$.

A branch decomposition represents the retained interaction edges as leaves of
a binary tree~\cite{jou2016bwm,robertson1991branch}.
Cutting a tree edge divides the interactions into two subproblems; residue positions
shared by both form its separator. A DP table stores the sum of subproblem
Boltzmann weights for each separator assignment; once the separator RCs are
assigned, child subproblems can be summed independently.
Let $\mathcal T$ denote the edges used for DP on the rooted branch decomposition tree.
Write $p_0$ for the root edge, $M(p)$ for the separator, $\lambda(p)$ for the
$\lambda$-set of residue positions, and $\operatorname{ch}(p)$ for the
child edges.
Each residue position is assigned once. Separators satisfy running intersection,
and every energy factor and hard constraint is allocated exactly once to
a bag $M(p)\cup\lambda(p)$ containing its scope. The root separator is empty.
Construction details are given in Appendix Section~\ref{SI-sec:sampling-proof}.

Let $m$ and $\ell$ be RC assignments to $M(p)$ and $\lambda(p)$, respectively,
with $\ell\in\Lambda_p=\prod_{i\in\lambda(p)}R_i$. Define the local factor
$\psi_p(m,\ell)=\ind_p(m,\ell)e^{-E_p(m,\ell)/RT}$, where $E_p$ sums
the energies allocated to the bag and $\ind_p$ is the product of its
zero/one constraint factors. For each child $f$, let $m_f(m,\ell)$ be the
restriction of $(m,\ell)$ to $M(f)$. The DP table entries satisfy the
sum-product recurrence
\begin{equation}
 q_p(m)=\sum_{\ell\in\Lambda_p}\psi_p(m,\ell)
          \prod_{f\in \operatorname{ch}(p)}q_f(m_f(m,\ell)),\qquad
 \qe=q_{p_0}(\varnothing).
 \label{eq:dp}
\end{equation}

At each separator assignment $m$ reached during top-down sampling, $q_p(m)>0$,
and PACK* samples $\ell$ with probability
\begin{equation}
 P_p(\ell\mid m)=\frac{\psi_p(m,\ell)
                     \prod\limits_{f\in\operatorname{ch}(p)}q_f(m_f(m,\ell))}{q_p(m)}.
 \label{eq:dp-conditional}
\end{equation}
Because each child DP table sums over all remaining RC assignments in its
subtree given its separator assignment, this rule gives the conditional
probability under the proposal distribution $\pe$.
Independent top-down traversals implement ancestral
sampling~\cite{wainwright2008graphical}, yielding i.i.d. RC assignments for
estimating $q_t$. Exactness assumes exact arithmetic.

\begin{proposition}[Exact normalization and independent sampling]
\label{lem:sample}
Under the stated conditions for factor allocation, with $0<q_\eta<\infty$,
Eq.~\eqref{eq:dp} computes $q_\eta$ exactly. Applying the conditional sampling
rule in Eq.~\eqref{eq:dp-conditional} recursively in top-down order generates
a complete RC assignment from $P_\eta$. Independent traversals using
fresh randomness yield independent complete RC assignments.
\end{proposition}

\vspace{-0.1in}
\subsection{Learning and selecting proposal energy functions}
\label{sec:proposal-learning}
\vspace{-0.05in}
PACK* learns and selects $\ee$ to reduce the number of
$E_t$ evaluations needed to attain $\varepsilon$.
Using initial draws from $P_m$, PACK* initializes corrections from termwise
energy differences at the minimized conformations. Regularized weighted
regression jointly refines the constant, 1-body, and retained 2-body
corrections to approximate $E_t(c)-E_m(c)$. Unobserved corrections remain zero
(Fig.~\ref{fig:pipeline}D).

Two-fold cross-validation compares values of the 1-body and 2-body shrinkage
parameters using held-out importance weight statistics, accounting for weight
variability and tail behavior. Among fits passing screening, PACK* selects
the one with the lowest predicted sample count needed to attain
$\varepsilon'\le\varepsilon$. Fitting, screening, fallback selection, and
sample count prediction are detailed in Appendix Section~\ref{SI-sec:proposal-fitting}.

Pilot sampling assesses whether $\varepsilon$ is predicted to be
attainable within the sample budget; failed pilot samples supply additional
training observations for refitting within budget. After a passing pilot,
$\ee$ remains unchanged, and independent validation determines a final
sample count $n$. Conditional on all preceding observations and choices,
a fresh batch of $n$ i.i.d. RC assignments from $\pe$ supplies both
a confidence interval $[\qlo,\qhi]$ and a hypothesis test of the conditional mean excess
assumption. PACK* reports $[\qlo,\qhi]$ only if validation succeeds, the final
hypothesis test does not reject the assumption, and $\varepsilon'\le\varepsilon$.
Appendix Algorithm~\ref{SI-alg:packstar} gives the
complete procedure.
\vspace{-0.1in}
\section{PAC-style guarantees for partition function estimation}
\vspace{-0.1in}
\label{sec:theory}

Here our goal is $q_t$ itself instead of
the partition function ratios in conventional free energy perturbation (FEP)
applications. The standard importance reweighting identity~\cite{zwanzig1954,RobertC04}
connects them:
\begin{equation}
 q_t=\qe\E_{P_\eta}[w].
 \label{eq:identity}
\end{equation}
For $n$ independent RC assignments $c_1,\ldots,c_n$ from $\pe$, let
$\bar w=n^{-1}\sum_i w(c_i)$ and $\widehat q_t=\qe\bar w$.
The estimator $\widehat q_t$ is conditionally unbiased for a given
$\pe$ (Appendix Lemmas~\ref{SI-lem:a2}--\ref{SI-lem:a3}). This identity requires no
bound on the accuracy of the learned energy correction.

We separate each weight into a bounded part and an excess part.
Choose a positive clipping threshold $B$ before final
sampling and define
\begin{equation}
 y(c)=\min\!\left(\frac{w(c)}{B},1\right),\qquad
 u(c)=\max\!\left(\frac{w(c)}{B}-1,0\right).
 \label{eq:bulk-tail}
\end{equation}
Then $w(c)=B(y(c)+u(c))$, and the importance reweighting identity gives
\[
 q_t=\qe B\bigl(\E_{P_\eta}[y]+\E_{P_\eta}[u]\bigr).
\]
The bounded part permits concentration bounds; bounding $\E_{P_\eta}[u]$
requires an additional assumption.
Write $\pi=\Prob_{P_\eta}(u>0)$ for the exceedance probability.
For a specified conditional mean excess bound $m_u>0$, assume
\begin{equation}
 \E_{P_\eta}[u\mid u>0]\le m_u\quad\text{when }\pi>0.
 \label{eq:severity}
\end{equation}
Then $0\le\E[u]\le m_u\pi$; when $\pi=0$, $\E[u]=0$.

The choice of $\eb=\ef=\delta/6$ per molecular state gives total failure
probability at most $\delta$ for one specified sequence.
For $n\ge2$ final samples, let $[\lb,\ub]\subseteq[0,1]$ be the two-sided
empirical Bernstein confidence interval for $\E[y]$~\cite{maurer2009empirical},
with coverage probability at least $1-\eb$.
Let $\pup$ be the one-sided Clopper--Pearson upper confidence bound for
$\pi$~\cite{clopper1934}, with coverage probability at least $1-\ef$.
Their explicit formulas are given in the appendix
(Eqs.~\eqref{SI-eq:si-eb}--\eqref{SI-eq:si-cp}).
Let $h$ denote all observations and choices made before final sampling,
including energy function fitting, pilot sampling, and validation.

\begin{theorem}[PAC-style guarantees after learning]
\label{thm:fs}
Consider $E_t$ with deterministic evaluation. Condition on the
history $h$ before final sampling, which determines $\pe$ satisfying
$\pe(c)>0$ for every $c\in C$ with $P_t(c)>0$, its exact
normalizing constant $\qe$, $B$, $m_u$,
$n\ge2$, and error probabilities
$\eb,\ef>0$ with $\eb+\ef<1$. If Eq.~\eqref{eq:severity} holds and the final
samples are conditionally independent with distribution $P_\eta$, then
\begin{equation}
 [\qlo,\qhi]=[\qe B\lb,\ \qe B(\ub+m_u\pup)],\qquad
 \Prob(q_t\in[\qlo,\qhi]\mid h)\ge1-\eb-\ef.
 \label{eq:fs-interval}
\end{equation}
Consequently, conditional on $h$, the probability of both returning
$[\qlo,\qhi]$ with $\varepsilon'\le\varepsilon$ and having lower endpoint relative
estimation error greater than $\varepsilon$ is at most $\eb+\ef$.
\end{theorem}
Under the conditional mean excess assumption, $\pi\le\pup$ implies
$\E[u]\le m_u\pup$. Combining this bound with the empirical Bernstein interval
for $\E[y]$ and applying a union bound gives the conditional coverage bound.
All learning precedes the independent final batch, so conditioning on $h$
avoids a union bound over the proposal distributions considered during
learning. When the theorem's premises hold almost surely, averaging over $h$
gives an unconditional lower bound on coverage probability.

The bound controls the joint probability of return and excessive
relative estimation error, rather than the error probability conditional on
return. Each attempt uses one final batch of predetermined size for both the
interval and hypothesis test.

\setlength{\textfloatsep}{10pt plus 2pt minus 2pt}
\setlength{\intextsep}{6pt plus 2pt minus 2pt}
\setlength{\abovecaptionskip}{3pt}
\begin{figure}[p]
\centering
\includegraphics[width=\textwidth]{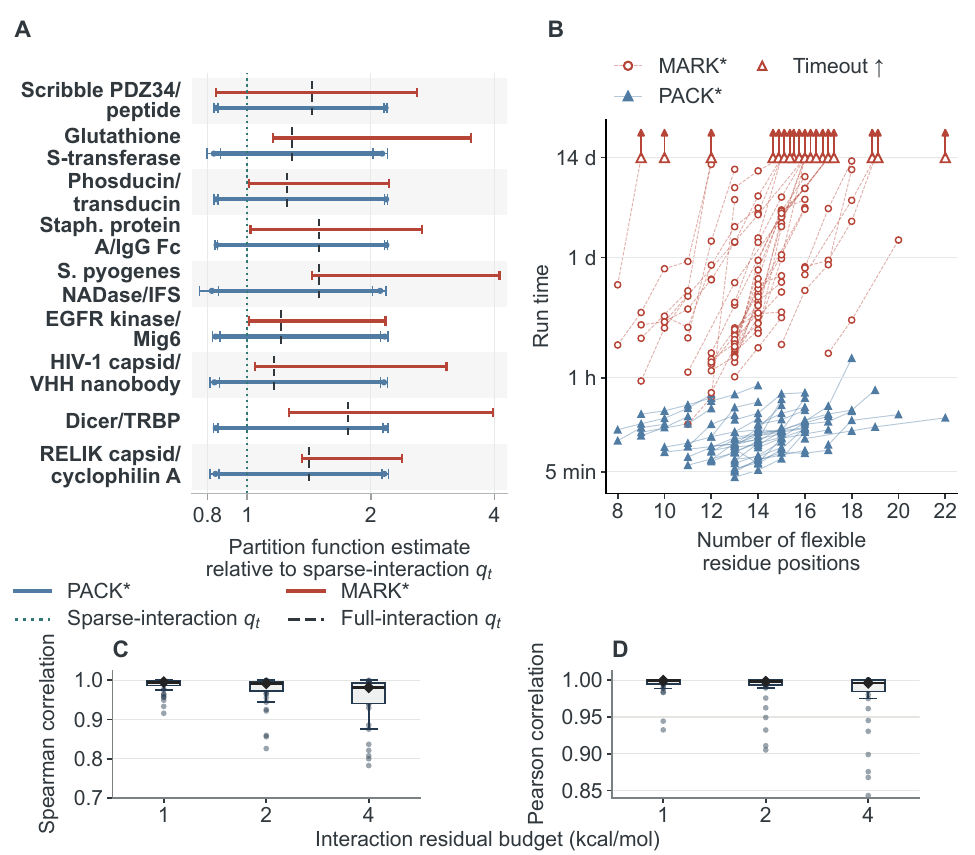}
\par\vspace{-3pt}
\renewcommand{\small}{\fontsize{10}{10.6}\selectfont}
\caption[PACK* estimates are accurate and efficient.]{\textbf{PACK* estimates are accurate and efficient.}
\textbf{(A)} Each row represents one of nine protein conformational spaces
with 5--8 flexible residue positions, small enough for exhaustive enumeration.
Here $q_t$ denotes the partition function.
Black dashed segments: exact $q_t$ computed with all physical interactions (full-interaction $q_t$).
Green dotted vertical line at 1: exact $q_t$ after interaction sparsification (sparse-interaction $q_t$).
All partition function values and interval endpoints are divided by the
sparse-interaction $q_t$ for that row; the horizontal axis is logarithmic.
Across 100 independently seeded runs per row, every PACK* interval contained
the exhaustive full-interaction and sparse-interaction $q_t$ values.
Blue bars: median lower and upper endpoints of these
100 intervals; endpoint error bars: their minima and maxima.
Red bars: MARK* intervals.
\textbf{(B)} \textbf{Runtimes for \stat{runtimedesigns} protein design problems from
\stat{runtimesystems} protein--ligand complexes.}
Blue filled triangles: completed PACK* runs.
Red open circles: completed MARK* runs.
Red open triangles with upward vertical arrows: MARK* runs that did not
complete within 14 days.
Blue solid lines connect PACK* runs for the same complex as flexibility
increases; red dashed lines connect MARK* runs.
The interaction residual budget (RB) controls sparsification by limiting
the strengths of removed residue-pair interactions. PACK* used RB=1 kcal/mol.
\textbf{(C)} Spearman correlations between PACK* and MARK* sequence rankings
for \stat{common1systems} protein--ligand complexes at RB=1, 2, and 4 kcal/mol.
\textbf{(D)} Pearson correlations between PACK* and MARK*
$\log_{10}K^*$ binding scores.
In C--D, each point represents one complex. Boxes show the interquartile
range; lines within boxes and black diamonds mark medians. Whiskers extend
to the most extreme observations within 1.5 interquartile ranges beyond
the box.}
\label{fig:census-reference}
\end{figure}

\begin{corollary}[Binding score intervals from partition functions for molecular states]
\label{cor:kstar}
When all three molecular state intervals contain their respective partition
functions and both denominator lower bounds are positive,
\begin{equation}
 \frac{\qPL^{\ominus}}{\qP^{\oplus}\qL^{\oplus}}
 \le K^*\le
 \frac{\qPL^{\oplus}}{\qP^{\ominus}\qL^{\ominus}}.
 \label{eq:kstar-bounds}
\end{equation}
The joint failure probability is at most the sum of $\eb+\ef$ over the three
molecular states $\mathrm P$, $\mathrm L$, and $\mathrm{PL}$.
\end{corollary}
\vspace{-0.1in}

\section{Experiments}
\vspace{-0.1in}
\label{sec:experiments}
We tested partition function intervals against exhaustive enumeration,
binding score agreement with MARK*, and runtime as flexibility increased
(Fig.~\ref{fig:census-reference}). A separate experiment examined mutation-induced
side-chain conformational changes and contact formation at a binding interface.
The binding calculations used $\varepsilon=0.683$ for each molecular state,
with $\delta=0.05$ as the overall failure probability bound across the three
states. The interaction residual
budget (RB, kcal/mol) limits the sum of maximum absolute 2-body energies over
deleted physical interactions, using both reference energy matrices.
Appendix Sections~\ref{SI-sec:s0-choice} and \ref{SI-sec:experiment-details}
give the statistical assumptions, hypothesis tests, and protocols.
\vspace{-0.1in}
\subsection{Agreement with exhaustive enumeration}
\vspace{-0.1in}
We exhaustively enumerated 9 systems with 5--8 flexible residue positions and
1.0--10.1 million RC assignments, and compared PACK* against the exact
partition function for the sparse-interaction energy function used in each run. Across 100 independent seeds per system,
all 900 PACK* intervals contained the exhaustive sparse-interaction partition
function and satisfied $\varepsilon'\le\varepsilon$.
Across the nine systems, the median $\varepsilon'$ over the 100 seeds ranged
from 0.6111 to 0.6142 (Fig.~\ref{fig:census-reference}A).
All PACK* and MARK* intervals also contained the corresponding exhaustive
value with all interactions.
\vspace{-0.1in}
\subsection{Binding score agreement under interaction sparsification}
\label{sec:binding-score-agreement}
\vspace{-0.1in}
A 38-complex sweep tested how interaction sparsification affected sequence
ranking relative to MARK*. PACK* used RB values of 1, 2, and 4~kcal/mol and constant, 1-body, and
2-body learned energy corrections, with $\tau=1$ kcal/mol. RB changed the
retained physical interaction graph in these runs. We compared the same
\stat{common1pairs} complex--sequence pairs across \stat{common1systems}
complexes at every RB value, requiring finite scores and partition function
estimates for all three molecular states in each run.
For each complex, we computed Spearman and Pearson correlations across sequences.
At RB=1~kcal/mol, the median correlations across complexes were
\stat{common1spearmanshort} and \stat{common1pearson}, respectively.
Increasing RB to 4~kcal/mol reduced median rank correlation to
\stat{common4spearmanshort}.
The median Spearman correlation at RB=1~kcal/mol was comparable to that between
deterministic A*--K* and MARK* (0.997; Appendix Section~\ref{SI-sec:deterministic-ranking}
and Appendix Table~\ref{SI-tab:deterministic-ranking}), supporting the practical reliability
of PACK*'s sequence rankings.
\vspace{-0.1in}
\subsection{Computational costs as flexibility increases}
\label{sec:computational-costs}
\vspace{-0.1in}
We evaluated how the runtimes of PACK* and MARK* scaled with the number
of flexible residue positions, using \stat{runtimedesigns} design problems from
\stat{runtimesystems} protein--ligand complexes
(Fig.~\ref{fig:census-reference}B). Runs for the same design problem used the same
structure, mutable and flexible residue positions, sequence list, and accuracy.
MARK* used full interactions and 64 CPU, its maximum supported parallel
CPU count. PACK* used RB=1~kcal/mol and 128 CPU with zero or four GPUs.
Both PACK* and MARK* performed constrained local minimization on CPUs
because under our available hardware resources, CPU execution provided
higher aggregate CCD throughput than GPU execution.
PACK* used GPUs only for its DP computations.
For the \stat{runtimepairs} problems with completed timings
from both methods, PACK* was \stat{runtimeratiomin1}--\stat{runtimeratiomax1}
times faster under these resource and interaction settings.
\stat{runtimecensored} MARK* runs reached the 14-day limit, yielding speedup lower bounds of
\stat{runtimeboundmin1}--\stat{runtimeboundmax1}
(Fig.~\ref{fig:census-reference}B; Appendix Table~\ref{SI-tab:frontier-pairs}).

\vspace{-0.1in}
\subsection{Generation and Analysis of Mutation-Dependent Conformational Ensembles}
\label{sec:mi-analysis}
\begin{figure}[!htbp]
\centering
\includegraphics[width=\textwidth]{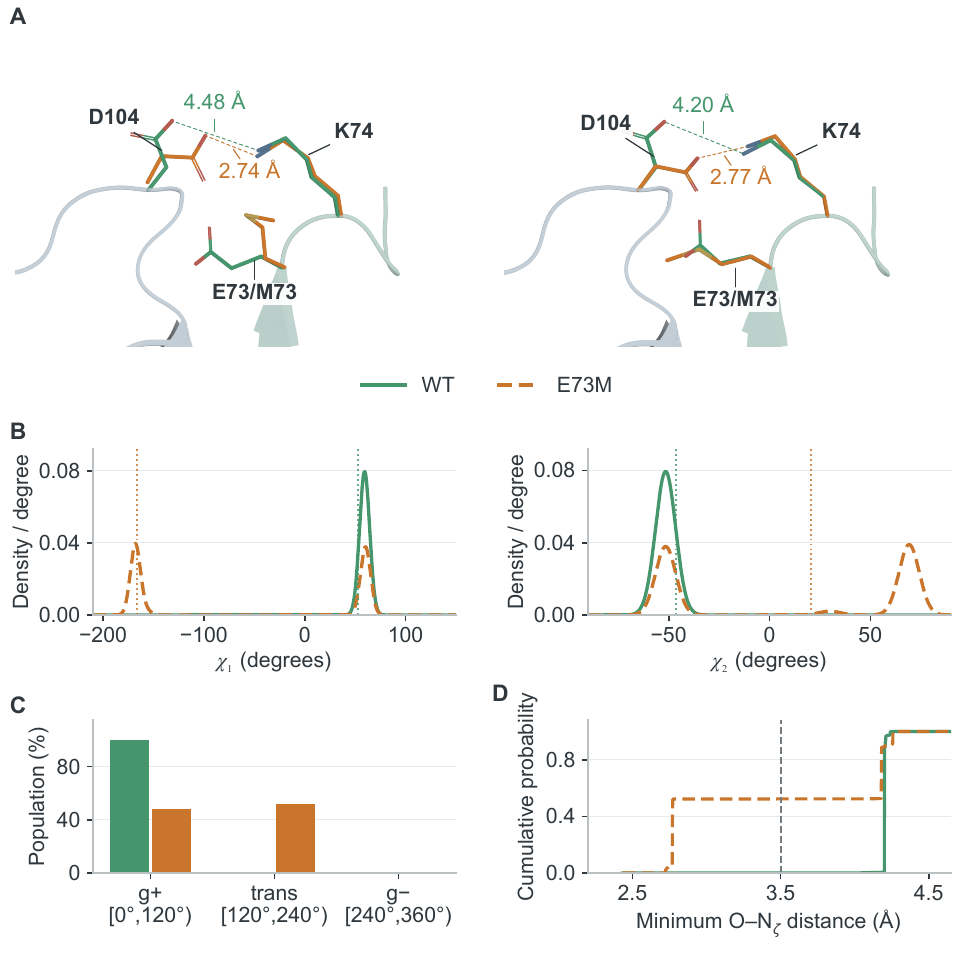}

\par\vspace{-11pt}
\caption{PACK* generated importance-weighted conformational ensembles and
identified D104 and K74 as the residues with the largest and second-largest
mutation-induced conformational shifts in SHV-1--BLIP.
D104 distributions are shown here; K74 angular distributions are in
Appendix Fig.~\ref{SI-fig:within-rc-redistribution}. WT denotes wild type.
\textbf{(A) Structural context.} Left: crystal side chains of WT
(PDB ID: 2G2U) and the BLIP E73M mutant (PDB ID: 3C4P).
Dashed lines show $d$, the minimum distance from K74 $\mathrm{N}_{\zeta}$
to either D104 carboxylate oxygen. Right: lowest-energy sampled WT
conformation and lowest-energy sampled E73M conformation satisfying
$d\le 3.5$~\AA. Carbon atoms: green, WT; orange, E73M.
Oxygen: red; nitrogen: blue; sulfur: gold.
\textbf{(B) D104 angles.} D104 $\sidechi{1}$ and $\sidechi{2}$ distributions
after constrained local minimization across all residue conformations (RCs),
computed using importance-weighted periodic kernel density estimation
(KDE)~\cite{dimarzio2011torus} with $5^{\circ}$ bandwidth.
Green solid curves: WT; orange dashed curves: E73M.
Green vertical dotted lines: WT crystal angles; orange vertical dotted
lines: E73M crystal angles. $\sidechi{2}$ has a $180^{\circ}$ period because
the terminal oxygens are equivalent.
\textbf{(C) D104 $\sidechi{1}$ basins.} Importance-weighted populations
in the gauche+ (g$^+$), trans, and gauche$-$ (g$^-$) basins.
Angle ranges are shown on the horizontal axis, with $\sidechi{1}$ taken modulo $360^{\circ}$.
\textbf{(D) Contact geometry.} Importance-weighted cumulative distribution
of $d$. Vertical dashed line: $d=3.5$~\AA\ contact threshold.
Panels C and D show statistics computed from importance-weighted samples
without KDE smoothing.}
\label{fig:mi-analysis}
\end{figure}

\vspace{-0.1in}
We used PACK* to generate importance-weighted conformational ensembles
for the SHV-1 $\beta$-lactamase--BLIP complex in its WT and BLIP E73M forms.
We then compared their side-chain conformational distributions to
characterize the response to mutation E73M. Reynolds et
al.~\cite{reynolds2008shv} observed a D104--K74 salt bridge in the E73M
crystal structure that was absent in WT and missed by their original
design calculation. From the WT to the E73M crystal structure (PDB IDs
2G2U and 3C4P, respectively), the minimum distance between a D104 carboxylate oxygen and
the terminal K74 $\mathrm{N}_{\zeta}$ atom shortened from
4.48 to 2.74~\AA\ (Fig.~\ref{fig:mi-analysis}A).
We compared both sequences structurally using 13 flexible
residue positions. Constrained minimization 
retained all residue interactions.

PACK* identified D104 and K74 as the residues with the largest and second-largest
mutation-induced shifts in their side-chain conformational distributions,
respectively (Fig.~\ref{fig:mi-analysis}B; Appendix Table~\ref{SI-tab:shv-jsd-ranking}).
We computed the side-chain angular distributions using importance-weighted periodic
kernel density estimation (KDE)~\cite{dimarzio2011torus} with a
bandwidth of $5^{\circ}$.
Vertical green and orange dotted lines in Fig.~\ref{fig:mi-analysis}B show
the corresponding D104 dihedral angles in the experimental WT (2G2U) and
E73M (3C4P) crystal structures, respectively.
The mutation-induced shifts in the D104 and K74 side-chain conformational
distributions revealed two distinct scales of conformational response,
consistent with the structural observations~\cite{reynolds2008shv}:
a larger reorientation of D104 and more localized
adjustments of K74. The E73M mutation increased the population of D104
conformations in the trans $\sidechi{1}$ basin relative to the
wild-type-dominant gauche+ (g$^+$) basin, favoring conformations in which
the D104 carboxylate group approaches the K74 ammonium group
(Fig.~\ref{fig:mi-analysis}C,D). In contrast, K74 exhibited
changes in its side-chain dihedral angles primarily within existing
rotamer basins, with little redistribution of probability between basins.

To quantify the proximity of these residues, we defined $d$ for each
conformation as the minimum distance between the K74 $\mathrm{N}_{\zeta}$
atom and either carboxylate oxygen of D104,
$\mathrm{O}_{\delta 1}$ or $\mathrm{O}_{\delta 2}$:
\[
 d=\min\bigl\{
 \operatorname{dist}(\mathrm{D104}\,\mathrm{O}_{\delta 1},\mathrm{K74}\,\mathrm{N}_{\zeta}),
 \operatorname{dist}(\mathrm{D104}\,\mathrm{O}_{\delta 2},\mathrm{K74}\,\mathrm{N}_{\zeta})
 \bigr\},
\]
and classified conformations with $d\le 3.5$~\AA\ as satisfying a salt-bridge contact
proxy~\cite{lenguyen2025saltbridges}.
Accordingly, the weighted population of the contact proxy increased from
0.000464\% in WT to 52.24\% in E73M.
This increase agreed qualitatively with the experimentally observed
formation of the D104--K74 salt bridge~\cite{reynolds2008shv}.
A closer examination of the D104 dihedral angles showed that the increased
salt-bridge contact population was primarily associated with $\sidechi{1}$
reorientation: 52.24\% of the E73M ensemble adopted $\sidechi{1}$ angles within
$9^{\circ}$ of the mutant crystal value of $-166.31^{\circ}$.
The $\sidechi{2}$ distribution also shifted toward the crystallographic angle,
with 49.44\% of the E73M population concentrated near $69.2^{\circ}$,
compared with the WT peak near $-51.5^{\circ}$, although only 2.80\% lay
within $9^{\circ}$ of the mutant crystal angle of $20.46^{\circ}$,
accounting for terminal oxygen symmetry.
This incomplete recovery of the precise $\sidechi{2}$ geometry may reflect
limitations of the underlying biophysical model, including the
backbone, rather than random-seed variability in PACK* sampling, since similar
angular distributions were obtained across ten random seeds.
These PACK* results robustly connected mutation-induced conformational
redistribution with contact propensity.
Appendix Section~\ref{SI-sec:mi-analysis} gives the basin definitions,
K74 angle distributions, complete statistics, and bandwidth sensitivity.

\vspace{-0.1in}
\section{Conclusion}
\vspace{-0.1in}
Previously, neural Boltzmann generators learned expressive generative models
to approximate target
equilibrium ensembles~\cite{noe2019boltzmann,zheng2024equilibrium,akhoundsadegh2024iterated,plainer2025consistent}.
In contrast, PACK* takes a complementary approach: it learns corrections
that preserve the factorization of the proposal energy function $\ee$
and exploits that factorization to normalize the proposal exactly and
generate independent RC assignments. Constrained local minimization
and importance reweighting then produce conformational ensembles and
finite-sample confidence intervals for the target partition function
under a mild tail assumption.

In comprehensive benchmarks, PACK* quickly produced partition function intervals
that contained the exact target values and achieved strong agreement with MARK* sequence rankings.
In particular, PACK* accelerated calculations by up to three orders of
magnitude. While MARK* relies on a priority queue that is less amenable
to GPU acceleration~\cite{zhou2014parallel}, PACK* supports concurrent
batches of constrained minimizations
and GPU-accelerated dynamic programming (DP). This architectural
reformulation enables the observed speedups by combining fewer
expensive target energy evaluations with parallel computation.

By generating importance-weighted ensembles for WT and E73M SHV-1--BLIP
complexes, PACK* inferred two distinct conformational responses to the
BLIP E73M mutation: redistribution between D104 rotamer basins and
angular adjustments primarily within K74 rotamer basins. In the
generated E73M ensemble, approximately 52\% of the population satisfied
the D104--K74 salt-bridge contact proxy, compared with nearly zero percent in WT.
These ensemble predictions were consistent with the experimentally
observed salt bridge and illustrate how PACK* can generate and analyze
conformational ensembles to explain mutation-induced changes in protein
interactions.
\clearpage
\bibliographystyle{splncs04}
\bibliography{refs}

\PackstarBeginAppendix
\newgeometry{
    left=1.05in,
    right=1.05in,
    top=1.05in,
    bottom=1.10in
}

\title{Optional Appendix:\\Complete Proofs and Supporting Results}
\titlerunning{Optional Appendix: Complete Proofs}
\author{Liyang Zhu\inst{1} \and Jonathan D. Jou\inst{2} \and
Allen C. McBride\inst{1} \and
Bruce R. Donald\inst{1,3,4,5}\thanks{Corresponding author: brd+recomb27@cs.duke.edu.
This work was supported by the National Institutes of Health under grant
R35-GM144042 to B.R.D. The PACK* implementation is free and open source and is
available at \url{https://github.com/liyang868/OSPREY3}.}}
\authorrunning{L. Zhu, J. D. Jou, A. C. McBride, and B. R. Donald}
\institute{Department of Computer Science, Duke University, Durham, NC
\and Computational Chemistry, Schr\"odinger, Inc., New York, NY
\and Department of Chemistry, Duke University, Durham, NC
\and Department of Biochemistry, Duke University, Durham, NC
\and Department of Mathematics, Duke University, Durham, NC}

\maketitle
Here we provide an optional appendix. This appendix gives complete
proofs of exact computation of proposal partition functions, independent sampling, and the
statistical guarantees for a specified target partition function over a finite set of residue
conformation (RC) assignments (Appendix Sections~\ref{sec:sampling-proof}--\ref{sec:fs-proof}).
Appendix Section~\ref{sec:structural-proof} details computational costs and proposal energy
function selection, and Appendix Section~\ref{sec:free-energy-proof} derives free energy and
binding score bounds. Appendix Section~\ref{sec:experiment-details} gives experimental protocols
and additional results, including mutation-induced changes in residue side-chain dihedral-angle
distributions (Appendix Section~\ref{sec:mi-analysis}).

\section{Algorithm and the separation of sampling stages}
\label{sec:algorithm-overview}
Algorithm~\ref{alg:packstar} describes adaptive fitting of the proposal energy function and
estimation using clipped weights for one sequence and molecular state.
The user specifies the flexible residue positions, their allowed RCs, and
the physical interaction model. The chosen sparsification parameters
determine the retained interactions.
The initial and fitted proposal energy functions share a branch decomposition;
resource assessment and energy function selection are detailed in
Appendix Section~\ref{sec:structural-proof}.
The clipping parameters are $\kappa=\exp(\tau/RT)$ and $B=\mu\kappa$,
where $\tau$ is specified in kcal/mol. In the implementation, fitting
observations determine the positive, dimensionless $\mu$ through the
weighted mean in Appendix Section~\ref{sec:proposal-fitting}.
Changing $\mu$ or $\tau$ changes the threshold $B$, the exceedance
probability, and the interval width and predicted sample count.
The statistical guarantees apply when the conditional mean excess
assumption holds at the selected threshold.

\renewcommand{\thealgorithm}{S\arabic{algorithm}}
\begin{algorithm}[!htbp]
\caption{PACK* for one sequence and molecular state}
\label{alg:packstar}
\begin{algorithmic}[1]
\Require Physical energy function and set of allowed RCs; resource budgets; user-specified
accuracy $\varepsilon$; error probabilities; clipping parameter $\tau$;
conditional mean excess bound $m_u$
\Statex \textbf{Prepare the target energy function and initial proposal energy function}
\State Select retained interactions; specify the target energy function $E_t$ and allowed set $C$
\State Construct the branch decomposition and select a feasible root
\State Compute the initial DP partition function bounds; return that interval if $\varepsilon'\le\varepsilon$
\State Draw RC assignments from the initial proposal distribution and evaluate their target energies
\Statex \textbf{Fit the proposal energy function within the refit budget}
\Repeat
  \State For each pair of 1-body and 2-body shrinkage parameter values, initialize and refine the energy corrections on each training fold using Appendix Section~\ref{sec:proposal-fitting}
  \State Assess each pair of parameter values on the two held-out folds using the weight moments and sample count prediction in Appendix Section~\ref{sec:sample-size-prediction}
  \State For each pair of parameter values, repeat the initialization and regression using all training observations; apply the screening criteria in Appendix Section~\ref{sec:fit-screening}
  \State Select the passing fit with the lowest predicted sample count, using the tie rules in Appendix Section~\ref{sec:fit-screening}; if none passes, use the fallback procedure there
  \State If neither is available, stop without reporting a partition function interval
  \State Reuse the retained graph decomposition and root; normalize by DP
  \State Draw a pilot sample from the resulting proposal distribution and evaluate target energies
  \State Assess whether the clipped mean lower bound is positive and the predicted interval satisfies $\varepsilon'\le\varepsilon$ at the sample cap
  \If{either pilot condition is unmet}
    \State Stop without reporting a partition function interval if the refit budget is exhausted
    \State Add the pilot samples and target energies to the training data for the next fit
  \EndIf
\Until{both pilot conditions hold, or the attempt ends without a partition function interval}
\Statex \textbf{Set the final sampling parameters}
\State Draw separate validation samples; stop if validation fails
\State Choose the final sample count $n$ from validation, with $n\ge2$ and a margin below the user-specified accuracy
\Statex \textbf{Construct the partition function interval}
\State Draw exactly $n$ fresh independent RC assignments and evaluate $E_t$
\State Compute one hypothesis test p-value from the final batch excesses
\State Stop without reporting a partition function interval if the p-value is at most 0.10
\State Compute the interval of Theorem~\ref{thm:a4} from their importance weights
\State Report validity and whether $\varepsilon'\le\varepsilon$ separately
\end{algorithmic}
\end{algorithm}

The proposal energy function, $\mu$, and $\kappa$ remain unchanged during
validation and final sampling.
All structural choices, fitting, pilot decisions, validation, and the final
sample count precede the independent final batch. The hypothesis test and interval
use the same final batch importance weights. Repeated assignments may share
a deterministic target energy evaluation, with multiplicities preserved.
Applying the procedure to the three
molecular states gives the intervals used for binding score propagation.

\section{Notation and ancestral sampling on a branch decomposition}
\label{sec:sampling-proof}
\subsection{Notation}
All symbols agree with the main text. Assignments are lowercase;
scalars are lowercase except for the clipping threshold $B$.
Sets are uppercase except for the BWM* $\lambda$-sets; collections of sets are
calligraphic. Energy functions use uppercase $E$, whereas $\E$ denotes
expectation. Conventional $R$, $T$, and $K^*$ retain their usual notation.
Partition functions are $q$ and probability mass functions $P$.
The indicator $\ind_A(c)$ equals one when $c\in A$ and zero otherwise.
A finite allowed RC assignment set $C$ and a constrained local
minimization protocol, implemented here by cyclic coordinate descent (CCD),
define the target energy function throughout the procedure. Temperature
is positive, and the target partition function satisfies $0<q_t<\infty$.

\begin{table}[htbp]
\centering
\caption{Notation for the method and proofs.}
\small
\begin{tabular}{@{}ll@{}}
\toprule
Symbol & Meaning \\
\midrule
$V,\ R_i,\ C$ & Residue position set, RC set, allowed RC assignment set \\
$\mathcal T,\ p,\ p_0$ & Computation tree edges, edge index, root edge \\
$M(p),\ \lambda(p)$ & Separator positions and local residue positions \\
$\operatorname{ch}(p)$ & Child edges after tree compression \\
$m,\ \ell,\ \Lambda_p$ & Separator assignment, local assignment, local domain \\
$\ind_p,\ \psi_p$ & Local constraint indicator and Boltzmann factor \\
$\mathcal S_2$ & Collection of retained 2-body scopes $\{i,j\}\subseteq V$ \\
$E_t$ & Target energy function \\
$E_m,\ E_r$ & Pairwise-minimized energy, rigid energy \\
$E_\eta$ & Proposal energy function \\
$q_t,\ q_m,\ q_r,\ q_\eta$ & Corresponding partition functions \\
$P_\eta(c)$ & Probability of RC assignment $c$ under the proposal distribution \\
$w$ & Importance weight \\
$\xi$ & Difference between target energy and proposal energy \\
$B$ & Positive clipping threshold for raw weights, $B=\mu\kappa$ \\
$\mu$ & Positive divisor used to scale importance weights \\
$\kappa$ & Threshold applied to $w/\mu$, with $\kappa=\exp(\tau/RT)$ \\
$m_u$ & Conditional mean excess bound \\
$\tau$ & Parameter in kcal/mol determining $\kappa$ \\
$y,\ u,\ \pi$ & Clipped component, excess component, exceedance probability \\
$q^{\ominus},\ q^{\oplus}$ & Partition function interval endpoints \\
$\varepsilon$ & User-specified accuracy \\
$\varepsilon'$ & Computed value $1-q^{\ominus}/q^{\oplus}$ \\
$\delta$ & Overall failure probability allowance \\
$\delta_y,\delta_\pi$ & Error probabilities for mean and binomial bounds \\
\bottomrule
\end{tabular}
\end{table}
Here $q_\eta$ is the proposal partition function (normalizing constant).
The conditional mean excess bound is $m_u=20$. The positive constant $\mu$
is dimensionless; $\tau$ is measured in kcal/mol, and
$\kappa=\exp(\tau/RT)$ is dimensionless. RB is a separate interaction
residual budget; changing it changes the target energy function defined on the sparsified
interaction graph, while $\tau$ and $\kappa$ specify clipping.
Counts of final samples are $n$, whereas $d=|V|$
denotes the number of residue positions.
For an interval with finite positive upper endpoint, define
$\varepsilon'=1-q^{\ominus}/q^{\oplus}$.
The user-specified accuracy $0<\varepsilon<1$ is met when
$\varepsilon'\le\varepsilon$.

\subsection{Factor allocation}
Following BWM*~\cite{jou2016bwm}, the $\lambda$-set $\lambda(p)$ contains
the residue positions newly assigned at tree edge $p$.
In the rooted graph construction, before compression, an ordinary internal
edge with child edges $c_1,c_2$ uses
$\lambda(p)=(M(c_1)\cup M(c_2))\setminus M(p)$.
For a leaf corresponding to a retained interaction $\{i,j\}$,
$\lambda(p)=\{i,j\}\setminus M(p)$. Rooting subdivides a chosen
decomposition edge: both child separators equal that edge's middle set,
and the artificial root edge has an empty separator. An internal node
with only one child uses the same rule with that child's separator.
The implementation stores a local DP table at edges with nonempty
$\lambda$-sets. Edges with empty $\lambda$-sets are traversed to reach the
next descendants with DP tables. This compression of the rooted
decomposition tree preserves the retained physical interactions and leaves
the separator and $\lambda$-set of every surviving edge unchanged.
Let $\mathcal T$ be the resulting computation tree edges, with root edge
$p_0$ and child edge set $\operatorname{ch}(p)$. The root is retained for
the final combination even if its $\lambda$-set is empty.
The $\lambda$-sets assign every residue position exactly once.
Each separator $M(p)$
contains residue positions already assigned on the ancestral path, and the root
separator is empty. Separators have the running intersection property:
if two separators contain the same residue position, every separator along the
path connecting them in the branch decomposition tree also contains that residue position.
Every energy factor is allocated
exactly once to a bag $M(p)\cup\lambda(p)$ containing its scope; let $E_p$ be
the sum of the allocated energy terms. Allocate a constant offset to one bag only.
Let $m$ be an RC assignment to $M(p)$ and $\ell$ an RC assignment to
all residue positions in $\lambda(p)$.
Hard constraints are represented by zero/one factors, each allocated to a
bag containing its scope. Their product at a bag is
\begin{equation}
 \ind_p(m,\ell)=\begin{cases}
 1,&\text{if all hard constraints allocated to this bag hold},\\
 0,&\text{otherwise}.
 \end{cases}
 \label{eq:si-local-constraint}
\end{equation}
The collection of these constraints must enforce exactly $C$.
An unrepresented constraint across subtrees would invalidate the recurrence in Eq.~\eqref{eq:si-dp}.

The local assignment domain is
$\Lambda_p=\prod_{i\in\lambda(p)}R_i$; when $\lambda(p)$ is empty this set
contains only the empty assignment. The residue positions in $\lambda(p)$ are disjoint
from $M(p)$. All hard restrictions are enforced by the allocated factors
$\ind_p$. This agrees with the
BWM* enumeration of all local assignments for each separator assignment;
each separator assignment $m$ gives values $E_p(m,\ell)$ and
$\ind_p(m,\ell)$ for $\ell$ in the same domain
$\Lambda_p$~\cite{jou2016bwm}.
For a child edge $f$, let $m_f(m,\ell)$ be the restriction of
$(m,\ell)$ to $M(f)$. Define the local Boltzmann factor by
\begin{equation}
 \psi_p(m,\ell)=\ind_p(m,\ell)e^{-E_p(m,\ell)/RT}.
 \label{eq:si-local-factor}
\end{equation}
For any RC assignment $c$ specifying all positions in this bag, write
$\psi_p(c)=\psi_p(c|_{M(p)},c|_{\lambda(p)})$ and use $E_p(c)$
for the same restriction in $E_p$. Thus the edge index $p$ labels a
factor, while $m$ and $\ell$ specify its RC arguments.
Let $E(c)=\sum_{p\in\mathcal T}E_p(c)$ be the energy represented by
these factors. Since every factor is allocated once,
\begin{equation}
 \prod_{p\in\mathcal T}\psi_p(c)=\ind_C(c)e^{-E(c)/RT}.
 \label{eq:si-factor-product}
\end{equation}
Equation~\eqref{eq:si-factor-product} gives the Boltzmann weight of each
allowed complete RC assignment. The same tree edges and local domains
are counted in Appendix Section~\ref{sec:structural-proof}.

\begin{auditedproposition}{\getrefnumber{MAIN-lem:sample}}
\label{lem:a1}
Assume the factorization above has positive finite total mass. The
recurrence
\begin{equation}
 q_p(m)=\sum_{\ell\in\Lambda_p}\psi_p(m,\ell)
          \prod_{f\in \operatorname{ch}(p)}q_f(m_f(m,\ell))
 \label{eq:si-dp}
\end{equation}
computes $q=\sum_{c\in C}e^{-E(c)/RT}=q_{p_0}(\varnothing)$.
At separators with positive mass, top-down sampling with probabilities
\begin{equation}
 P_p(\ell\mid m)=\frac{\psi_p(m,\ell)
                     \prod_{f\in \operatorname{ch}(p)}q_f(m_f(m,\ell))}{q_p(m)}
 \label{eq:si-conditional}
\end{equation}
generates a complete RC assignment $c$ with probability $e^{-E(c)/RT}/q$.
\end{auditedproposition}
\begin{proof}
Let $\mathcal T_p$ contain $p$ and its descendant edges in the computation
tree. For a given separator assignment $m$, let $C_p(m)$ be the compatible
assignments to the residue positions in
$\bigcup_{f\in\mathcal T_p}\lambda(f)$, satisfying the constraints
allocated in this subtree. We prove by induction that
\begin{equation}
 q_p(m)=\sum_{c_p\in C_p(m)}
          \prod_{f\in\mathcal T_p}\psi_f(m\cup c_p).
 \label{eq:si-dp-induction}
\end{equation}
The union $m\cup c_p$ combines assignments on disjoint sets of residue
positions and supplies the arguments of every factor in this subtree.
At a terminal edge, the product contains only $\psi_p$, so
Eq.~\eqref{eq:si-dp} is exactly Eq.~\eqref{eq:si-dp-induction}.

For the induction step, assume Eq.~\eqref{eq:si-dp-induction} holds for
each child edge $f$ and every assignment to $M(f)$.
Consider a given local assignment $\ell$. Running intersection ensures
that $m$ and $\ell$ determine every child separator assignment
$m_f(m,\ell)$. The residue positions introduced in distinct child
subtrees are disjoint. Substituting the inductive hypothesis for the
child DP entries in Eq.~\eqref{eq:si-dp}, their product enumerates each
combination of compatible child assignments once. Multiplication by
$\psi_p(m,\ell)$ includes the factors allocated to the current bag and
gives zero when a constraint there fails. Summing over $\ell$ in
Eq.~\eqref{eq:si-dp} therefore includes every assignment in $C_p(m)$
once, proving the induction step.
At the root, $M(p_0)=\varnothing$ and $C_{p_0}(\varnothing)=C$, so
Eqs.~\eqref{eq:si-dp-induction} and~\eqref{eq:si-factor-product} give
$q_{p_0}(\varnothing)=\sum_{c\in C}e^{-E(c)/RT}=q$.

For $c\in C$ with positive weight, multiply the conditional probabilities
in Eq.~\eqref{eq:si-conditional} along its traversal. Substitution gives
\begin{align}
 \Prob(\text{output}=c)
 &=\prod_{p\in\mathcal T}P_p(c_{_{\lambda(p)}}\mid c_{_{M(p)}})\nonumber\\
 &=\prod_{p\in\mathcal T}
   \frac{\psi_p(c)\prod_{f\in\operatorname{ch}(p)}q_f(c|_{M(f)})}
        {q_p(c|_{M(p)})}\nonumber\\
 &=\frac{\prod_{p\in\mathcal T}\psi_p(c)}{q_{p_0}(\varnothing)}
   =\frac{e^{-\sum_p E_p(c)/RT}}{q}
   =\frac{e^{-E(c)/RT}}{q}.\label{eq:si-telescope}
\end{align}
Each nonroot edge has exactly one parent, so its DP entry occurs once
in that parent's numerator and once in its own denominator. These
entries cancel, leaving only the root denominator. The last equality
uses Eq.~\eqref{eq:si-factor-product} for $c\in C$.
A branch with zero mass is never selected, so its conditional distribution need
not be defined. Independent random choices for distinct complete traversals
give independent complete RC assignments.\qed
\end{proof}

In Algorithm~\ref{alg:si-sample}, $c_j$ is a partial RC assignment: it
maps the residue positions assigned so far to their selected RCs.
The initial map has no assigned positions. Each local assignment
$\ell$ extends this map on the positions in $\lambda(p)$.
\begin{algorithm}[t]
\caption{Ancestral sampling of RC assignments from precomputed DP tables with compatible factors}
\label{alg:si-sample}
\begin{algorithmic}[1]
\Require DP tables with entries $q_p(m)$, root $p_0$, sample size $n$
\For{$j=1,\ldots,n$}
 \State Initialize $c_j$ with no assigned residue positions
 \For{each edge $p$ in top-down order}
 \State Read separator assignment $m$ from the assigned RC labels of $c_j$
  \State Sample an RC assignment $\ell\in\Lambda_p$ according to Eq.~\eqref{eq:si-conditional}
 \State For each $i\in\lambda(p)$, set $c_j(i)=\ell_i$
 \EndFor
 \State Emit $c_j$, preserving repeated complete RC assignments
\EndFor
\end{algorithmic}
\end{algorithm}

\section{Importance sampling and conditioning on training data}
\label{sec:identity-proof}
\begin{supplemma}
\label{lem:a2}
Let $E_\eta=E_m+\eta$ be a given proposal energy function whose factorization
and allocated hard constraints satisfy the assumptions of Proposition~\ref{lem:a1}.
Assume the proposal distribution assigns positive probability to every RC assignment
with positive target Boltzmann weight. With
\begin{equation}
 P_\eta(c)=\frac{e^{-E_\eta(c)/RT}}{q_\eta},\qquad
 \xi(c)=E_t(c)-E_\eta(c),\qquad w(c)=e^{-\xi(c)/RT},
 \label{eq:si-weight}
\end{equation}
we have $q_t=q_\eta\E_{P_\eta}[w]$. The estimator $q_\eta\bar w$ from an
independent final sample is conditionally unbiased for $q_t$.
\end{supplemma}
\begin{proof}
Direct substitution and cancellation give
\begin{equation}
 \E_{P_\eta}[w]
 =\sum_{c\in C}\frac{e^{-E_\eta(c)/RT}}{q_\eta}
                       e^{(E_\eta(c)-E_t(c))/RT}
 =\frac1{q_\eta}\sum_{c\in C}e^{-E_t(c)/RT}
 =\frac{q_t}{q_\eta}.
 \label{eq:si-identity}
\end{equation}
Support ensures that no positive Boltzmann weight is omitted from the target partition function. Linearity of
expectation yields $\E[q_\eta\bar w]=q_t$ for new samples. The normalizing constant
has no Monte Carlo error conditional on learning, although its value may
vary across fitted proposal energy functions.
\qed
\end{proof}

\begin{supplemma}
\label{lem:a3}
Let $H$ denote the random complete history before final sampling and $h$ a
realized history, including all observations and choices of proposal energy
function, decomposition, $\mu$, $\kappa$,
conditional mean excess bound $m_u$, error probabilities, and
final count $n(h)\ge2$. Suppose conditional on $H=h$ the final samples are
independent with the selected proposal distribution. For a result for a batch
of predetermined size, let $F$ denote its failure event. For each realized
history $h$ satisfying the premises, $\Prob(F\mid H=h)\le\alpha(h)$.
The bound $\alpha(h)$ may differ across histories.
If those premises hold almost surely, then marginal failure probability is
at most $\E_H[\alpha(H)]$.
\end{supplemma}
\begin{proof}
Conditioning on $H=h$ determines every selected object and leaves precisely
the independent sampling experiment required by the result for a
predetermined sample size. The tower property gives
\[
 \Prob(F)=\E_H[\Prob(F\mid H)]\le\E_H[\alpha(H)].
\]
The expectation averages over the distribution of the history before final sampling.
For a specified error probability $\alpha_0$, it suffices that
$\E_H[\alpha(H)]\le\alpha_0$; a pointwise bound
$\alpha(h)\le\alpha_0$ is sufficient.
Here $\alpha_0$ denotes the desired marginal failure-probability bound for the
result being applied. For Theorem~\ref{thm:a4} with the same error probabilities
at every history, $\alpha_0=\eb+\ef$.\qed
\end{proof}

Training, cross-validation, pilot sampling, and validation
may all contribute to $h$. These data are excluded from the final sample. Selecting among many proposal energy functions through $h$ alone requires no union bound over their count for the
single final theorem. Selection using final statistics would not satisfy
the hypothesis. Conditional on the observations and choices before final
sampling, the final RC assignments are fresh independent draws from the
selected proposal distribution.
Energy evaluations may be reused with the sampled assignments
and their multiplicities preserved.

\section{A confidence interval using clipped importance weights}
\label{sec:fs-proof}
For the weight $w$ in Lemma~\ref{lem:a2}, choose a positive, dimensionless
$\mu$, $\kappa>0$, and conditional mean excess bound $m_u=20$.
The scaled weight $w/\mu$ is clipped at $\kappa$, corresponding to a
threshold for raw weights of $B=\mu\kappa=\mu\exp(\tau/RT)$. Define
\begin{align}
 y&=\min\left(w/B,1\right),&
 u&=\pos{w/B-1},\label{eq:si-fs-vars}\\
 \pi&=\Prob_{P_\eta}(u>0),&
 q_t&=q_\eta B\bigl(\E[y]+\E[u]\bigr).\nonumber
\end{align}
Here $u(c)=0$ when $w(c)\le B$, so
$\pi=\Prob_{P_\eta}(w>B)$.
The component means $\E[y]$ and $\E[u]$ contribute
$q_\eta B\E[y]$ and $q_\eta B\E[u]$ to $q_t$, respectively.
For an independent final sample of size $n\ge2$, let $y_i=y(c_i)$ and
$u_i=u(c_i)$. For error probabilities $0<\eb,\ef<1$, let
\begin{align}
 \bar y&=\frac1n\sum_{i=1}^n y_i,&
 \widehat v_y&=\frac1{n-1}\sum_{i=1}^n(y_i-\bar y)^2,&
 t_y&=\log(4/\eb),\nonumber\\
 d_n&=\sqrt{\frac{2\widehat v_y t_y}n}
                  +\frac{7t_y}{3(n-1)},&
 \lb&=\max\{0,\bar y-d_n\},&
 \ub&=\min\{1,\bar y+d_n\}.\label{eq:si-eb}
\end{align}
For $k=\sum_i\ind\{u_i>0\}$, we use the one-sided Clopper--Pearson upper
confidence bound for $\pi$~\cite{clopper1934,thulin2014exact}:
\begin{equation}
 \pup(k)=\begin{cases}
 1-\ef^{1/n},&k=0,\\
 \operatorname{Beta}^{-1}(1-\ef;k+1,n-k),&0<k<n,\\
 1,&k=n.
 \end{cases}
 \label{eq:si-cp}
\end{equation}
Here $\operatorname{Beta}^{-1}(v;a,b)$ is the $v$-quantile of a beta
distribution with shape parameters $a,b$.

\begin{auditedtheorem}{\getrefnumber{MAIN-thm:fs}}
\label{thm:a4}
Condition on the history $h$ before final sampling, satisfying
Proposition~\ref{lem:a1} and Lemmas~\ref{lem:a2}--\ref{lem:a3} and determining
the proposal distribution, $\mu,\kappa$, the conditional mean excess bound $m_u$,
the error probabilities, and $n$ as above. If
$\E[u\mid u>0]\le m_u$ whenever $\pi>0$, then
\begin{equation}
 \Prob\!\left(q_t\in
  \left[q_\eta B\lb,\ q_\eta B(\ub+m_u\pup)\right]\,\middle|\,h\right)
 \ge1-\eb-\ef.
 \label{eq:si-fs-coverage}
\end{equation}
\end{auditedtheorem}
\begin{proof}
We first bound the clipped mean, then the exceedance probability, and
combine the two bounds.

\emph{Clipped mean.} The one-sided empirical Bernstein bound for a variable in
$[0,1]$ uses radius
$\sqrt{2\widehat v_y\log(2/\alpha)/n}\allowbreak
+7\log(2/\alpha)/(3(n-1))$ with failure probability at most
$\alpha$~\cite{maurer2009empirical}. Apply it to $y$ with
$\alpha=\eb/2$ for the upper mean bound, and to $1-y$ with the same
error probability for the lower bound. Their sample variances coincide.
Intersecting with the deterministic range $[0,1]$ gives
\begin{equation}
 \Prob(\E[y]\notin[\lb,\ub]\mid h)\le\eb.
 \label{eq:si-bulk-event}
\end{equation}
The variance in this inequality equals the unbiased sample variance:
$[n(n-1)]^{-1}\sum_{i<j}(y_i-y_j)^2=\widehat v_y$.

\newpage
\emph{Exceedance probability.} Conditional on $h$, $k$ is binomial with parameters
$n$ and $\pi$. For $0\le j<n$ let
\[
 f_j(v)=\sum_{i=0}^j\binom ni v^i(1-v)^{n-i}.
\]
This is continuous and decreasing from one to zero on $[0,1]$.
Define $v_j$ by $f_j(v_j)=\ef$ and put $v_n=1$.
The beta/binomial identity gives
$v_j=\operatorname{Beta}^{-1}(1-\ef;j+1,n-j)$;
$j=0$ yields $(1-v_0)^n=\ef$. These are precisely
Eq.~\eqref{eq:si-cp}.
For any given true $\pi$, $\pi>v_j$ implies $f_j(\pi)<\ef$.
The indices satisfying this inequality form an initial segment. If its
largest element is $j_\star$, then
\[
 \Prob_\pi(\pi>v_k)=\Prob_\pi(k\le j_\star)
                    =f_{j_\star}(\pi)<\ef;
\]
if the segment is empty the probability is zero. Thus
$\Prob(\pi>\pup\mid h)\le\ef$. Discreteness can make the bound
conservative. In particular, $k=0$ still gives a positive upper endpoint.

\emph{Combined bounds.} When both confidence bounds hold,
\[
 0\le\E[u]\le m_u\pi\le m_u\pup.
\]
For $\pi>0$ this follows by conditioning on the exceedance event; for
$\pi=0$ the excess is zero almost surely. Since $w/B=y+u$, adding
this inequality to the bounds on $\E[y]$ and multiplying by $q_\eta B>0$ proves the
interval. The union bound gives failure at most $\eb+\ef$ even though
both statistics use the same batch.\qed
\end{proof}

The implementation uses $\eb=\ef=\delta/6$ per molecular state;
the failure probability for each molecular state is at most $\delta/3$. For three
specified molecular states, the union bound is $\delta$ under the stated assumptions.

If $D$ is the set of final outcomes that meet additional validity criteria and
satisfy $\varepsilon'\le\varepsilon$, then
\[
 \Prob\bigl(D\cap\{q_t\notin[q^{\ominus},q^{\oplus}]\}\mid h\bigr)
 \le\eb+\ef.
\]
When $[q^{\ominus},q^{\oplus}]$ contains $q_t$ and has a finite positive upper endpoint,
the relative estimation error of its lower endpoint satisfies
\[
 0\le (q_t-q^{\ominus})/q_t\le\varepsilon'.
\]
Thus if $D$ also requires $\varepsilon'\le\varepsilon$,
\[
 \Prob\!\left(D\cap\{|q^{\ominus}-q_t|/q_t>\varepsilon\}
       \,\middle|\,h\right)\le\eb+\ef.
\]
The bound applies to the relative estimation error of the lower endpoint $q^{\ominus}$
under the design with a predetermined final sample size.

\section{A hypothesis test of the conditional mean excess assumption}
\label{sec:diagnostic-proof}
\begin{suppproposition}
\label{prop:a5}
Condition on all observations and choices before the final draws, which determine the final batch size, proposal distribution,
$\mu$, $\kappa$, and a positive conditional mean excess bound $m_u$.
Suppose the final samples are independent draws from this distribution and
$\E[u\mid u>0]\le m_u$ whenever $\pi>0$.
For the final excesses $u_i=u(c_i)$ in Eq.~\eqref{eq:si-fs-vars}, let
\[
 k=\sum_{i=1}^n\ind\{u_i>0\},\qquad
 m=\frac1k\sum_{i:\,u_i>0}u_i\quad\text{when }k>0.
\]
For $k>0$, define the hypothesis test p-value as $\min(1,m_u/m)$.
For $k=0$, set the p-value to one and mark the hypothesis test inconclusive.
For every significance level $a\in(0,1)$, the
probability that this p-value is at most $a$ is at most $a$.
\end{suppproposition}
\begin{proof}
Under this conditioning, let $I=\{i:u_i>0\}$.
For every nonempty $I$ with positive probability, the excesses indexed by
$I$ are independent draws from the distribution conditional on positive
excess, so $\E[m\mid I]\le m_u$.
For $0<a<1$, the definition of the p-value gives
\[
 \min(1,m_u/m)\le a
 \quad\Longleftrightarrow\quad m_u/m\le a
 \quad\Longleftrightarrow\quad m\ge m_u/a.
\]
Thus $m_u/a$ is the rejection threshold for the observed mean excess.
Markov's inequality gives
\[
 \Prob(\text{p-value}\le a\mid I)
 =\Prob(m\ge m_u/a\mid I)
 \le\frac{\E[m\mid I]}{m_u/a}\le a.
\]
When $I$ is empty, the p-value is one. Averaging over $I$ proves the
stated conditional bound.\qed
\end{proof}
The hypothesis test is evaluated once on the final batch of predetermined size,
using the interval's importance weights and a significance level for each attempt
of 0.10.
Nonrejection does not establish the conditional mean excess assumption.
The test filters the return event while keeping the proposal distribution, final sample
count, and error probabilities of Theorem~\ref{thm:a4} unchanged.

\section{Computational work, storage, and energy function selection}
\label{sec:structural-proof}
Following the TESS accounting of BWM*~\cite{jou2016bwm}, let $m_p$ and
$\ell_p$ denote the numbers of RC assignments to $M(p)$ and $\lambda(p)$,
respectively:
\begin{equation}
 m_p=\prod_{i\in M(p)}|R_i|,\qquad
 \ell_p=\prod_{i\in\lambda(p)}|R_i|.
 \label{eq:si-dp-work}
\end{equation}
Our sum-product recurrence visits
$\omega=\sum_{p\in\mathcal T}m_p\ell_p$ dense separator/local assignment
combinations. BWM* bounds ordinary internal bags by
$\lfloor3w/2\rfloor$, where $w$ is the width of the chosen branch
decomposition. In our rooted construction, root and unary bags have size
at most $w$, interaction leaves contain at most two positions, and
empty-$\lambda$ compression preserves surviving bags. Therefore, with
$\bar r=\max_i|R_i|$,
\begin{equation}
 \omega\le |\mathcal T|\bar r^{\max\{2,\lfloor3w/2\rfloor\}}.
 \label{eq:si-branchwidth-cost}
\end{equation}
Each position is introduced once, giving $|\mathcal T|\le |V|+1$, including
the root. At each computation edge $p$, the two dense 8-byte separator
arrays occupy $h_p=16m_p$ bytes. These formulas account for assignment
combinations and the two arrays; energy-factor evaluation and other
allocations contribute additional time and memory.

\subsection{Selecting the proposal energy function on the retained graph}
\label{sec:proposal-fitting}
The proposal energy functions fitted with different values of the 1-body and
2-body shrinkage parameters use
constant, 1-body, and 2-body energy corrections on the
retained interaction graph. Fitting changes factor values while preserving
their scopes, so the decomposition, root, and dense DP table dimensions can
be reused.

For each observed energy matrix entry $j$, collect the training observations
whose RC assignments use that entry. For each such observation, evaluate the
corresponding 1-body or 2-body energy term at the minimized coordinates
and subtract the pairwise-minimized matrix entry. Average these differences
over the collected observations, retaining repeated draws with their
multiplicities. If an entry occurs in $n_j$ distinct training RC assignments,
multiply its mean by $n_j/(n_j+s_j)$. Here $s_j\ge0$ is the shrinkage parameter
for entry $j$. All 1-body entries share one parameter value, and all
2-body entries share another. Larger values shrink the initial corrections
more strongly toward zero and increase the regression penalty.
Unobserved entries stay zero.
Let $E_0$ denote the proposal energy function defined by these initial corrections.
Let $x_{ij}=1$ if RC assignment $c_i$ uses observed matrix entry $j$ in its
energy calculation, and $x_{ij}=0$ otherwise.
Write $g_i$ for the probability of $c_i$ under the proposal distribution that
generated that observation. The joint refinement uses
\[
 v_i=\frac{e^{-E_0(c_i)/RT}/g_i}
             {\sum_l e^{-E_0(c_l)/RT}/g_l},\qquad
 z_i=E_t(c_i)-E_0(c_i),\qquad
 m_j=\sum_i v_i x_{ij}.
\]
Let $\gamma$ be a constant energy adjustment and $\theta_j$ the additional
change to entry $j$ relative to its initialized value in $E_0$.
For training assignment $c_i$, $\gamma+\sum_j x_{ij}\theta_j$ approximates
$z_i=E_t(c_i)-E_0(c_i)$. Cyclic coordinate descent approximately minimizes
the following convex objective
\[
 \sum_i v_i\left(z_i-\gamma-\sum_j x_{ij}\theta_j\right)^2
 +\sum_j m_j\frac{s_j}{n_j}\theta_j^2,
 \qquad |\theta_j|\le h_\theta.
\]
Only observed entries have parameters, so $n_j\ge1$. The fitted energy is
$E_0(c)+\gamma+\sum_j x_j(c)\theta_j$. The implementation uses at most
40 sweeps, stopping early when both the constant shift and the largest
coordinate change are below $10^{-8}$ kcal/mol.
Each regression adjustment $\theta_j$ is constrained to
$[-h_\theta,h_\theta]$, with $h_\theta=3$ kcal/mol.
Each coordinate update is clipped to this interval.
The constant $\gamma$ is neither penalized nor capped.
Source probabilities include their normalizing constants, and repeated
draws retain their multiplicities in this regression.
The fitted energy corrections are used directly.

Two-fold cross-validation compares values of the 1-body and 2-body
shrinkage parameters using held-out
weight moments, including $\rho$ in Eq.~\eqref{eq:si-rho}, and the predicted
sample count for the user-specified accuracy. Each fit excludes its
validation fold; during refitting, earlier learning batches are available
to both fits and the newest batch supplies the held-out observations.

\paragraph{Sample size prediction.}
\label{sec:sample-size-prediction}
We predict $\varepsilon'$ at each proposed sample count
and select a count that meets the user-specified accuracy with a margin. Within each held-out
fold, reweight its $N$ observations to the distribution defined by the fitted
proposal energy function using
\[
 \beta_i=
 \frac{e^{-E_\eta(c_i)/RT}/g_i}
      {\sum_l e^{-E_\eta(c_l)/RT}/g_l},
 \qquad
 \mu=\sum_i\beta_i w(c_i).
\]
This data-dependent $\mu$ estimates the mean importance weight under the
fitted proposal distribution and determines the clipping threshold
$B=\mu\kappa$.
The parameter $\tau$ is specified before final sampling, with
$\kappa=\exp(\tau/RT)$.
Compute $y_i$ and $u_i$ using this fold's $\mu$ and the specified
$\kappa$. Define the estimates
\[
 \widehat b=\sum_i\beta_i y_i,\qquad
 \widehat v=\frac{\sum_i\beta_i(y_i-\widehat b)^2}
                  {1-\sum_i\beta_i^2},\qquad
 \widehat\pi=\sum_i\beta_i\ind\{u_i>0\}.
\]
To combine folds, multiply each fold's normalized observation weights by
its fraction of all held-out observations and recompute these moments.
For the selected fit using all training data, the same formula for the weighted mean on its
fitting data determines $\mu$ before its pilot; $\mu$ remains unchanged through
validation and final sampling. Pilot and validation samples drawn directly
from the selected proposal distribution provide ordinary sample means, sample variances, and exceedance
frequencies under that unchanged distribution and $\mu$.

For a proposed final count $n$, use the radius in Eq.~\eqref{eq:si-eb}
with variance $1.3\widehat v$ and center $\widehat b$, and evaluate
Eq.~\eqref{eq:si-cp} using $\lceil n\widehat\pi\rceil$ as the predicted number
of exceedances. Substitution into the interval formula in
Theorem~\ref{thm:a4} gives the predicted $\varepsilon'$.
An integer bisection search selects a count predicted to satisfy
$\varepsilon'\le0.9\varepsilon$. If the sample cap permits
$\varepsilon'\le\varepsilon$ but not this margin, the cap is selected;
if even $\varepsilon'\le\varepsilon$ is unattainable at the cap, the screen
fails. Independent validation supplies the moments used to determine the actual
final sample count.
The final sample cap was 4000 in the partition function benchmarks;
the SHV-1--BLIP ensemble study used its separately stated cap
(Appendix Section~\ref{sec:mi-analysis}).
These predictions guide planning and energy function selection. Theorem~\ref{thm:a4}
gives conditional coverage for the fresh final batch under its assumptions;
the actual final interval determines whether the user-specified accuracy is met.

\paragraph{Screening fitted energy functions and choosing a fallback.}
\label{sec:fit-screening}
Each fitted proposal energy function is screened against four criteria:
\begin{enumerate}
 \item The reweighting, moment, and probability mass calculations yield
 finite values in each held-out fold, in the pooled assessment, and
 after fitting all training data.
 \item Reweighting from the generating proposal distributions to the
 distribution defined by the fitted energy function meets the
 minimum effective sample size fraction in each fold, the pooled assessment,
 and the refit using all training data. For $N$ observations with normalized weights
 $\beta_i$, this fraction is $1/(N\sum_i\beta_i^2)$.
 \item At the maximum final sample count, the pooled prediction has a
 positive lower endpoint for the clipped mean and satisfies
 $\varepsilon'\le\varepsilon$.
 \item The refit using all training data limits the increase in estimated probability
 assigned to RC assignments that use poorly sampled 1-body or 2-body
 entries. An entry is poorly sampled when it occurs in fewer than five
 distinct fitting assignments. Separately for 1-body and 2-body entries,
 compare the reweighted fraction of observations using at least one
 such entry with their unweighted fraction; the ratio must not exceed 1.25.
 When the unweighted fraction is zero, the reweighted fraction must also be zero.
\end{enumerate}
The last item is an implementation criterion for energy function selection.
It is enforced on the refit using all training data; its values from individual folds
are descriptive statistics.
The criteria for predicted $\varepsilon'$ and a positive lower endpoint apply to the pooled assessment.
Fits meeting these criteria are ranked by the predicted sample count, then
the predicted $\varepsilon'$ at the sample cap, then the second moment, with further
deterministic rules for resolving ties. Fitting all training data and applying
these criteria precede construction of DP tables for the selected proposal energy function and
pilot sampling.

If no fit meets these criteria, PACK* tests the proposal energy function fitted with zero
1-body and 2-body shrinkage as a fallback for pilot sampling. It must have finite
screening statistics and meet a preset minimum
effective sample size fraction in cross-validation, including each fold, and
in the refit using all training data, using the same effective sample size definition above.
If this fit fails either criterion, the attempt
ends without a partition function interval.

The energy function selected by screening or fallback defines the proposal
distribution used to draw fresh pilot samples and evaluate their target energies.
Pilot samples grow from 100 to 400 RC assignments under the same proposal
distribution.
The pilot sample passes when its lower bound
for the clipped mean is positive and the predicted interval satisfies
$\varepsilon'\le\varepsilon$ at the maximum final sample count. PACK* then proceeds to
independent validation. Otherwise, PACK* adds the pilot observations
to the training data and refits if budget remains, with at most eight refits
after the initial fit; reaching the refit limit
without a sample that satisfies the pilot criteria
ends the attempt. An error during fitting or assessment of a fitted energy function also ends
the attempt.

Independent validation uses 400 fresh RC assignments and requires a positive
lower bound for the clipped mean and predicted $\varepsilon'\le\varepsilon$
at the maximum permitted sample count; otherwise the attempt ends.
Its observations determine the final count using the sample size prediction
rule above. All these choices precede the independent final sample.

Reusing structural cost estimates requires an unchanged graph, set of allowed RCs,
rooting convention, and work and storage accounting. Each fitted proposal energy function requires
recomputed DP values consistent with its energy terms. Total cost includes DP construction,
sampling, energy function fitting, decomposition and resource assessment, and constrained local minimization.

\subsection{Importance weight moments and estimator variance}
For target distribution $P_t(c)=e^{-E_t(c)/RT}/q_t$, define
\begin{equation}
 \rho=\frac{\E_{P_\eta}[w^2]}{\E_{P_\eta}[w]^2}.
 \label{eq:si-rho}
\end{equation}
Lemma~\ref{lem:a2} implies $w(c)/\E[w]=P_t(c)/P_\eta(c)$ on the support of the proposal distribution.
Consequently
\begin{align}
 \rho&=\sum_{c\in C}\frac{P_t(c)^2}{P_\eta(c)},\nonumber\\
 \rho-1&=\sum_{c\in C}\frac{(P_t(c)-P_\eta(c))^2}{P_\eta(c)},
 \label{eq:si-overlap}\\
 \frac{\Var(q_\eta\bar w)}{q_t^2}&=\frac{\rho-1}{n}.
 \nonumber
\end{align}
For the middle identity, expand the square and use
$\sum_c P_t(c)=\sum_c P_\eta(c)=1$; for the last, use independence of
the $n$ samples. Terms for which both probabilities are zero
can be omitted. Thus smaller $\rho$ reduces the relative
variance of the importance sampling estimator.
The held-out energy function selection metric $\log\rho$ estimates this objective.

\section{Free energy and binding score intervals}
\label{sec:free-energy-proof}
\begin{auditedcorollary}{\getrefnumber{MAIN-cor:kstar}}
\label{prop:a12}

A confidence interval containing the partition function gives the free energy interval
\begin{equation}
 f_t=-RT\log q_t\in
        [-RT\log q^{\oplus},-RT\log q^{\ominus}].
 \label{eq:si-free-energy-interval}
\end{equation}
When all three molecular state intervals contain their respective partition
functions and both denominator lower bounds are positive,
\begin{equation}
 K^*\in\left[
 \frac{\qPL^{\ominus}}
      {\qP^{\oplus}\qL^{\oplus}},\;
 \frac{\qPL^{\oplus}}
      {\qP^{\ominus}\qL^{\ominus}}\right].
 \label{eq:si-kstar-interval}
\end{equation}
\end{auditedcorollary}
\begin{proof}

The function $z\mapsto-RT\log z$ is decreasing for $z>0$, proving
Eq.~\eqref{eq:si-free-energy-interval}. Its upper endpoint is $+\infty$
when $q^{\ominus}=0$.

For positive numbers the binding ratio increases with its numerator and
decreases with each denominator. Replacing the numerator by its lower
bound and both denominators by their upper bounds gives the left endpoint
of Eq.~\eqref{eq:si-kstar-interval}; the opposite substitutions give the
right endpoint. A union bound on the constituent molecular state partition function intervals controls
failure without an independence assumption. Any standard state factors
must be specified consistently at both endpoints.\qed
\end{proof}

\section{Choice of the conditional mean excess bound}
\label{sec:s0-choice}
Finite-sample confidence bounds must account for RC assignments that have not
been observed. An arbitrarily large importance weight can occur on an event
rare enough that a sample of predetermined size misses it with high probability while its
contribution to the mean is arbitrarily large. A finite upper confidence bound
therefore needs a distributional assumption or an independently derived
bound restricting this unobserved contribution. Wang and
Ramdas~\cite{wang2023catoni}, Section 10.1, discuss the corresponding
impossibility of unrestricted mean inference.

Moment-based methods provide one route. Wang and
Ramdas~\cite{wang2023catoni} construct confidence sequences using a known
variance bound and extend their construction to bounded central moments
of order strictly greater than one and at most two.
Wei et al.~\cite{wei2024catoni} sharpen the latter construction under a
known bound on the corresponding central moment.
Block and Shetty~\cite{block2026partition} characterize partition function
estimation through their notion of \emph{integrated coverage}, which controls probability mass under the target distribution in
regions with large ratios of target density to proposal density, and relate this
quantity to f-divergence bounds. Their results apply to general proposal
distributions and target distributions and include heavy-tailed regimes. In each case,
the statistical guarantee uses quantitative information about contributions that a
finite sample may miss.

PACK* places this assumption on the conditional mean of the excess
component $u$ given an exceedance, using the bound $m_u$ in
Theorem~\ref{thm:a4}. The clipped component $y$ is bounded, and the
exceedance probability $\pi$ is estimated from the final sample.
When $\pi\le\pup$, the assumption gives $\E[u]\le m_u\pup$.
Multiplying by $q_\eta B$ gives the contribution
$q_\eta Bm_u\pup$ to the upper endpoint in
Eq.~\eqref{eq:si-fs-coverage}.
Holding the other quantities unchanged, increasing $m_u$ enlarges the upper
endpoint in proportion to $\pup$.
Even when no exceedance is observed, Eq.~\eqref{eq:si-cp} supplies a
positive upper bound on its probability and hence a bound on
unseen excess contributions.

Improving the proposal distribution's overlap with the target distribution
can tighten the confidence interval by reducing the
frequency of large importance weights.

For the experiments, we set the conditional mean excess bound to 20
before the final batch of each reported attempt. This is the population
assumption used by Theorem~\ref{thm:a4}; individual excesses may exceed it.

\subsection{Observational units and hypothesis testing}
Each observation below is one distinct molecular state computation with an unchanged
proposal distribution and a final sample of predetermined size. Its empirical conditional mean excess is the sum of
positive normalized excesses divided by their count. Repeated sampled
assignments retain their multiplicities; a molecular state reused by several
binding scores is counted only once. Molecular states with at least one exceedance
enter the summaries, and hypothesis test decisions follow the rule for each attempt
of Proposition~\ref{prop:a5}. Molecular states without exceedances have no empirical
estimate of the conditional mean excess.

\begin{table}[htbp]
\centering
\caption{Empirical conditional mean excess of positive normalized excesses by
interaction residual budget (RB, kcal/mol). Only molecular states with at least
one positive excess are included; the mean excess is dimensionless.
Nonrejected denotes a hypothesis test p-value above 0.10.}
\label{tab:s0-diagnostics}
\small
\begin{tabular}{lrrrrrr}
\toprule
RB & Molecular states & Rejected & Remaining & Median & 95th pct. & Maximum \\
\midrule
1 & 56 & 3 & 53 & 0.706 & 21.867 & 31.345 \\
2 & 58 & 1 & 57 & 1.427 & 25.765 & 55.012 \\
4 & 68 & 0 & 68 & 1.906 & 20.455 & 61.830 \\
\bottomrule
\end{tabular}

\par\smallskip
\parbox{.95\linewidth}{\footnotesize Median, percentile, and maximum refer
to the nonrejected molecular states.}
\end{table}

\subsection{Empirical excess summaries}
We analyzed final samples from \stat{tailtotal} molecular state computations
across the three RB settings. Of these, \stat{tailtested} had at least one
positive excess, and \stat{tailremaining} were not rejected by the hypothesis test.
The observed conditional mean excess was at most 20 in \stat{tailatmost20} of
\stat{tailremaining} nonrejected molecular states with positive excesses
(\stat{tailkeptpercent}\%), and in \stat{tailatmost20} of all \stat{tailtested}
states with positive excesses (\stat{tailallpercent}\%). Among nonrejected states,
the median was \stat{tailmedian}, the 95th percentile \stat{tailp95}, and the
maximum \stat{tailmaximum}. Observed means above 20 occurred at all three RB values.

Table~\ref{tab:rejected-states} gives the four rejected molecular states.

\begin{table}[htbp]
\centering
\caption{Molecular states rejected by the hypothesis test of the conditional mean excess assumption
at significance level 0.10. RB is the interaction residual budget in kcal/mol.
Count is the number of positive normalized excesses; Mean is their dimensionless
arithmetic mean. Sequence labels give chain, residue number, and amino acid
identity at each mutable residue position.}
\label{tab:rejected-states}
\scriptsize
\setlength{\tabcolsep}{3pt}
\begin{tabular}{@{}p{.30\textwidth}p{.17\textwidth}rrrr@{}}
\toprule
System (PDB ID) & Molecular state & RB & Count & Mean & p-value\\
\midrule
\raggedright Fibronectin F1/FnBPA-5 (3CAL) & \raggedright C147 TRP\newline complex & 1 & 13 & 1445.129 & 0.01384\\
\raggedright Cohesin/dockerin (4U3S) & \raggedright B280 PHE,\newline B277 CYS\newline complex & 1 & 1 & $4.0783\times10^9$ & $4.904\times10^{-9}$\\
\raggedright Scribble PDZ34/peptide (4WYU) & \raggedright A94 ASN,\newline A70 MET\newline complex & 1 & 13 & 298.156 & 0.06708\\
\raggedright Fibronectin F1/FnBPA-5 (2RL0) & \raggedright A85 MET\newline protein & 2 & 22 & 210.904 & 0.09483\\
\bottomrule
\end{tabular}
\end{table}

The summaries are descriptive, especially for the \stat{tailsingle} of
\stat{tailremaining} nonrejected molecular states with only one positive excess.
The hypothesis test rejects only when the observed mean is at least 200, so
it can retain means above 20.

\section{Experimental records and additional results}
\label{sec:experiment-details}

\subsection{Comparison with exhaustive enumeration}
For each of \stat{refsystems} systems, exhaustive calculations summed the Boltzmann weights
of every RC assignment under both the full-interaction and sparse-interaction energy functions,
evaluated by constrained local minimization. Each PACK*
run and MARK* control used the same structure and set of allowed RCs as the
corresponding exhaustive enumeration. Each system used 100 independently
seeded PACK* runs (seeds 10000--10099).
All \stat{refruns} intervals with exhaustive benchmarks met the user-specified
accuracy $\varepsilon=0.683$.
Table~\ref{tab:census-results} reports natural log partition functions for
the full-interaction and sparse-interaction partition functions,
inclusion counts for the same 100 PACK* intervals, and
$\varepsilon'=1-q^{\ominus}/q^{\oplus}$.
All \stat{refsystems} MARK* controls attained the user-specified accuracy and contained their
corresponding exhaustive full-interaction partition functions.

The sparse enumerations retained the interaction edges and set of allowed RCs
used by the RB=1 runs. All \stat{refruns} PACK* intervals also contained the
corresponding exhaustively enumerated sparse-interaction partition function.

\begin{table}[t]
\centering
\caption{Full-interaction and sparse-interaction partition functions and PACK* interval inclusion
counts. Both $\ln q$ columns use natural logarithms. Full incl. and Sparse incl.
count the same 100 PACK* intervals against each enumerated value; the last column gives their
median $\varepsilon'$. Sparse-interaction enumeration computes the PACK* target partition function;
the exhaustive full-interaction partition function provides the comparison.}
\label{tab:census-results}
\scriptsize
\setlength{\tabcolsep}{1.6pt}
\begin{tabular}{p{.25\linewidth}rrrrrrr}
\toprule
System (PDB ID) & Pos. & Assignments & \shortstack{Full-interaction\\$\ln q$} & \shortstack{Sparse-interaction\\$\ln q$} & Full incl. & Sparse incl. & $\varepsilon'$ \\
\midrule
{\raggedright Scribble PDZ34/\allowbreak{}peptide\newline (PDB 4WYU)\par} & 7 & 1,969,920 & 229.000639 & 228.636308 & 100/100 & 100/100 & 0.6136 \\
{\raggedright Glutathione S-transferase\newline (PDB 1GWC)\par} & 6 & 1,190,700 & 294.745642 & 294.493119 & 100/100 & 100/100 & 0.6116 \\
{\raggedright Phosducin/\allowbreak{}transducin\newline (PDB 1A0R)\par} & 6 & 1,058,400 & 295.913104 & 295.689884 & 100/100 & 100/100 & 0.6142 \\
{\raggedright Staph. protein A/\allowbreak{}IgG Fc\newline (PDB 4WWI)\par} & 6 & 1,008,000 & 245.666585 & 245.262465 & 100/100 & 100/100 & 0.6142 \\
{\raggedright S. pyogenes NADase/\allowbreak{}IFS\newline (PDB 4KT6)\par} & 5 & 1,005,480 & 222.916872 & 222.513585 & 100/100 & 100/100 & 0.6111 \\
{\raggedright EGFR kinase/\allowbreak{}Mig6\newline (PDB 2RFE)\par} & 5 & 1,005,480 & 217.519892 & 217.332552 & 100/100 & 100/100 & 0.6139 \\
{\raggedright HIV-1 capsid/\allowbreak{}VHH nanobody\newline (PDB 2XXM)\par} & 5 & 1,005,480 & 227.518908 & 227.368945 & 100/100 & 100/100 & 0.6136 \\
{\raggedright Dicer/\allowbreak{}TRBP\newline (PDB 4WYQ)\par} & 7 & 10,054,800 & 304.706957 & 304.140952 & 100/100 & 100/100 & 0.6137 \\
{\raggedright RELIK capsid/\allowbreak{}cyclophilin A\newline (PDB 2XGY)\par} & 8 & 10,080,000 & 340.819765 & 340.475545 & 100/100 & 100/100 & 0.6129 \\
\bottomrule
\end{tabular}

\end{table}

For each of the \stat{refsystems} systems with an exhaustive benchmark, the Wilson 95\% confidence interval for the proportion of
runs whose intervals contain either enumerated value, evaluated separately, was
[0.963,1.000]~\cite{wilson1927}.
The observational units are 100 independently seeded runs per system;
these are pointwise confidence intervals.
\paragraph{Interaction residual budget.}
For each residue pair, we take the maximum absolute 2-body energy over all
allowed RC pairs and over both the rigid and pairwise-minimized energy matrices.
Starting from the full residue interaction graph, edges are considered for deletion
in increasing order of this maximum, with more distant residue pairs first
when maxima are tied. An edge is deleted only if the sum of these maxima
over all deleted edges remains within the interaction residual budget (RB).
Thus RB bounds the absolute change in the sum of either the rigid or pairwise-minimized energy matrix entries
for any allowed RC assignment.
Table~\ref{tab:reference-edges} gives the sparsification statistics for
all \stat{refsystems} systems.
\begin{table}[htbp]
\centering
\caption{Interaction sparsification statistics for the exhaustive benchmark
systems at interaction residual budget RB=1 kcal/mol.}
\label{tab:reference-edges}
\small
\begin{tabular}{p{.39\linewidth}rrr}
\toprule
System (PDB ID) & Retained/total & Removed & \shortstack{Interaction residual\\(kcal/mol)} \\
\midrule
{\raggedright Scribble PDZ34/\allowbreak{}peptide\newline (PDB 4WYU)\par} & 12/21 & 9 & 0.747128 \\
{\raggedright Glutathione S-transferase\newline (PDB 1GWC)\par} & 13/15 & 2 & 0.635471 \\
{\raggedright Phosducin/\allowbreak{}transducin\newline (PDB 1A0R)\par} & 11/15 & 4 & 0.685342 \\
{\raggedright Staph. protein A/\allowbreak{}IgG Fc\newline (PDB 4WWI)\par} & 10/15 & 5 & 0.873983 \\
{\raggedright S. pyogenes NADase/\allowbreak{}IFS\newline (PDB 4KT6)\par} & 7/10 & 3 & 0.877327 \\
{\raggedright EGFR kinase/\allowbreak{}Mig6\newline (PDB 2RFE)\par} & 7/10 & 3 & 0.481804 \\
{\raggedright HIV-1 capsid/\allowbreak{}VHH nanobody\newline (PDB 2XXM)\par} & 9/10 & 1 & 0.239846 \\
{\raggedright Dicer/\allowbreak{}TRBP\newline (PDB 4WYQ)\par} & 15/21 & 6 & 0.811235 \\
{\raggedright RELIK capsid/\allowbreak{}cyclophilin A\newline (PDB 2XGY)\par} & 14/28 & 14 & 0.964262 \\
\bottomrule
\end{tabular}

\end{table}
These statistics summarize sparsification at the energy matrix level;
a uniform bound on the energy difference between the full-interaction and sparse-interaction
energy functions, each evaluated by separate constrained local minimization, would require additional analysis.

\subsection{Sequence matching and eligibility criteria for the RB comparison}
\setcounter{secnumdepth}{3}
\subsubsection{Matching and eligibility.}
The interaction residual budget (RB) controls graph sparsification;
it is distinct from $\tau$, which was set to 1 kcal/mol,
and the conditional mean excess bound $m_u$, which was set to 20. The sweep used constant, 1-body,
and 2-body learned energy corrections, seed 42, fresh energy matrices, and
joint refinement by weighted regression. CPU counts,
GPU allocations, Java heaps, and DP host budgets varied by system and RB.
The force field, constrained local minimization protocol, and user-specified accuracy were shared across RB settings.
The sweep measured binding score agreement and the number of sequences with
reported scores under the resource allocations for each system.

Within a system, outputs were paired using the full sequence identifier.
One system--sequence pair is one biological system paired with one sequence,
the observational unit for binding score comparisons.
Eligibility requires partition function estimates for all three molecular states and finite logarithmic
binding scores in both methods. We applied the same criteria
before computing Spearman rank correlations, Pearson correlations, or absolute binding free energy differences.
For each system, the common eligible sequence set is the intersection across
the MARK* baseline and all three RB runs. We required at least three
pairs and nonconstant scores within each system; the systems and
sequences satisfying these requirements form the common cohort.
Spearman rank correlations used average ranks for ties in the
binding scores rounded to six decimal places. Table~\ref{tab:rb-common} summarizes systems
with equal weight.
For the sensitivity analysis, we instead intersected eligible sequences
separately for each RB value and MARK*.

\begin{table}[t]
\centering
\caption{Agreement between PACK* and MARK* binding scores on each system's
common eligible sequence intersection across MARK* and all three interaction
residual budgets (RB, kcal/mol). Pairs counts system--sequence observations.
Correlations are medians of the correlations within each system, with equal
system weight. Energy diff. (kcal/mol) is the median across systems of their
median absolute binding free energy differences from MARK*, using
$\Delta\Delta G=-RT\ln(K^*_{\mathrm{P}}/K^*_{\mathrm{M}})$ at 298.15 K,
where P and M denote PACK* and MARK*, respectively.}
\label{tab:rb-common}
\small
\begin{tabular}{rrrrrr}
\toprule
RB & Systems & Pairs & Spearman & Pearson & Energy diff. (kcal/mol) \\
\midrule
1 & 36 & 873 & 0.9955 & 0.9993 & 0.1446 \\
2 & 36 & 873 & 0.9917 & 0.9980 & 0.2657 \\
4 & 36 & 873 & 0.9814 & 0.9960 & 0.3639 \\
\bottomrule
\end{tabular}

\end{table}

\begin{table}[t]
\centering
\caption{Agreement between PACK* and MARK* binding scores using a separate
eligible sequence intersection for each interaction residual budget (RB,
kcal/mol) and MARK*. Pairs counts system--sequence observations. Correlations
are medians of the correlations within each system, with equal system weight.
Energy diff. (kcal/mol) is the median across systems of their median absolute
binding free energy differences from MARK*, using
$\Delta\Delta G=-RT\ln(K^*_{\mathrm{P}}/K^*_{\mathrm{M}})$ at 298.15 K,
where P and M denote PACK* and MARK*, respectively. Sequence sets can differ
between RB values.}
\label{tab:rb-pairwise}
\small
\begin{tabular}{rrrrrr}
\toprule
RB & Systems & Pairs & Spearman & Pearson & Energy diff. (kcal/mol) \\
\midrule
1 & 36 & 903 & 0.9954 & 0.9992 & 0.1458 \\
2 & 36 & 899 & 0.9915 & 0.9977 & 0.2608 \\
4 & 36 & 901 & 0.9835 & 0.9971 & 0.3682 \\
\bottomrule
\end{tabular}

\end{table}

\begin{figure}[!htbp]
\centering
\includegraphics[width=\textwidth,height=.83\textheight,keepaspectratio]{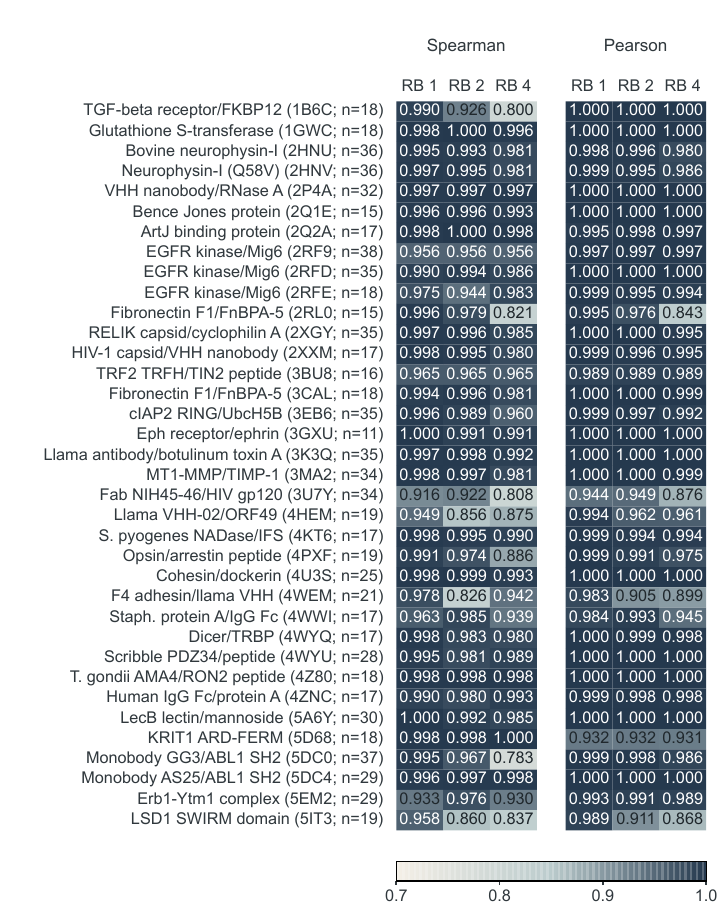}
\caption{Spearman and Pearson correlations between PACK* and MARK* binding
scores within each system, on the common eligible sequence intersection across
MARK* and all three interaction residual budgets (RB, kcal/mol). Each row gives
the system name, PDB ID, and number of system--sequence observations.
Residue positions are listed in Table~\ref{tab:sequence-comparison-designs}. The two panels share the
same color scale (0.7--1.0). Correlations are rounded to three decimal places;
values displayed as 1.000 may therefore be slightly less than 1.}
\label{fig:rb-system-correlations}
\end{figure}

Main Figure~\ref{MAIN-fig:census-reference}C--D visualizes the summaries for the common cohort;
Figure~\ref{fig:rb-system-correlations} gives each system's values and sample size.
These comparisons describe jointly available binding scores and include the
effects of changing the retained physical interaction graph with RB.

\subsubsection{Baseline and cohort definition.}
The baseline is the earlier MARK* benchmark campaign,
separate from the \stat{refsystems} controls on reduced spaces. The system identifiers
and mutable/flexible lists were the same as those in the current design specification.
The comparison measures agreement with these baseline MARK* binding scores.

Runs for all 38 protein--ligand complexes finished at each RB value: all input
sequences were processed and the runs ended normally with recorded runtime,
even when some sequences yielded no binding score. The common
correlation cohort contains 36 systems; the two systems without baseline
MARK* scores are phosducin/transducin
(PDB ID: 1A0R) and TRF2 TRFH/Apollo peptide (PDB ID: 3BUA).
The molecular states rejected by the hypothesis test in Table~\ref{tab:rejected-states} do not
yield eligible partition function estimates.

Table~\ref{tab:sequence-comparison-designs} lists the shared protein--ligand
systems, PDB structures, and mutable and flexible residue positions for
this comparison and the deterministic baseline comparison in
Appendix Section~\ref{sec:deterministic-ranking}.
\begingroup
\small
\setlength{\tabcolsep}{4pt}
\begin{longtable}{p{.32\linewidth}p{.16\linewidth}p{.43\linewidth}}
\caption{Shared design definitions for the PACK* versus MARK* and A*--K* versus MARK* sequence comparisons. Each identifier gives a chain letter and residue number in the benchmark structure. Mutable residue positions also have side-chain flexibility; the final column lists the additional flexible residue positions. The same definitions apply to all three RB settings. Eligible sequence counts are given in the comparison figures.}\label{tab:sequence-comparison-designs}\\
\toprule
Protein/ligand (PDB ID) & Mutable residue positions & Additional flexible residue positions \\
\midrule
\endfirsthead
\toprule
Protein/ligand (PDB ID) & Mutable residue positions & Additional flexible residue positions \\
\midrule
\endhead
\bottomrule
\endfoot
\raggedright Phosducin/\allowbreak{}transducin\newline (1A0R) & \raggedright G390 & \raggedright\arraybackslash G387,\allowbreak{} G392,\allowbreak{} G386,\allowbreak{} G393,\allowbreak{} B280,\allowbreak{} B281,\allowbreak{} B283,\allowbreak{} B279,\allowbreak{} B284 \\
\raggedright TGF-$\beta$ receptor/\allowbreak{}FKBP12\newline (1B6C) & \raggedright E965 & \raggedright\arraybackslash F994,\allowbreak{} F995,\allowbreak{} E903,\allowbreak{} E892,\allowbreak{} E957,\allowbreak{} E925,\allowbreak{} E963,\allowbreak{} E947,\allowbreak{} E922,\allowbreak{} E942,\allowbreak{} E940,\allowbreak{} E967,\allowbreak{} E948 \\
\raggedright Glutathione S-transferase\newline (1GWC) & \raggedright C514 & \raggedright\arraybackslash B314,\allowbreak{} B309,\allowbreak{} B313,\allowbreak{} B311,\allowbreak{} B317,\allowbreak{} C510,\allowbreak{} C505,\allowbreak{} C446,\allowbreak{} C503,\allowbreak{} C511,\allowbreak{} C498,\allowbreak{} C500 \\
\raggedright Bovine neurophysin-I\newline (2HNU) & \raggedright D309,\allowbreak{} D318 & \raggedright\arraybackslash D311,\allowbreak{} D271,\allowbreak{} D316,\allowbreak{} D306,\allowbreak{} D305,\allowbreak{} C192,\allowbreak{} C182,\allowbreak{} C233,\allowbreak{} C181 \\
\raggedright Neurophysin-I (Q58V)\newline (2HNV) & \raggedright C237,\allowbreak{} C228 & \raggedright\arraybackslash D273,\allowbreak{} D262,\allowbreak{} D314,\allowbreak{} D263,\allowbreak{} C235,\allowbreak{} C224,\allowbreak{} C225,\allowbreak{} C230,\allowbreak{} C190 \\
\raggedright VHH nanobody/\allowbreak{}RNase A\newline (2P4A) & \raggedright D476,\allowbreak{} D475 & \raggedright\arraybackslash B226,\allowbreak{} B225,\allowbreak{} B228,\allowbreak{} A111,\allowbreak{} A69,\allowbreak{} A71 \\
\raggedright Bence Jones protein\newline (2Q1E) & \raggedright D418 & \raggedright\arraybackslash D427,\allowbreak{} D365,\allowbreak{} D363,\allowbreak{} D420,\allowbreak{} D416,\allowbreak{} D425,\allowbreak{} C309,\allowbreak{} C316,\allowbreak{} C314 \\
\raggedright ArtJ binding protein\newline (2Q2A) & \raggedright B385 & \raggedright\arraybackslash B403,\allowbreak{} B400,\allowbreak{} B383,\allowbreak{} B362,\allowbreak{} B363,\allowbreak{} B391,\allowbreak{} B394,\allowbreak{} B390,\allowbreak{} A144,\allowbreak{} A156,\allowbreak{} A149,\allowbreak{} A153 \\
\raggedright EGFR kinase/\allowbreak{}Mig6\newline (2RF9) & \raggedright D586,\allowbreak{} D585 & \raggedright\arraybackslash D582,\allowbreak{} D578,\allowbreak{} B499,\allowbreak{} B503 \\
\raggedright EGFR kinase/\allowbreak{}Mig6\newline (2RFD) & \raggedright B547,\allowbreak{} B544 & \raggedright\arraybackslash B552,\allowbreak{} B543,\allowbreak{} B541,\allowbreak{} B540,\allowbreak{} B531,\allowbreak{} B530,\allowbreak{} A238 \\
\raggedright EGFR kinase/\allowbreak{}Mig6\newline (2RFE) & \raggedright B488 & \raggedright\arraybackslash F1138,\allowbreak{} B500,\allowbreak{} B499,\allowbreak{} B504,\allowbreak{} B506,\allowbreak{} B475,\allowbreak{} B471,\allowbreak{} B452,\allowbreak{} B524,\allowbreak{} B505,\allowbreak{} B496,\allowbreak{} B491 \\
\raggedright Fibronectin F1/\allowbreak{}FnBPA-5\newline (2RL0) & \raggedright A85 & \raggedright\arraybackslash G95,\allowbreak{} G97,\allowbreak{} A75,\allowbreak{} A65,\allowbreak{} A63,\allowbreak{} A49,\allowbreak{} A48,\allowbreak{} A51,\allowbreak{} A50,\allowbreak{} A87,\allowbreak{} A77,\allowbreak{} A76 \\
\raggedright RELIK capsid/\allowbreak{}cyclophilin A\newline (2XGY) & \raggedright B189,\allowbreak{} B184 & \raggedright\arraybackslash B248,\allowbreak{} B242,\allowbreak{} B277,\allowbreak{} B192,\allowbreak{} B190,\allowbreak{} B186,\allowbreak{} B183,\allowbreak{} B250,\allowbreak{} B251,\allowbreak{} A83,\allowbreak{} A76 \\
\raggedright HIV-1 capsid/\allowbreak{}VHH nanobody\newline (2XXM) & \raggedright B107 & \raggedright\arraybackslash B115,\allowbreak{} B117,\allowbreak{} B109,\allowbreak{} B173,\allowbreak{} B105,\allowbreak{} B167,\allowbreak{} B165,\allowbreak{} A30,\allowbreak{} A4,\allowbreak{} A6,\allowbreak{} A5 \\
\raggedright TRF2 TRFH/\allowbreak{}TIN2 peptide\newline (3BU8) & \raggedright B243 & \raggedright\arraybackslash D431,\allowbreak{} A34,\allowbreak{} A35 \\
\raggedright TRF2 TRFH/\allowbreak{}Apollo peptide\newline (3BUA) & \raggedright C447,\allowbreak{} C450 & \raggedright\arraybackslash D638,\allowbreak{} D641,\allowbreak{} D642,\allowbreak{} D645,\allowbreak{} D623,\allowbreak{} D619,\allowbreak{} D616,\allowbreak{} D620,\allowbreak{} C417,\allowbreak{} C414,\allowbreak{} C410,\allowbreak{} C421,\allowbreak{} C452,\allowbreak{} C443,\allowbreak{} C444,\allowbreak{} C446 \\
\raggedright Fibronectin F1/\allowbreak{}FnBPA-5\newline (3CAL) & \raggedright C147 & \raggedright\arraybackslash D210,\allowbreak{} D212,\allowbreak{} C126,\allowbreak{} C128,\allowbreak{} C113,\allowbreak{} C137,\allowbreak{} C111,\allowbreak{} C139,\allowbreak{} C120,\allowbreak{} C149 \\
\raggedright cIAP2 RING/\allowbreak{}UbcH5B\newline (3EB6) & \raggedright A44,\allowbreak{} A45 & \raggedright\arraybackslash B128,\allowbreak{} A55,\allowbreak{} A29,\allowbreak{} A19,\allowbreak{} A42,\allowbreak{} A41,\allowbreak{} A37 \\
\raggedright Eph receptor/\allowbreak{}ephrin\newline (3GXU) & \raggedright A30 & \raggedright\arraybackslash B249,\allowbreak{} B250,\allowbreak{} B251 \\
\raggedright Llama antibody/\allowbreak{}botulinum toxin A\newline (3K3Q) & \raggedright C444,\allowbreak{} C440 & \raggedright\arraybackslash B254,\allowbreak{} B316,\allowbreak{} B253,\allowbreak{} B263,\allowbreak{} B358,\allowbreak{} B284,\allowbreak{} C416,\allowbreak{} C456,\allowbreak{} C454,\allowbreak{} C446,\allowbreak{} C436,\allowbreak{} C420,\allowbreak{} C413,\allowbreak{} C437,\allowbreak{} C439 \\
\raggedright MT1-MMP/\allowbreak{}TIMP-1\newline (3MA2) & \raggedright A316,\allowbreak{} A254 & \raggedright\arraybackslash B341,\allowbreak{} B343,\allowbreak{} A276,\allowbreak{} A290,\allowbreak{} A291,\allowbreak{} A271,\allowbreak{} A277,\allowbreak{} A287,\allowbreak{} A318,\allowbreak{} A274,\allowbreak{} A255,\allowbreak{} A315,\allowbreak{} A286 \\
\raggedright Fab NIH45-46/\allowbreak{}HIV gp120\newline (3U7Y) & \raggedright G384,\allowbreak{} G382 & \raggedright\arraybackslash G385,\allowbreak{} L587,\allowbreak{} L649 \\
\raggedright Llama VHH-02/\allowbreak{}ORF49\newline (4HEM) & \raggedright C372 & \raggedright\arraybackslash B226,\allowbreak{} C326,\allowbreak{} C303,\allowbreak{} C318,\allowbreak{} C360,\allowbreak{} C377,\allowbreak{} C322,\allowbreak{} C332,\allowbreak{} C328,\allowbreak{} C330,\allowbreak{} C353 \\
\raggedright S. pyogenes NADase/\allowbreak{}IFS\newline (4KT6) & \raggedright D712 & \raggedright\arraybackslash D822,\allowbreak{} D815,\allowbreak{} D759,\allowbreak{} C512,\allowbreak{} C511,\allowbreak{} C509,\allowbreak{} C519,\allowbreak{} C515,\allowbreak{} C518 \\
\raggedright Opsin/\allowbreak{}arrestin peptide\newline (4PXF) & \raggedright A249 & \raggedright\arraybackslash B328,\allowbreak{} B330,\allowbreak{} A246,\allowbreak{} A245,\allowbreak{} A226,\allowbreak{} A225,\allowbreak{} A138,\allowbreak{} A229 \\
\raggedright Cohesin/\allowbreak{}dockerin\newline (4U3S) & \raggedright B280,\allowbreak{} B277 & \raggedright\arraybackslash B309,\allowbreak{} B276,\allowbreak{} B273,\allowbreak{} B274,\allowbreak{} A32,\allowbreak{} A76,\allowbreak{} A145,\allowbreak{} A97,\allowbreak{} A94,\allowbreak{} A93,\allowbreak{} A95 \\
\raggedright F4 adhesin/\allowbreak{}llama VHH\newline (4WEM) & \raggedright B347 & \raggedright\arraybackslash B357,\allowbreak{} B278,\allowbreak{} A42,\allowbreak{} A4 \\
\raggedright Staph. protein A/\allowbreak{}IgG Fc\newline (4WWI) & \raggedright B60 & \raggedright\arraybackslash E382,\allowbreak{} E384,\allowbreak{} E540,\allowbreak{} E378,\allowbreak{} E538,\allowbreak{} B62,\allowbreak{} B64,\allowbreak{} B65,\allowbreak{} B58 \\
\raggedright Dicer/\allowbreak{}TRBP\newline (4WYQ) & \raggedright E394 & \raggedright\arraybackslash D288,\allowbreak{} D289,\allowbreak{} D285,\allowbreak{} E367,\allowbreak{} E371,\allowbreak{} E373,\allowbreak{} E365,\allowbreak{} E391,\allowbreak{} E390 \\
\raggedright Scribble PDZ34/\allowbreak{}peptide\newline (4WYU) & \raggedright A94,\allowbreak{} A70 & \raggedright\arraybackslash B212,\allowbreak{} B213,\allowbreak{} A12,\allowbreak{} A90,\allowbreak{} A96,\allowbreak{} A93,\allowbreak{} A92,\allowbreak{} A72,\allowbreak{} A69,\allowbreak{} A88 \\
\raggedright T. gondii AMA4/\allowbreak{}RON2 peptide\newline (4Z80) & \raggedright D967,\allowbreak{} D970 & \raggedright\arraybackslash D972,\allowbreak{} D971,\allowbreak{} D969,\allowbreak{} B621,\allowbreak{} B620,\allowbreak{} B646,\allowbreak{} B758,\allowbreak{} B756,\allowbreak{} B649,\allowbreak{} B760,\allowbreak{} B563,\allowbreak{} B663 \\
\raggedright Human IgG Fc/\allowbreak{}protein A\newline (4ZNC) & \raggedright E534 & \raggedright\arraybackslash F716,\allowbreak{} F733,\allowbreak{} F731,\allowbreak{} F718,\allowbreak{} F731,\allowbreak{} E499,\allowbreak{} E528,\allowbreak{} E497 \\
\raggedright LecB lectin/\allowbreak{}mannoside\newline (5A6Y) & \raggedright D455,\allowbreak{} D452 & \raggedright\arraybackslash D356,\allowbreak{} D359,\allowbreak{} D431,\allowbreak{} D355,\allowbreak{} D429,\allowbreak{} D450,\allowbreak{} D453,\allowbreak{} D396,\allowbreak{} C274,\allowbreak{} D363,\allowbreak{} C247,\allowbreak{} C275,\allowbreak{} C334,\allowbreak{} C331 \\
\raggedright KRIT1 ARD-FERM\newline (5D68) & \raggedright A313 & \raggedright\arraybackslash C903,\allowbreak{} C937,\allowbreak{} C905,\allowbreak{} C910,\allowbreak{} C935,\allowbreak{} C914,\allowbreak{} A310,\allowbreak{} A309,\allowbreak{} A265,\allowbreak{} A261,\allowbreak{} A262 \\
\raggedright Monobody GG3/\allowbreak{}ABL1 SH2\newline (5DC0) & \raggedright B116,\allowbreak{} B138 & \raggedright\arraybackslash B181,\allowbreak{} B137,\allowbreak{} A51,\allowbreak{} A52 \\
\raggedright Monobody AS25/\allowbreak{}ABL1 SH2\newline (5DC4) & \raggedright A95,\allowbreak{} A28 & \raggedright\arraybackslash B176,\allowbreak{} B172,\allowbreak{} B173,\allowbreak{} B174,\allowbreak{} A25,\allowbreak{} A26,\allowbreak{} A97,\allowbreak{} A21,\allowbreak{} A46 \\
\raggedright Erb1-Ytm1 complex\newline (5EM2) & \raggedright B614,\allowbreak{} B612 & \raggedright\arraybackslash B606,\allowbreak{} B609,\allowbreak{} B607,\allowbreak{} B537,\allowbreak{} A105,\allowbreak{} A103 \\
\raggedright LSD1 SWIRM domain\newline (5IT3) & \raggedright A41 & \raggedright\arraybackslash B134,\allowbreak{} B153,\allowbreak{} A37,\allowbreak{} A34,\allowbreak{} A38 \\
\end{longtable}
\endgroup

\begin{figure}[!htbp]
\centering
\includegraphics[width=.84\textwidth,height=.83\textheight,keepaspectratio]{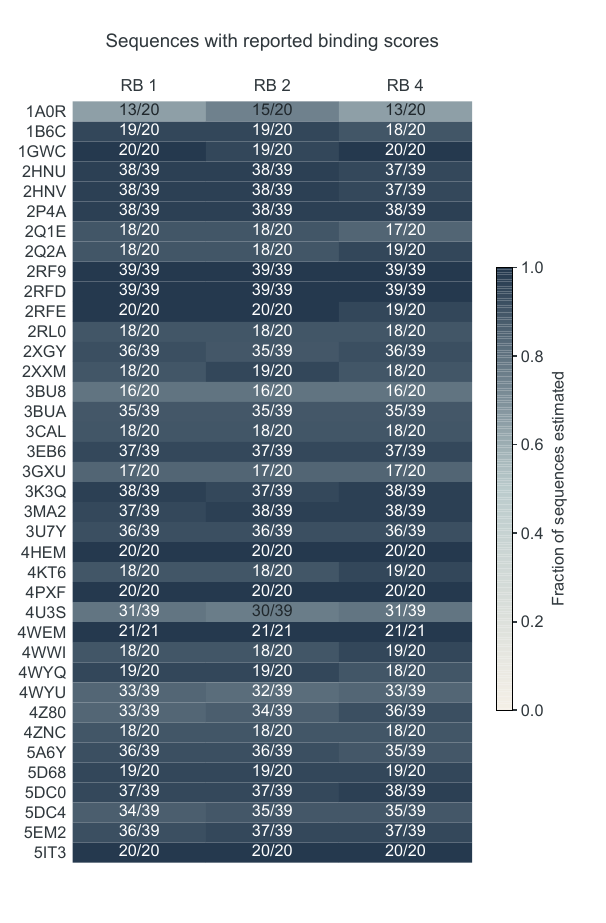}
\caption{Sequences with reported binding scores in finished runs at each
interaction residual budget (RB, kcal/mol). Each cell is
the number with partition function estimates for the protein, ligand, and complex molecular states, divided
by the full sequence count. Runs for all 38 protein--ligand complexes finished
at all three RB values.
These counts describe outputs; correlation eligibility
additionally requires finite scores and the appropriate MARK* intersection.}
\label{fig:rb-estimated-sequences}
\end{figure}

\setcounter{secnumdepth}{2}
\subsection{Sequence ranking agreement between deterministic baselines}
\label{sec:deterministic-ranking}
\begin{figure}[!htbp]
\centering
\includegraphics[width=\textwidth,height=.83\textheight,keepaspectratio]{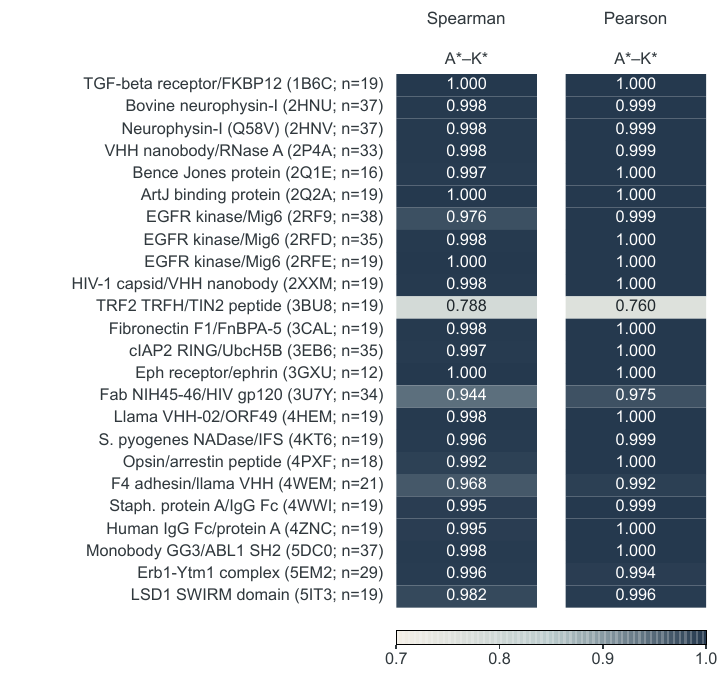}
\caption{Spearman and Pearson correlations between A*--K* and MARK* binding
scores within each system. Each row gives the system name, PDB ID, and
number of system--sequence observations. Residue positions are listed
in Table~\ref{tab:sequence-comparison-designs}. The two panels use the same
color scale (0.7--1.0) as Figure~\ref{fig:rb-system-correlations}; numeric
entries preserve small differences near 1.}
\label{fig:deterministic-system-correlations}
\end{figure}

As a reference for interpreting sequence ranking agreement, we revisited our
earlier comparison of the deterministic A*--K* and MARK* implementations on
benchmark systems with continuous side-chain flexibility.
Both methods used user-specified accuracy
$\varepsilon=0.683$ for each molecular state. Within each system, sequences were paired by
identifier, and only sequences with finite binding scores in both methods
were included. Spearman and Pearson correlations were computed within each system and
summarized with equal system weight. Across 24 systems and 591
system--sequence pairs, the median Spearman correlation was 0.997, and 19 of
the 24 systems had correlations of at least 0.99.
Table~\ref{tab:deterministic-ranking} summarizes the correlations and binding
score differences; Figure~\ref{fig:deterministic-system-correlations} gives
each system's correlations and sample size, using the same presentation as
Figure~\ref{fig:rb-system-correlations}. The shared systems and residue
positions are listed once in Table~\ref{tab:sequence-comparison-designs}.
This historical comparison and the current PACK*--MARK* analysis used their
respective eligible sequence sets.

\begin{table}[htbp]
\centering
\caption{Agreement between A*--K* and MARK* binding scores. Pairs counts
system--sequence observations with finite binding scores in both methods.
Correlations are medians of the correlations within each system, with equal
system weight. Log score diff. is the median across systems of their median
absolute $\log_{10}$ binding score differences from MARK*.}
\label{tab:deterministic-ranking}
\small
\begin{tabular}{rrrrr}
\toprule
Systems & Pairs & Spearman & Pearson & Log score diff. \\
\midrule
24 & 591 & 0.9973 & 0.9997 & 0.1600 \\
\bottomrule
\end{tabular}

\end{table}

\subsection{Runtime comparison protocol}
A protein design problem specifies a structure, mutable and flexible
residue positions, and a sequence list.
Runs were paired by design problem and full sequence list, with the same mutable
and flexible residue positions, structures, user-specified accuracy, and constrained local minimization protocol.
We included a timing when the full sequence list had been processed
with consistent output handling and total runtime was available.
We computed speedups as MARK* runtime divided by PACK* runtime
for the \stat{runtimepairs} design problems with completed timings
from both methods (Table~\ref{tab:frontier-pairs}).

The cohort contains \stat{runtimedesigns} protein design problems based on
\stat{runtimesystems} protein--ligand complexes.
Figure~\ref{MAIN-fig:census-reference}B shows the available timings from
this cohort. MARK* ran on AMD EPYC 9554 processors, using 64 CPU threads.
PACK* used the same CPUs with 128 threads and either zero or four NVIDIA RTX PRO 6000 GPUs. These allocations and the interaction choices
below define the runtime comparison. The ratio for a design problem
uses measured times from both methods. 14-day MARK* timeouts
are shown as time lower bounds and upward arrows. Dividing each program-time
lower bound by the corresponding PACK* time gives the speedup lower bound,
rounded down; startup and scheduler teardown are excluded from timeout bounds.
Coincident timeout arrows are offset horizontally for visibility;
the underlying numbers of flexible residue positions are unchanged.
Table~\ref{tab:runtime-designs} specifies each design problem by its
source PDB structure and its mutable and flexible residue positions.
The PDB ID and total number of mutable and flexible residue positions identify
the corresponding row in Table~\ref{tab:frontier-pairs}.

MARK* used full interactions; PACK* used the sparse-interaction energy function at RB=1 kcal/mol.
Total runtime was measured from program startup
through completion.

\begingroup
\scriptsize
\begin{longtable}{rp{0.17\linewidth}p{0.66\linewidth}}
\caption{Protein design problems used in the runtime comparison. Source structures and protein--ligand interfaces identify the systems; each row lists the total number of mutable and flexible residue positions and their chain identifiers and residue numbers. These positions define the corresponding row in Table~\ref{tab:frontier-pairs}.}\label{tab:runtime-designs}\\
\toprule
Positions & Mutable residues & Flexible residues \\
\midrule
\endfirsthead
\toprule
Positions & Mutable residues & Flexible residues \\
\midrule
\endhead
\bottomrule
\endfoot
\multicolumn{3}{p{0.94\linewidth}}{\textbf{Phosducin/transducin (PDB 1A0R)}}\\*
8 & G390 & G387,\allowbreak{} G392,\allowbreak{} B280,\allowbreak{} B281,\allowbreak{} B283,\allowbreak{} B279,\allowbreak{} B284 \\
9 & G390 & G387,\allowbreak{} G392,\allowbreak{} G386,\allowbreak{} B280,\allowbreak{} B281,\allowbreak{} B283,\allowbreak{} B279,\allowbreak{} B284 \\
10 & G390 & G387,\allowbreak{} G392,\allowbreak{} G386,\allowbreak{} G393,\allowbreak{} B280,\allowbreak{} B281,\allowbreak{} B283,\allowbreak{} B279,\allowbreak{} B284 \\
\multicolumn{3}{p{0.94\linewidth}}{\textbf{TGF-beta receptor/FKBP12 (PDB 1B6C)}}\\*
16 & E965 & F994,\allowbreak{} F995,\allowbreak{} E903,\allowbreak{} E892,\allowbreak{} E957,\allowbreak{} E925,\allowbreak{} E963,\allowbreak{} E947,\allowbreak{} E922,\allowbreak{} E942,\allowbreak{} E940,\allowbreak{} E967,\allowbreak{} E948,\allowbreak{} E893,\allowbreak{} E964 \\
17 & E965 & F994,\allowbreak{} F995,\allowbreak{} E903,\allowbreak{} E892,\allowbreak{} E957,\allowbreak{} E925,\allowbreak{} E963,\allowbreak{} E947,\allowbreak{} E922,\allowbreak{} E942,\allowbreak{} E940,\allowbreak{} E967,\allowbreak{} E948,\allowbreak{} E893,\allowbreak{} E964,\allowbreak{} E956 \\
18 & E965 & F994,\allowbreak{} F995,\allowbreak{} E903,\allowbreak{} E892,\allowbreak{} E957,\allowbreak{} E925,\allowbreak{} E963,\allowbreak{} E947,\allowbreak{} E922,\allowbreak{} E942,\allowbreak{} E940,\allowbreak{} E967,\allowbreak{} E948,\allowbreak{} E893,\allowbreak{} E964,\allowbreak{} E956,\allowbreak{} E966 \\
19 & E965 & F994,\allowbreak{} F995,\allowbreak{} E903,\allowbreak{} E892,\allowbreak{} E957,\allowbreak{} E925,\allowbreak{} E963,\allowbreak{} E947,\allowbreak{} E922,\allowbreak{} E942,\allowbreak{} E940,\allowbreak{} E967,\allowbreak{} E948,\allowbreak{} E893,\allowbreak{} E964,\allowbreak{} E956,\allowbreak{} E966,\allowbreak{} E962 \\
\multicolumn{3}{p{0.94\linewidth}}{\textbf{Glutathione S-transferase (PDB 1GWC)}}\\*
13 & C514 & B314,\allowbreak{} B309,\allowbreak{} B313,\allowbreak{} B311,\allowbreak{} B317,\allowbreak{} C510,\allowbreak{} C505,\allowbreak{} C446,\allowbreak{} C503,\allowbreak{} C511,\allowbreak{} C498,\allowbreak{} C500 \\
14 & C514 & B314,\allowbreak{} B309,\allowbreak{} B313,\allowbreak{} B311,\allowbreak{} B317,\allowbreak{} C510,\allowbreak{} C505,\allowbreak{} C446,\allowbreak{} C503,\allowbreak{} C511,\allowbreak{} C498,\allowbreak{} C500,\allowbreak{} C497 \\
15 & C514 & B314,\allowbreak{} B309,\allowbreak{} B313,\allowbreak{} B311,\allowbreak{} B317,\allowbreak{} C510,\allowbreak{} C505,\allowbreak{} C446,\allowbreak{} C503,\allowbreak{} C511,\allowbreak{} C498,\allowbreak{} C500,\allowbreak{} C497,\allowbreak{} C445 \\
16 & C514 & B314,\allowbreak{} B309,\allowbreak{} B313,\allowbreak{} B311,\allowbreak{} B317,\allowbreak{} C510,\allowbreak{} C505,\allowbreak{} C446,\allowbreak{} C503,\allowbreak{} C511,\allowbreak{} C498,\allowbreak{} C500,\allowbreak{} C497,\allowbreak{} C445,\allowbreak{} C447 \\
\multicolumn{3}{p{0.94\linewidth}}{\textbf{Bovine neurophysin-I (PDB 2HNU)}}\\*
16 & D309,\allowbreak{} D318 & D311,\allowbreak{} D271,\allowbreak{} D316,\allowbreak{} D306,\allowbreak{} D305,\allowbreak{} C192,\allowbreak{} C182,\allowbreak{} C233,\allowbreak{} C181,\allowbreak{} D319,\allowbreak{} C193,\allowbreak{} C191,\allowbreak{} C180,\allowbreak{} C234 \\
18 & D309,\allowbreak{} D318 & D311,\allowbreak{} D271,\allowbreak{} D316,\allowbreak{} D306,\allowbreak{} D305,\allowbreak{} C192,\allowbreak{} C182,\allowbreak{} C233,\allowbreak{} C181,\allowbreak{} D319,\allowbreak{} C193,\allowbreak{} C191,\allowbreak{} C180,\allowbreak{} C234,\allowbreak{} D307,\allowbreak{} C232 \\
19 & D309,\allowbreak{} D318 & D311,\allowbreak{} D271,\allowbreak{} D316,\allowbreak{} D306,\allowbreak{} D305,\allowbreak{} C192,\allowbreak{} C182,\allowbreak{} C233,\allowbreak{} C181,\allowbreak{} D319,\allowbreak{} C193,\allowbreak{} C191,\allowbreak{} C180,\allowbreak{} C234,\allowbreak{} D307,\allowbreak{} C232,\allowbreak{} D308 \\
22 & D309,\allowbreak{} D318 & D311,\allowbreak{} D271,\allowbreak{} D316,\allowbreak{} D306,\allowbreak{} D305,\allowbreak{} C192,\allowbreak{} C182,\allowbreak{} C233,\allowbreak{} C181,\allowbreak{} D319,\allowbreak{} C193,\allowbreak{} C191,\allowbreak{} C180,\allowbreak{} C234,\allowbreak{} D307,\allowbreak{} C232,\allowbreak{} D308,\allowbreak{} D267,\allowbreak{} C179,\allowbreak{} C197 \\
\multicolumn{3}{p{0.94\linewidth}}{\textbf{Neurophysin-I (Q58V) (PDB 2HNV)}}\\*
13 & C237,\allowbreak{} C228 & D273,\allowbreak{} D262,\allowbreak{} D314,\allowbreak{} D263,\allowbreak{} C235,\allowbreak{} C224,\allowbreak{} C225,\allowbreak{} C230,\allowbreak{} C190,\allowbreak{} C238,\allowbreak{} D261 \\
14 & C237,\allowbreak{} C228 & D273,\allowbreak{} D262,\allowbreak{} D314,\allowbreak{} D263,\allowbreak{} C235,\allowbreak{} C224,\allowbreak{} C225,\allowbreak{} C230,\allowbreak{} C190,\allowbreak{} C238,\allowbreak{} D261,\allowbreak{} D274 \\
15 & C237,\allowbreak{} C228 & D273,\allowbreak{} D262,\allowbreak{} D314,\allowbreak{} D263,\allowbreak{} C235,\allowbreak{} C224,\allowbreak{} C225,\allowbreak{} C230,\allowbreak{} C190,\allowbreak{} C238,\allowbreak{} D261,\allowbreak{} D274,\allowbreak{} D315 \\
\multicolumn{3}{p{0.94\linewidth}}{\textbf{VHH nanobody/RNase A (PDB 2P4A)}}\\*
17 & D476,\allowbreak{} D475 & B226,\allowbreak{} B225,\allowbreak{} B228,\allowbreak{} A111,\allowbreak{} A69,\allowbreak{} A71,\allowbreak{} B224,\allowbreak{} A112,\allowbreak{} B229,\allowbreak{} A68,\allowbreak{} B227,\allowbreak{} A70,\allowbreak{} D474,\allowbreak{} D477,\allowbreak{} B231 \\
18 & D476,\allowbreak{} D475 & B226,\allowbreak{} B225,\allowbreak{} B228,\allowbreak{} A111,\allowbreak{} A69,\allowbreak{} A71,\allowbreak{} B224,\allowbreak{} A112,\allowbreak{} B229,\allowbreak{} A68,\allowbreak{} B227,\allowbreak{} A70,\allowbreak{} D474,\allowbreak{} D477,\allowbreak{} B231,\allowbreak{} C356 \\
20 & D476,\allowbreak{} D475 & B226,\allowbreak{} B225,\allowbreak{} B228,\allowbreak{} A111,\allowbreak{} A69,\allowbreak{} A71,\allowbreak{} B224,\allowbreak{} A112,\allowbreak{} B229,\allowbreak{} A68,\allowbreak{} B227,\allowbreak{} A70,\allowbreak{} D474,\allowbreak{} D477,\allowbreak{} B231,\allowbreak{} C356,\allowbreak{} D469,\allowbreak{} C357 \\
\multicolumn{3}{p{0.94\linewidth}}{\textbf{Bence Jones protein (PDB 2Q1E)}}\\*
13 & D418 & D427,\allowbreak{} D365,\allowbreak{} D363,\allowbreak{} D420,\allowbreak{} D416,\allowbreak{} D425,\allowbreak{} C309,\allowbreak{} C316,\allowbreak{} C314,\allowbreak{} C315,\allowbreak{} D364,\allowbreak{} D426 \\
14 & D418 & D427,\allowbreak{} D365,\allowbreak{} D363,\allowbreak{} D420,\allowbreak{} D416,\allowbreak{} D425,\allowbreak{} C309,\allowbreak{} C316,\allowbreak{} C314,\allowbreak{} C315,\allowbreak{} D364,\allowbreak{} D426,\allowbreak{} D362 \\
16 & D418 & D427,\allowbreak{} D365,\allowbreak{} D363,\allowbreak{} D420,\allowbreak{} D416,\allowbreak{} D425,\allowbreak{} C309,\allowbreak{} C316,\allowbreak{} C314,\allowbreak{} C315,\allowbreak{} D364,\allowbreak{} D426,\allowbreak{} D362,\allowbreak{} D366,\allowbreak{} D428 \\
\multicolumn{3}{p{0.94\linewidth}}{\textbf{ArtJ binding protein (PDB 2Q2A)}}\\*
13 & B385 & B403,\allowbreak{} B400,\allowbreak{} B383,\allowbreak{} B362,\allowbreak{} B363,\allowbreak{} B391,\allowbreak{} B394,\allowbreak{} B390,\allowbreak{} A144,\allowbreak{} A156,\allowbreak{} A149,\allowbreak{} A153 \\
14 & B385 & B403,\allowbreak{} B400,\allowbreak{} B383,\allowbreak{} B362,\allowbreak{} B363,\allowbreak{} B391,\allowbreak{} B394,\allowbreak{} B390,\allowbreak{} A144,\allowbreak{} A156,\allowbreak{} A149,\allowbreak{} A153,\allowbreak{} B404 \\
15 & B385 & B403,\allowbreak{} B400,\allowbreak{} B383,\allowbreak{} B362,\allowbreak{} B363,\allowbreak{} B391,\allowbreak{} B394,\allowbreak{} B390,\allowbreak{} A144,\allowbreak{} A156,\allowbreak{} A149,\allowbreak{} A153,\allowbreak{} B404,\allowbreak{} B402 \\
\multicolumn{3}{p{0.94\linewidth}}{\textbf{EGFR kinase/Mig6 (PDB 2RF9)}}\\*
13 & D586,\allowbreak{} D585 & D582,\allowbreak{} D578,\allowbreak{} B499,\allowbreak{} B503,\allowbreak{} B504,\allowbreak{} B498,\allowbreak{} D581,\allowbreak{} D583,\allowbreak{} D587,\allowbreak{} D579,\allowbreak{} D584 \\
14 & D586,\allowbreak{} D585 & D582,\allowbreak{} D578,\allowbreak{} B499,\allowbreak{} B503,\allowbreak{} B504,\allowbreak{} B498,\allowbreak{} D581,\allowbreak{} D583,\allowbreak{} D587,\allowbreak{} D579,\allowbreak{} D584,\allowbreak{} B502 \\
15 & D586,\allowbreak{} D585 & D582,\allowbreak{} D578,\allowbreak{} B499,\allowbreak{} B503,\allowbreak{} B504,\allowbreak{} B498,\allowbreak{} D581,\allowbreak{} D583,\allowbreak{} D587,\allowbreak{} D579,\allowbreak{} D584,\allowbreak{} B502,\allowbreak{} B500 \\
\multicolumn{3}{p{0.94\linewidth}}{\textbf{EGFR kinase/Mig6 (PDB 2RFD)}}\\*
9 & B547,\allowbreak{} B544 & B543,\allowbreak{} B541,\allowbreak{} B540,\allowbreak{} B531,\allowbreak{} B530,\allowbreak{} A238,\allowbreak{} B545 \\
10 & B547,\allowbreak{} B544 & B543,\allowbreak{} B541,\allowbreak{} B540,\allowbreak{} B531,\allowbreak{} B530,\allowbreak{} A238,\allowbreak{} B545,\allowbreak{} B553 \\
11 & B547,\allowbreak{} B544 & B543,\allowbreak{} B541,\allowbreak{} B540,\allowbreak{} B531,\allowbreak{} B530,\allowbreak{} A238,\allowbreak{} B545,\allowbreak{} B553,\allowbreak{} A239 \\
12 & B547,\allowbreak{} B544 & B543,\allowbreak{} B541,\allowbreak{} B540,\allowbreak{} B531,\allowbreak{} B530,\allowbreak{} A238,\allowbreak{} B545,\allowbreak{} B553,\allowbreak{} A239,\allowbreak{} A237 \\
\multicolumn{3}{p{0.94\linewidth}}{\textbf{EGFR kinase/Mig6 (PDB 2RFE)}}\\*
12 & B488 & F1138,\allowbreak{} B500,\allowbreak{} B499,\allowbreak{} B504,\allowbreak{} B506,\allowbreak{} B475,\allowbreak{} B471,\allowbreak{} B452,\allowbreak{} B524,\allowbreak{} B505,\allowbreak{} B496 \\
13 & B488 & F1138,\allowbreak{} B500,\allowbreak{} B499,\allowbreak{} B504,\allowbreak{} B506,\allowbreak{} B475,\allowbreak{} B471,\allowbreak{} B452,\allowbreak{} B524,\allowbreak{} B505,\allowbreak{} B496,\allowbreak{} B491 \\
14 & B488 & F1138,\allowbreak{} B500,\allowbreak{} B499,\allowbreak{} B504,\allowbreak{} B506,\allowbreak{} B475,\allowbreak{} B471,\allowbreak{} B452,\allowbreak{} B524,\allowbreak{} B505,\allowbreak{} B496,\allowbreak{} B491,\allowbreak{} B501 \\
\multicolumn{3}{p{0.94\linewidth}}{\textbf{Fibronectin F1/FnBPA-5 (PDB 2RL0)}}\\*
12 & A85 & G95,\allowbreak{} G97,\allowbreak{} A75,\allowbreak{} A65,\allowbreak{} A63,\allowbreak{} A49,\allowbreak{} A48,\allowbreak{} A51,\allowbreak{} A50,\allowbreak{} A87,\allowbreak{} A77 \\
13 & A85 & G95,\allowbreak{} G97,\allowbreak{} A75,\allowbreak{} A65,\allowbreak{} A63,\allowbreak{} A49,\allowbreak{} A48,\allowbreak{} A51,\allowbreak{} A50,\allowbreak{} A87,\allowbreak{} A77,\allowbreak{} A76 \\
15 & A85 & G95,\allowbreak{} G97,\allowbreak{} A75,\allowbreak{} A65,\allowbreak{} A63,\allowbreak{} A49,\allowbreak{} A48,\allowbreak{} A51,\allowbreak{} A50,\allowbreak{} A87,\allowbreak{} A77,\allowbreak{} A76,\allowbreak{} G96,\allowbreak{} G94 \\
16 & A85 & G95,\allowbreak{} G97,\allowbreak{} A75,\allowbreak{} A65,\allowbreak{} A63,\allowbreak{} A49,\allowbreak{} A48,\allowbreak{} A51,\allowbreak{} A50,\allowbreak{} A87,\allowbreak{} A77,\allowbreak{} A76,\allowbreak{} G96,\allowbreak{} G94,\allowbreak{} G98 \\
\multicolumn{3}{p{0.94\linewidth}}{\textbf{RELIK capsid/cyclophilin A (PDB 2XGY)}}\\*
11 & B189,\allowbreak{} B184 & B248,\allowbreak{} B242,\allowbreak{} B277,\allowbreak{} B192,\allowbreak{} B190,\allowbreak{} B186,\allowbreak{} B183,\allowbreak{} A83,\allowbreak{} A76 \\
12 & B189,\allowbreak{} B184 & B248,\allowbreak{} B242,\allowbreak{} B277,\allowbreak{} B192,\allowbreak{} B190,\allowbreak{} B186,\allowbreak{} B183,\allowbreak{} B250,\allowbreak{} A83,\allowbreak{} A76 \\
13 & B189,\allowbreak{} B184 & B248,\allowbreak{} B242,\allowbreak{} B277,\allowbreak{} B192,\allowbreak{} B190,\allowbreak{} B186,\allowbreak{} B183,\allowbreak{} B250,\allowbreak{} B251,\allowbreak{} A83,\allowbreak{} A76 \\
\multicolumn{3}{p{0.94\linewidth}}{\textbf{HIV-1 capsid/VHH nanobody (PDB 2XXM)}}\\*
12 & B107 & B115,\allowbreak{} B117,\allowbreak{} B109,\allowbreak{} B173,\allowbreak{} B105,\allowbreak{} B167,\allowbreak{} B165,\allowbreak{} A30,\allowbreak{} A4,\allowbreak{} A6,\allowbreak{} A5 \\
13 & B107 & B115,\allowbreak{} B117,\allowbreak{} B109,\allowbreak{} B173,\allowbreak{} B105,\allowbreak{} B167,\allowbreak{} B165,\allowbreak{} A30,\allowbreak{} A4,\allowbreak{} A6,\allowbreak{} A5,\allowbreak{} B106 \\
14 & B107 & B115,\allowbreak{} B117,\allowbreak{} B109,\allowbreak{} B173,\allowbreak{} B105,\allowbreak{} B167,\allowbreak{} B165,\allowbreak{} A30,\allowbreak{} A4,\allowbreak{} A6,\allowbreak{} A5,\allowbreak{} B106,\allowbreak{} B168 \\
15 & B107 & B115,\allowbreak{} B117,\allowbreak{} B109,\allowbreak{} B173,\allowbreak{} B105,\allowbreak{} B167,\allowbreak{} B165,\allowbreak{} A30,\allowbreak{} A4,\allowbreak{} A6,\allowbreak{} A5,\allowbreak{} B106,\allowbreak{} B168,\allowbreak{} B104 \\
\multicolumn{3}{p{0.94\linewidth}}{\textbf{TRF2 TRFH/TIN2 peptide (PDB 3BU8)}}\\*
13 & B243 & D431,\allowbreak{} A34,\allowbreak{} A35,\allowbreak{} D430,\allowbreak{} B242,\allowbreak{} A36,\allowbreak{} D432,\allowbreak{} B244,\allowbreak{} A33,\allowbreak{} B246,\allowbreak{} A37,\allowbreak{} A32 \\
14 & B243 & D431,\allowbreak{} A34,\allowbreak{} A35,\allowbreak{} D430,\allowbreak{} B242,\allowbreak{} A36,\allowbreak{} D432,\allowbreak{} B244,\allowbreak{} A33,\allowbreak{} B246,\allowbreak{} A37,\allowbreak{} A32,\allowbreak{} B240 \\
15 & B243 & D431,\allowbreak{} A34,\allowbreak{} A35,\allowbreak{} D430,\allowbreak{} B242,\allowbreak{} A36,\allowbreak{} D432,\allowbreak{} B244,\allowbreak{} A33,\allowbreak{} B246,\allowbreak{} A37,\allowbreak{} A32,\allowbreak{} B240,\allowbreak{} A31 \\
16 & B243 & D431,\allowbreak{} A34,\allowbreak{} A35,\allowbreak{} D430,\allowbreak{} B242,\allowbreak{} A36,\allowbreak{} D432,\allowbreak{} B244,\allowbreak{} A33,\allowbreak{} B246,\allowbreak{} A37,\allowbreak{} A32,\allowbreak{} B240,\allowbreak{} A31,\allowbreak{} A38 \\
\multicolumn{3}{p{0.94\linewidth}}{\textbf{TRF2 TRFH/Apollo peptide (PDB 3BUA)}}\\*
14 & C447,\allowbreak{} C450 & D638,\allowbreak{} D641,\allowbreak{} D642,\allowbreak{} D645,\allowbreak{} D623,\allowbreak{} D619,\allowbreak{} D616,\allowbreak{} D620,\allowbreak{} C417,\allowbreak{} C414,\allowbreak{} C410,\allowbreak{} C421 \\
15 & C447,\allowbreak{} C450 & D638,\allowbreak{} D641,\allowbreak{} D642,\allowbreak{} D645,\allowbreak{} D623,\allowbreak{} D619,\allowbreak{} D616,\allowbreak{} D620,\allowbreak{} C417,\allowbreak{} C414,\allowbreak{} C410,\allowbreak{} C421,\allowbreak{} C452 \\
16 & C447,\allowbreak{} C450 & D638,\allowbreak{} D641,\allowbreak{} D642,\allowbreak{} D645,\allowbreak{} D623,\allowbreak{} D619,\allowbreak{} D616,\allowbreak{} D620,\allowbreak{} C417,\allowbreak{} C414,\allowbreak{} C410,\allowbreak{} C421,\allowbreak{} C452,\allowbreak{} C443 \\
\multicolumn{3}{p{0.94\linewidth}}{\textbf{Fibronectin F1/FnBPA-5 (PDB 3CAL)}}\\*
12 & C147 & D212,\allowbreak{} C126,\allowbreak{} C128,\allowbreak{} C113,\allowbreak{} C137,\allowbreak{} C111,\allowbreak{} C139,\allowbreak{} C120,\allowbreak{} C149,\allowbreak{} C125,\allowbreak{} C148 \\
13 & C147 & D212,\allowbreak{} C126,\allowbreak{} C128,\allowbreak{} C113,\allowbreak{} C137,\allowbreak{} C111,\allowbreak{} C139,\allowbreak{} C120,\allowbreak{} C149,\allowbreak{} C125,\allowbreak{} C148,\allowbreak{} C127 \\
14 & C147 & D212,\allowbreak{} C126,\allowbreak{} C128,\allowbreak{} C113,\allowbreak{} C137,\allowbreak{} C111,\allowbreak{} C139,\allowbreak{} C120,\allowbreak{} C149,\allowbreak{} C125,\allowbreak{} C148,\allowbreak{} C127,\allowbreak{} C146 \\
15 & C147 & D212,\allowbreak{} C126,\allowbreak{} C128,\allowbreak{} C113,\allowbreak{} C137,\allowbreak{} C111,\allowbreak{} C139,\allowbreak{} C120,\allowbreak{} C149,\allowbreak{} C125,\allowbreak{} C148,\allowbreak{} C127,\allowbreak{} C146,\allowbreak{} C112 \\
\multicolumn{3}{p{0.94\linewidth}}{\textbf{cIAP2 RING/UbcH5B (PDB 3EB6)}}\\*
13 & A44,\allowbreak{} A45 & B128,\allowbreak{} A55,\allowbreak{} A29,\allowbreak{} A19,\allowbreak{} A42,\allowbreak{} A41,\allowbreak{} A37,\allowbreak{} A30,\allowbreak{} A38,\allowbreak{} A28,\allowbreak{} A54 \\
14 & A44,\allowbreak{} A45 & B128,\allowbreak{} A55,\allowbreak{} A29,\allowbreak{} A19,\allowbreak{} A42,\allowbreak{} A41,\allowbreak{} A37,\allowbreak{} A30,\allowbreak{} A38,\allowbreak{} A28,\allowbreak{} A54,\allowbreak{} A40 \\
15 & A44,\allowbreak{} A45 & B128,\allowbreak{} A55,\allowbreak{} A29,\allowbreak{} A19,\allowbreak{} A42,\allowbreak{} A41,\allowbreak{} A37,\allowbreak{} A30,\allowbreak{} A38,\allowbreak{} A28,\allowbreak{} A54,\allowbreak{} A40,\allowbreak{} B129 \\
16 & A44,\allowbreak{} A45 & B128,\allowbreak{} A55,\allowbreak{} A29,\allowbreak{} A19,\allowbreak{} A42,\allowbreak{} A41,\allowbreak{} A37,\allowbreak{} A30,\allowbreak{} A38,\allowbreak{} A28,\allowbreak{} A54,\allowbreak{} A40,\allowbreak{} B129,\allowbreak{} A56 \\
\multicolumn{3}{p{0.94\linewidth}}{\textbf{Eph receptor/ephrin (PDB 3GXU)}}\\*
13 & A30 & B249,\allowbreak{} B250,\allowbreak{} B251,\allowbreak{} B252,\allowbreak{} A29,\allowbreak{} A31,\allowbreak{} B248,\allowbreak{} B204,\allowbreak{} A40,\allowbreak{} B206,\allowbreak{} B205,\allowbreak{} B203 \\
14 & A30 & B249,\allowbreak{} B250,\allowbreak{} B251,\allowbreak{} B252,\allowbreak{} A29,\allowbreak{} A31,\allowbreak{} B248,\allowbreak{} B204,\allowbreak{} A40,\allowbreak{} B206,\allowbreak{} B205,\allowbreak{} B203,\allowbreak{} A39 \\
15 & A30 & B249,\allowbreak{} B250,\allowbreak{} B251,\allowbreak{} B252,\allowbreak{} A29,\allowbreak{} A31,\allowbreak{} B248,\allowbreak{} B204,\allowbreak{} A40,\allowbreak{} B206,\allowbreak{} B205,\allowbreak{} B203,\allowbreak{} A39,\allowbreak{} A41 \\
16 & A30 & B249,\allowbreak{} B250,\allowbreak{} B251,\allowbreak{} B252,\allowbreak{} A29,\allowbreak{} A31,\allowbreak{} B248,\allowbreak{} B204,\allowbreak{} A40,\allowbreak{} B206,\allowbreak{} B205,\allowbreak{} B203,\allowbreak{} A39,\allowbreak{} A41,\allowbreak{} B239 \\
\multicolumn{3}{p{0.94\linewidth}}{\textbf{Llama antibody/botulinum toxin A (PDB 3K3Q)}}\\*
15 & C444,\allowbreak{} C440 & B254,\allowbreak{} B316,\allowbreak{} B253,\allowbreak{} B263,\allowbreak{} B358,\allowbreak{} B284,\allowbreak{} C416,\allowbreak{} C456,\allowbreak{} C454,\allowbreak{} C446,\allowbreak{} C436,\allowbreak{} C420,\allowbreak{} C413 \\
16 & C444,\allowbreak{} C440 & B254,\allowbreak{} B316,\allowbreak{} B253,\allowbreak{} B263,\allowbreak{} B358,\allowbreak{} B284,\allowbreak{} C416,\allowbreak{} C456,\allowbreak{} C454,\allowbreak{} C446,\allowbreak{} C436,\allowbreak{} C420,\allowbreak{} C413,\allowbreak{} C437 \\
17 & C444,\allowbreak{} C440 & B254,\allowbreak{} B316,\allowbreak{} B253,\allowbreak{} B263,\allowbreak{} B358,\allowbreak{} B284,\allowbreak{} C416,\allowbreak{} C456,\allowbreak{} C454,\allowbreak{} C446,\allowbreak{} C436,\allowbreak{} C420,\allowbreak{} C413,\allowbreak{} C437,\allowbreak{} C439 \\
18 & C444,\allowbreak{} C440 & B254,\allowbreak{} B316,\allowbreak{} B253,\allowbreak{} B263,\allowbreak{} B358,\allowbreak{} B284,\allowbreak{} C416,\allowbreak{} C456,\allowbreak{} C454,\allowbreak{} C446,\allowbreak{} C436,\allowbreak{} C420,\allowbreak{} C413,\allowbreak{} C437,\allowbreak{} C439,\allowbreak{} C453 \\
\multicolumn{3}{p{0.94\linewidth}}{\textbf{MT1-MMP/TIMP-1 (PDB 3MA2)}}\\*
14 & A316,\allowbreak{} A254 & B341,\allowbreak{} B343,\allowbreak{} A276,\allowbreak{} A290,\allowbreak{} A291,\allowbreak{} A271,\allowbreak{} A277,\allowbreak{} A287,\allowbreak{} A318,\allowbreak{} A274,\allowbreak{} A255,\allowbreak{} A315 \\
15 & A316,\allowbreak{} A254 & B341,\allowbreak{} B343,\allowbreak{} A276,\allowbreak{} A290,\allowbreak{} A291,\allowbreak{} A271,\allowbreak{} A277,\allowbreak{} A287,\allowbreak{} A318,\allowbreak{} A274,\allowbreak{} A255,\allowbreak{} A315,\allowbreak{} A286 \\
16 & A316,\allowbreak{} A254 & B341,\allowbreak{} B343,\allowbreak{} A276,\allowbreak{} A290,\allowbreak{} A291,\allowbreak{} A271,\allowbreak{} A277,\allowbreak{} A287,\allowbreak{} A318,\allowbreak{} A274,\allowbreak{} A255,\allowbreak{} A315,\allowbreak{} A286,\allowbreak{} A317 \\
\multicolumn{3}{p{0.94\linewidth}}{\textbf{Fab NIH45-46/HIV gp120 (PDB 3U7Y)}}\\*
8 & G384,\allowbreak{} G382 & G385,\allowbreak{} L587,\allowbreak{} L649,\allowbreak{} L648,\allowbreak{} G386,\allowbreak{} L586 \\
9 & G384,\allowbreak{} G382 & G385,\allowbreak{} L587,\allowbreak{} L649,\allowbreak{} L648,\allowbreak{} G386,\allowbreak{} L586,\allowbreak{} L650 \\
10 & G384,\allowbreak{} G382 & G385,\allowbreak{} L587,\allowbreak{} L649,\allowbreak{} L648,\allowbreak{} G386,\allowbreak{} L586,\allowbreak{} L650,\allowbreak{} G383 \\
11 & G384,\allowbreak{} G382 & G385,\allowbreak{} L587,\allowbreak{} L649,\allowbreak{} L648,\allowbreak{} G386,\allowbreak{} L586,\allowbreak{} L650,\allowbreak{} G383,\allowbreak{} G381 \\
12 & G384,\allowbreak{} G382 & G385,\allowbreak{} L587,\allowbreak{} L649,\allowbreak{} L648,\allowbreak{} G386,\allowbreak{} L586,\allowbreak{} L650,\allowbreak{} G383,\allowbreak{} G381,\allowbreak{} L588 \\
\multicolumn{3}{p{0.94\linewidth}}{\textbf{Llama VHH-02/ORF49 (PDB 4HEM)}}\\*
12 & C372 & B226,\allowbreak{} C326,\allowbreak{} C303,\allowbreak{} C318,\allowbreak{} C360,\allowbreak{} C377,\allowbreak{} C322,\allowbreak{} C332,\allowbreak{} C328,\allowbreak{} C330,\allowbreak{} C353 \\
13 & C372 & B226,\allowbreak{} C326,\allowbreak{} C303,\allowbreak{} C318,\allowbreak{} C360,\allowbreak{} C377,\allowbreak{} C322,\allowbreak{} C332,\allowbreak{} C328,\allowbreak{} C330,\allowbreak{} C353,\allowbreak{} C331 \\
14 & C372 & B226,\allowbreak{} C326,\allowbreak{} C303,\allowbreak{} C318,\allowbreak{} C360,\allowbreak{} C377,\allowbreak{} C322,\allowbreak{} C332,\allowbreak{} C328,\allowbreak{} C330,\allowbreak{} C353,\allowbreak{} C331,\allowbreak{} C321 \\
\multicolumn{3}{p{0.94\linewidth}}{\textbf{S. pyogenes NADase/IFS (PDB 4KT6)}}\\*
9 & D712 & D822,\allowbreak{} D815,\allowbreak{} C512,\allowbreak{} C511,\allowbreak{} C509,\allowbreak{} C519,\allowbreak{} C515,\allowbreak{} C518 \\
10 & D712 & D822,\allowbreak{} D815,\allowbreak{} D759,\allowbreak{} C512,\allowbreak{} C511,\allowbreak{} C509,\allowbreak{} C519,\allowbreak{} C515,\allowbreak{} C518 \\
11 & D712 & D822,\allowbreak{} D815,\allowbreak{} D759,\allowbreak{} C512,\allowbreak{} C511,\allowbreak{} C509,\allowbreak{} C519,\allowbreak{} C515,\allowbreak{} C518,\allowbreak{} D816 \\
\multicolumn{3}{p{0.94\linewidth}}{\textbf{Opsin/arrestin peptide (PDB 4PXF)}}\\*
13 & A249 & B328,\allowbreak{} B330,\allowbreak{} A246,\allowbreak{} A245,\allowbreak{} A226,\allowbreak{} A225,\allowbreak{} A138,\allowbreak{} A229,\allowbreak{} A244,\allowbreak{} A247,\allowbreak{} A227,\allowbreak{} A139 \\
14 & A249 & B328,\allowbreak{} B330,\allowbreak{} A246,\allowbreak{} A245,\allowbreak{} A226,\allowbreak{} A225,\allowbreak{} A138,\allowbreak{} A229,\allowbreak{} A244,\allowbreak{} A247,\allowbreak{} A227,\allowbreak{} A139,\allowbreak{} A250 \\
15 & A249 & B328,\allowbreak{} B330,\allowbreak{} A246,\allowbreak{} A245,\allowbreak{} A226,\allowbreak{} A225,\allowbreak{} A138,\allowbreak{} A229,\allowbreak{} A244,\allowbreak{} A247,\allowbreak{} A227,\allowbreak{} A139,\allowbreak{} A250,\allowbreak{} A228 \\
\multicolumn{3}{p{0.94\linewidth}}{\textbf{Cohesin/dockerin (PDB 4U3S)}}\\*
13 & B280,\allowbreak{} B277 & B309,\allowbreak{} B276,\allowbreak{} B273,\allowbreak{} B274,\allowbreak{} A32,\allowbreak{} A76,\allowbreak{} A145,\allowbreak{} A97,\allowbreak{} A94,\allowbreak{} A93,\allowbreak{} A95 \\
14 & B280,\allowbreak{} B277 & B309,\allowbreak{} B276,\allowbreak{} B273,\allowbreak{} B274,\allowbreak{} A32,\allowbreak{} A76,\allowbreak{} A145,\allowbreak{} A97,\allowbreak{} A94,\allowbreak{} A93,\allowbreak{} A95,\allowbreak{} A144 \\
15 & B280,\allowbreak{} B277 & B309,\allowbreak{} B276,\allowbreak{} B273,\allowbreak{} B274,\allowbreak{} A32,\allowbreak{} A76,\allowbreak{} A145,\allowbreak{} A97,\allowbreak{} A94,\allowbreak{} A93,\allowbreak{} A95,\allowbreak{} A144,\allowbreak{} A146 \\
\multicolumn{3}{p{0.94\linewidth}}{\textbf{F4 adhesin/llama VHH (PDB 4WEM)}}\\*
13 & B347 & B357,\allowbreak{} B278,\allowbreak{} A42,\allowbreak{} A4,\allowbreak{} B356,\allowbreak{} A3,\allowbreak{} A43,\allowbreak{} A5,\allowbreak{} A41,\allowbreak{} B279,\allowbreak{} B277,\allowbreak{} A196 \\
14 & B347 & B357,\allowbreak{} B278,\allowbreak{} A42,\allowbreak{} A4,\allowbreak{} B356,\allowbreak{} A3,\allowbreak{} A43,\allowbreak{} A5,\allowbreak{} A41,\allowbreak{} B279,\allowbreak{} B277,\allowbreak{} A196,\allowbreak{} B276 \\
15 & B347 & B357,\allowbreak{} B278,\allowbreak{} A42,\allowbreak{} A4,\allowbreak{} B356,\allowbreak{} A3,\allowbreak{} A43,\allowbreak{} A5,\allowbreak{} A41,\allowbreak{} B279,\allowbreak{} B277,\allowbreak{} A196,\allowbreak{} B276,\allowbreak{} B346 \\
17 & B347 & B357,\allowbreak{} B278,\allowbreak{} A42,\allowbreak{} A4,\allowbreak{} B356,\allowbreak{} A3,\allowbreak{} A43,\allowbreak{} A5,\allowbreak{} A41,\allowbreak{} B279,\allowbreak{} B277,\allowbreak{} A196,\allowbreak{} B276,\allowbreak{} B346,\allowbreak{} A195,\allowbreak{} B355 \\
\multicolumn{3}{p{0.94\linewidth}}{\textbf{Staph. protein A/IgG Fc (PDB 4WWI)}}\\*
9 & B60 & E382,\allowbreak{} E384,\allowbreak{} E540,\allowbreak{} E378,\allowbreak{} E538,\allowbreak{} B62,\allowbreak{} B64,\allowbreak{} B65 \\
11 & B60 & E382,\allowbreak{} E384,\allowbreak{} E540,\allowbreak{} E378,\allowbreak{} E538,\allowbreak{} B62,\allowbreak{} B64,\allowbreak{} B65,\allowbreak{} B58,\allowbreak{} E541 \\
12 & B60 & E382,\allowbreak{} E384,\allowbreak{} E540,\allowbreak{} E378,\allowbreak{} E538,\allowbreak{} B62,\allowbreak{} B64,\allowbreak{} B65,\allowbreak{} B58,\allowbreak{} E541,\allowbreak{} B59 \\
\multicolumn{3}{p{0.94\linewidth}}{\textbf{Dicer/TRBP (PDB 4WYQ)}}\\*
10 & E394 & D288,\allowbreak{} D289,\allowbreak{} D285,\allowbreak{} E367,\allowbreak{} E371,\allowbreak{} E373,\allowbreak{} E365,\allowbreak{} E391,\allowbreak{} E390 \\
11 & E394 & D288,\allowbreak{} D289,\allowbreak{} D285,\allowbreak{} E367,\allowbreak{} E371,\allowbreak{} E373,\allowbreak{} E365,\allowbreak{} E391,\allowbreak{} E390,\allowbreak{} E372 \\
12 & E394 & D288,\allowbreak{} D289,\allowbreak{} D285,\allowbreak{} E367,\allowbreak{} E371,\allowbreak{} E373,\allowbreak{} E365,\allowbreak{} E391,\allowbreak{} E390,\allowbreak{} E372,\allowbreak{} E370 \\
13 & E394 & D288,\allowbreak{} D289,\allowbreak{} D285,\allowbreak{} E367,\allowbreak{} E371,\allowbreak{} E373,\allowbreak{} E365,\allowbreak{} E391,\allowbreak{} E390,\allowbreak{} E372,\allowbreak{} E370,\allowbreak{} D284 \\
14 & E394 & D288,\allowbreak{} D289,\allowbreak{} D285,\allowbreak{} E367,\allowbreak{} E371,\allowbreak{} E373,\allowbreak{} E365,\allowbreak{} E391,\allowbreak{} E390,\allowbreak{} E372,\allowbreak{} E370,\allowbreak{} D284,\allowbreak{} D286 \\
\multicolumn{3}{p{0.94\linewidth}}{\textbf{Scribble PDZ34/peptide (PDB 4WYU)}}\\*
11 & A94,\allowbreak{} A70 & B212,\allowbreak{} B213,\allowbreak{} A12,\allowbreak{} A90,\allowbreak{} A96,\allowbreak{} A93,\allowbreak{} A92,\allowbreak{} A72,\allowbreak{} A69 \\
12 & A94,\allowbreak{} A70 & B212,\allowbreak{} B213,\allowbreak{} A12,\allowbreak{} A90,\allowbreak{} A96,\allowbreak{} A93,\allowbreak{} A92,\allowbreak{} A72,\allowbreak{} A69,\allowbreak{} A88 \\
13 & A94,\allowbreak{} A70 & B212,\allowbreak{} B213,\allowbreak{} A12,\allowbreak{} A90,\allowbreak{} A96,\allowbreak{} A93,\allowbreak{} A92,\allowbreak{} A72,\allowbreak{} A69,\allowbreak{} A88,\allowbreak{} A95 \\
\multicolumn{3}{p{0.94\linewidth}}{\textbf{T. gondii AMA4/RON2 peptide (PDB 4Z80)}}\\*
13 & D967,\allowbreak{} D970 & D972,\allowbreak{} D971,\allowbreak{} B621,\allowbreak{} B620,\allowbreak{} B646,\allowbreak{} B758,\allowbreak{} B756,\allowbreak{} B649,\allowbreak{} B760,\allowbreak{} B563,\allowbreak{} B663 \\
14 & D967,\allowbreak{} D970 & D972,\allowbreak{} D971,\allowbreak{} D969,\allowbreak{} B621,\allowbreak{} B620,\allowbreak{} B646,\allowbreak{} B758,\allowbreak{} B756,\allowbreak{} B649,\allowbreak{} B760,\allowbreak{} B563,\allowbreak{} B663 \\
15 & D967,\allowbreak{} D970 & D972,\allowbreak{} D971,\allowbreak{} D969,\allowbreak{} B621,\allowbreak{} B620,\allowbreak{} B646,\allowbreak{} B758,\allowbreak{} B756,\allowbreak{} B649,\allowbreak{} B760,\allowbreak{} B563,\allowbreak{} B663,\allowbreak{} B759 \\
16 & D967,\allowbreak{} D970 & D972,\allowbreak{} D971,\allowbreak{} D969,\allowbreak{} B621,\allowbreak{} B620,\allowbreak{} B646,\allowbreak{} B758,\allowbreak{} B756,\allowbreak{} B649,\allowbreak{} B760,\allowbreak{} B563,\allowbreak{} B663,\allowbreak{} B759,\allowbreak{} B648 \\
\multicolumn{3}{p{0.94\linewidth}}{\textbf{Human IgG Fc/protein A (PDB 4ZNC)}}\\*
12 & E534 & F716,\allowbreak{} F733,\allowbreak{} F731,\allowbreak{} F718,\allowbreak{} E499,\allowbreak{} E528,\allowbreak{} E497,\allowbreak{} E533,\allowbreak{} E500,\allowbreak{} F732,\allowbreak{} E535 \\
13 & E534 & F716,\allowbreak{} F733,\allowbreak{} F731,\allowbreak{} F718,\allowbreak{} E499,\allowbreak{} E528,\allowbreak{} E497,\allowbreak{} E533,\allowbreak{} E500,\allowbreak{} F732,\allowbreak{} E535,\allowbreak{} E496 \\
14 & E534 & F716,\allowbreak{} F733,\allowbreak{} F731,\allowbreak{} F718,\allowbreak{} E499,\allowbreak{} E528,\allowbreak{} E497,\allowbreak{} E533,\allowbreak{} E500,\allowbreak{} F732,\allowbreak{} E535,\allowbreak{} E496,\allowbreak{} F717 \\
15 & E534 & F716,\allowbreak{} F733,\allowbreak{} F731,\allowbreak{} F718,\allowbreak{} E499,\allowbreak{} E528,\allowbreak{} E497,\allowbreak{} E533,\allowbreak{} E500,\allowbreak{} F732,\allowbreak{} E535,\allowbreak{} E496,\allowbreak{} F717,\allowbreak{} F730 \\
\multicolumn{3}{p{0.94\linewidth}}{\textbf{LecB lectin/mannoside (PDB 5A6Y)}}\\*
16 & D455,\allowbreak{} D452 & D356,\allowbreak{} D359,\allowbreak{} D431,\allowbreak{} D355,\allowbreak{} D429,\allowbreak{} D450,\allowbreak{} D453,\allowbreak{} D396,\allowbreak{} C274,\allowbreak{} D363,\allowbreak{} C247,\allowbreak{} C275,\allowbreak{} C334,\allowbreak{} C331 \\
17 & D455,\allowbreak{} D452 & D356,\allowbreak{} D359,\allowbreak{} D431,\allowbreak{} D355,\allowbreak{} D429,\allowbreak{} D450,\allowbreak{} D453,\allowbreak{} D396,\allowbreak{} C274,\allowbreak{} D363,\allowbreak{} C247,\allowbreak{} C275,\allowbreak{} C334,\allowbreak{} C331,\allowbreak{} D454 \\
18 & D455,\allowbreak{} D452 & D356,\allowbreak{} D359,\allowbreak{} D431,\allowbreak{} D355,\allowbreak{} D429,\allowbreak{} D450,\allowbreak{} D453,\allowbreak{} D396,\allowbreak{} C274,\allowbreak{} D363,\allowbreak{} C247,\allowbreak{} C275,\allowbreak{} C334,\allowbreak{} C331,\allowbreak{} D454,\allowbreak{} D451 \\
\multicolumn{3}{p{0.94\linewidth}}{\textbf{KRIT1 ARD-FERM (PDB 5D68)}}\\*
14 & A313 & C903,\allowbreak{} C937,\allowbreak{} C905,\allowbreak{} C910,\allowbreak{} C935,\allowbreak{} C914,\allowbreak{} A310,\allowbreak{} A309,\allowbreak{} A265,\allowbreak{} A261,\allowbreak{} A262,\allowbreak{} C934,\allowbreak{} C906 \\
15 & A313 & C903,\allowbreak{} C937,\allowbreak{} C905,\allowbreak{} C910,\allowbreak{} C935,\allowbreak{} C914,\allowbreak{} A310,\allowbreak{} A309,\allowbreak{} A265,\allowbreak{} A261,\allowbreak{} A262,\allowbreak{} C934,\allowbreak{} C906,\allowbreak{} C936 \\
16 & A313 & C903,\allowbreak{} C937,\allowbreak{} C905,\allowbreak{} C910,\allowbreak{} C935,\allowbreak{} C914,\allowbreak{} A310,\allowbreak{} A309,\allowbreak{} A265,\allowbreak{} A261,\allowbreak{} A262,\allowbreak{} C934,\allowbreak{} C906,\allowbreak{} C936,\allowbreak{} C911 \\
\multicolumn{3}{p{0.94\linewidth}}{\textbf{Monobody GG3/ABL1 SH2 (PDB 5DC0)}}\\*
10 & B116,\allowbreak{} B138 & B181,\allowbreak{} B137,\allowbreak{} A51,\allowbreak{} A52,\allowbreak{} B136,\allowbreak{} A53,\allowbreak{} B117,\allowbreak{} B115 \\
13 & B116,\allowbreak{} B138 & B181,\allowbreak{} B137,\allowbreak{} A51,\allowbreak{} A52,\allowbreak{} B136,\allowbreak{} A53,\allowbreak{} B117,\allowbreak{} B115,\allowbreak{} B139,\allowbreak{} B182,\allowbreak{} A50 \\
14 & B116,\allowbreak{} B138 & B181,\allowbreak{} B137,\allowbreak{} A51,\allowbreak{} A52,\allowbreak{} B136,\allowbreak{} A53,\allowbreak{} B117,\allowbreak{} B115,\allowbreak{} B139,\allowbreak{} B182,\allowbreak{} A50,\allowbreak{} B180 \\
16 & B116,\allowbreak{} B138 & B181,\allowbreak{} B137,\allowbreak{} A51,\allowbreak{} A52,\allowbreak{} B136,\allowbreak{} A53,\allowbreak{} B117,\allowbreak{} B115,\allowbreak{} B139,\allowbreak{} B182,\allowbreak{} A50,\allowbreak{} B180,\allowbreak{} B176,\allowbreak{} B179 \\
\multicolumn{3}{p{0.94\linewidth}}{\textbf{Monobody AS25/ABL1 SH2 (PDB 5DC4)}}\\*
11 & A95,\allowbreak{} A28 & B176,\allowbreak{} B172,\allowbreak{} B173,\allowbreak{} B174,\allowbreak{} A25,\allowbreak{} A26,\allowbreak{} A97,\allowbreak{} A21,\allowbreak{} A46 \\
13 & A95,\allowbreak{} A28 & B176,\allowbreak{} B172,\allowbreak{} B173,\allowbreak{} B174,\allowbreak{} A25,\allowbreak{} A26,\allowbreak{} A97,\allowbreak{} A21,\allowbreak{} A46,\allowbreak{} A24,\allowbreak{} A29 \\
15 & A95,\allowbreak{} A28 & B176,\allowbreak{} B172,\allowbreak{} B173,\allowbreak{} B174,\allowbreak{} A25,\allowbreak{} A26,\allowbreak{} A97,\allowbreak{} A21,\allowbreak{} A46,\allowbreak{} A24,\allowbreak{} A29,\allowbreak{} A47,\allowbreak{} B171 \\
17 & A95,\allowbreak{} A28 & B176,\allowbreak{} B172,\allowbreak{} B173,\allowbreak{} B174,\allowbreak{} A25,\allowbreak{} A26,\allowbreak{} A97,\allowbreak{} A21,\allowbreak{} A46,\allowbreak{} A24,\allowbreak{} A29,\allowbreak{} A47,\allowbreak{} B171,\allowbreak{} A27,\allowbreak{} B175 \\
\multicolumn{3}{p{0.94\linewidth}}{\textbf{Erb1-Ytm1 complex (PDB 5EM2)}}\\*
12 & B614,\allowbreak{} B612 & B606,\allowbreak{} B609,\allowbreak{} B607,\allowbreak{} B537,\allowbreak{} A105,\allowbreak{} A103,\allowbreak{} B538,\allowbreak{} B536,\allowbreak{} A104,\allowbreak{} A106 \\
13 & B614,\allowbreak{} B612 & B606,\allowbreak{} B609,\allowbreak{} B607,\allowbreak{} B537,\allowbreak{} A105,\allowbreak{} A103,\allowbreak{} B538,\allowbreak{} B536,\allowbreak{} A104,\allowbreak{} A106,\allowbreak{} B608 \\
14 & B614,\allowbreak{} B612 & B606,\allowbreak{} B609,\allowbreak{} B607,\allowbreak{} B537,\allowbreak{} A105,\allowbreak{} A103,\allowbreak{} B538,\allowbreak{} B536,\allowbreak{} A104,\allowbreak{} A106,\allowbreak{} B608,\allowbreak{} B611 \\
\multicolumn{3}{p{0.94\linewidth}}{\textbf{LSD1 SWIRM domain (PDB 5IT3)}}\\*
13 & A41 & B134,\allowbreak{} B153,\allowbreak{} A37,\allowbreak{} A34,\allowbreak{} A38,\allowbreak{} B133,\allowbreak{} B135,\allowbreak{} A39,\allowbreak{} A40,\allowbreak{} A35,\allowbreak{} A36,\allowbreak{} A33 \\
14 & A41 & B134,\allowbreak{} B153,\allowbreak{} A37,\allowbreak{} A34,\allowbreak{} A38,\allowbreak{} B133,\allowbreak{} B135,\allowbreak{} A39,\allowbreak{} A40,\allowbreak{} A35,\allowbreak{} A36,\allowbreak{} A33,\allowbreak{} A42 \\
15 & A41 & B134,\allowbreak{} B153,\allowbreak{} A37,\allowbreak{} A34,\allowbreak{} A38,\allowbreak{} B133,\allowbreak{} B135,\allowbreak{} A39,\allowbreak{} A40,\allowbreak{} A35,\allowbreak{} A36,\allowbreak{} A33,\allowbreak{} A42,\allowbreak{} B137 \\
16 & A41 & B134,\allowbreak{} B153,\allowbreak{} A37,\allowbreak{} A34,\allowbreak{} A38,\allowbreak{} B133,\allowbreak{} B135,\allowbreak{} A39,\allowbreak{} A40,\allowbreak{} A35,\allowbreak{} A36,\allowbreak{} A33,\allowbreak{} A42,\allowbreak{} B137,\allowbreak{} A31 \\
17 & A41 & B134,\allowbreak{} B153,\allowbreak{} A37,\allowbreak{} A34,\allowbreak{} A38,\allowbreak{} B133,\allowbreak{} B135,\allowbreak{} A39,\allowbreak{} A40,\allowbreak{} A35,\allowbreak{} A36,\allowbreak{} A33,\allowbreak{} A42,\allowbreak{} B137,\allowbreak{} A31,\allowbreak{} B136 \\
\end{longtable}

\begin{longtable}{lrrrrr}
\caption{Runtimes, speedups, and minimization ratios. Positions count flexible residue positions. Each PDB ID and position count identifies the design specified in Table~\ref{tab:runtime-designs}. Completed times cover processing of the full sequence list, including sequences for which no binding score was returned. A greater-than-or-equal sign denotes 14-day MARK* noncompletion and the resulting speedup lower bound. MARK* times at the 14-day limit are displayed as 336 h. Speedup lower bounds use program times excluding startup and teardown and are rounded down. The minimization ratio is the number of constrained local minimizations of complete RC assignments in MARK* divided by that in PACK*, summed over all sequences and molecular states, counting cached results once. For 14-day timeouts, its lower bound uses the last cumulative count recorded before termination.}\label{tab:frontier-pairs}\\
\toprule
PDB ID & Positions & MARK* (h) & PACK* (min) & Speedup & Minimization ratio \\
\midrule
\endfirsthead
\toprule
PDB ID & Positions & MARK* (h) & PACK* (min) & Speedup & Minimization ratio \\
\midrule
\endhead
\bottomrule
\endfoot
1A0R & 8 & 11.674 & 11.369 & 61.61 & 26.91 \\
1A0R & 9 & $\ge 336$ & 16.077 & $\ge 1253.94$ & $\ge 375.59$ \\
1A0R & 10 & $\ge 336$ & 17.586 & $\ge 1146.35$ & $\ge 390.95$ \\
1B6C & 16 & 18.782 & 19.602 & 57.49 & 21.41 \\
1B6C & 17 & 22.286 & 23.008 & 58.12 & 25.87 \\
1B6C & 18 & 62.395 & 25.534 & 146.62 & 59.15 \\
1B6C & 19 & $\ge 336$ & 43.275 & $\ge 465.86$ & $\ge 177.56$ \\
1GWC & 13 & 1.960 & 9.685 & 12.14 & 3.56 \\
1GWC & 14 & 3.556 & 10.121 & 21.08 & 5.70 \\
1GWC & 15 & 23.278 & 14.678 & 95.15 & 13.16 \\
1GWC & 16 & $\ge 336$ & 35.150 & $\ge 573.53$ & $\ge 2.22$ \\
2HNU & 16 & 15.254 & 12.024 & 76.11 & 19.93 \\
2HNU & 18 & 108.560 & 14.752 & 441.55 & 131.45 \\
2HNU & 19 & $\ge 336$ & 16.395 & $\ge 1229.66$ & $\ge 94.22$ \\
2HNU & 22 & $\ge 336$ & 20.831 & $\ge 967.80$ & $\ge 25.01$ \\
2HNV & 13 & 1.028 & 6.272 & 9.83 & 2.59 \\
2HNV & 14 & 2.210 & 7.681 & 17.26 & 5.10 \\
2HNV & 15 & 4.905 & 11.082 & 26.56 & 10.55 \\
2P4A & 17 & 1.910 & 10.229 & 11.21 & 3.98 \\
2P4A & 18 & 4.596 & 19.483 & 14.15 & 6.90 \\
2P4A & 20 & 38.095 & 22.688 & 100.75 & 17.93 \\
2Q1E & 13 & 3.279 & 5.010 & 39.27 & 13.35 \\
2Q1E & 14 & 9.269 & 6.923 & 80.33 & 32.62 \\
2Q1E & 16 & 89.599 & 8.800 & 610.89 & 81.69 \\
2Q2A & 13 & 3.737 & 6.976 & 32.15 & 8.02 \\
2Q2A & 14 & 5.444 & 7.466 & 43.75 & 13.32 \\
2Q2A & 15 & 5.672 & 9.515 & 35.77 & 19.14 \\
2RF9 & 13 & 1.808 & 10.232 & 10.60 & 2.78 \\
2RF9 & 14 & 4.999 & 11.923 & 25.16 & 3.27 \\
2RF9 & 15 & 9.635 & 13.688 & 42.24 & 7.34 \\
2RFD & 9 & 4.053 & 23.011 & 10.57 & 4.37 \\
2RFD & 10 & 4.390 & 25.021 & 10.53 & 4.88 \\
2RFD & 11 & 6.153 & 29.360 & 12.57 & 5.52 \\
2RFD & 12 & 13.537 & 35.257 & 23.04 & 5.58 \\
2RFE & 12 & 1.428 & 12.676 & 6.76 & 3.36 \\
2RFE & 13 & 3.048 & 14.364 & 12.73 & 8.33 \\
2RFE & 14 & 10.506 & 20.362 & 30.96 & 12.98 \\
2RL0 & 12 & 1.929 & 8.517 & 13.59 & 5.79 \\
2RL0 & 13 & 6.835 & 11.804 & 34.74 & 9.79 \\
2RL0 & 15 & 79.682 & 14.831 & 322.36 & 127.47 \\
2RL0 & 16 & 148.950 & 16.016 & 558.01 & 204.69 \\
2XGY & 11 & 0.293 & 5.964 & 2.95 & 0.89 \\
2XGY & 12 & 0.600 & 7.037 & 5.11 & 2.04 \\
2XGY & 13 & 110.735 & 8.274 & 803.01 & 530.90 \\
2XXM & 12 & 1.205 & 6.841 & 10.57 & 7.24 \\
2XXM & 13 & 1.760 & 10.815 & 9.76 & 9.92 \\
2XXM & 14 & 7.507 & 11.660 & 38.63 & 18.49 \\
2XXM & 15 & 12.498 & 33.943 & 22.09 & 31.33 \\
3BU8 & 13 & 1.509 & 6.700 & 13.51 & 3.66 \\
3BU8 & 14 & 8.508 & 10.648 & 47.94 & 14.26 \\
3BU8 & 15 & 25.748 & 12.450 & 124.09 & 34.36 \\
3BU8 & 16 & 85.910 & 18.365 & 280.68 & 97.02 \\
3BUA & 14 & 77.981 & 17.507 & 267.26 & 265.00 \\
3BUA & 15 & $\ge 336$ & 19.880 & $\ge 1014.10$ & $\ge 436.44 $\\
3BUA & 16 & $\ge 336$ & 25.416 & $\ge 793.19$ & $\ge 21.64$ \\
3CAL & 12 & 13.303 & 9.854 & 81.00 & 53.45 \\
3CAL & 13 & 35.950 & 10.427 & 206.86 & 113.50 \\
3CAL & 14 & 121.772 & 15.769 & 463.33 & 422.63 \\
3CAL & 15 & $\ge 336$ & 29.002 & $\ge 695.13$ & $\ge 1081.51$ \\
3EB6 & 13 & 2.148 & 4.343 & 29.67 & 8.48 \\
3EB6 & 14 & 6.862 & 5.238 & 78.60 & 14.76 \\
3EB6 & 15 & 59.291 & 7.056 & 504.20 & 92.43 \\
3EB6 & 16 & $\ge 336$ & 14.319 & $\ge 1407.87$ & $\ge 299.31$ \\
3GXU & 13 & 2.575 & 9.126 & 16.93 & 31.93 \\
3GXU & 14 & 13.691 & 6.192 & 132.67 & 138.04 \\
3GXU & 15 & 84.292 & 9.129 & 553.99 & 528.45 \\
3GXU & 16 & 111.315 & 16.111 & 414.54 & 735.40 \\
3K3Q & 15 & 2.410 & 14.921 & 9.69 & 5.05 \\
3K3Q & 16 & 10.055 & 18.943 & 31.85 & 16.75 \\
3K3Q & 17 & 19.839 & 28.389 & 41.93 & 29.34 \\
3K3Q & 18 & 244.855 & 100.635 & 145.99 & 171.41 \\
3MA2 & 14 & 3.236 & 17.337 & 11.20 & 6.28 \\
3MA2 & 15 & 54.370 & 23.491 & 138.87 & 62.43 \\
3MA2 & 16 & 332.509 & 27.593 & 723.02 & 198.22 \\
3U7Y & 8 & 2.380 & 15.234 & 9.38 & 3.21 \\
3U7Y & 9 & 5.611 & 17.796 & 18.92 & 6.98 \\
3U7Y & 10 & 17.854 & 20.597 & 52.01 & 11.25 \\
3U7Y & 11 & 21.640 & 26.860 & 48.34 & 19.91 \\
3U7Y & 12 & 280.936 & 38.633 & 436.31 & 210.81 \\
4HEM & 12 & 0.675 & 6.573 & 6.16 & 2.06 \\
4HEM & 13 & 2.262 & 8.427 & 16.11 & 5.44 \\
4HEM & 14 & 4.212 & 10.889 & 23.21 & 7.72 \\
4KT6 & 9 & 0.919 & 12.962 & 4.25 & 1.18 \\
4KT6 & 10 & 3.731 & 16.124 & 13.88 & 1.90 \\
4KT6 & 11 & 8.653 & 21.562 & 24.08 & 3.68 \\
4PXF & 13 & 7.347 & 8.419 & 52.36 & 13.76 \\
4PXF & 14 & 40.148 & 11.223 & 214.64 & 60.03 \\
4PXF & 15 & $\ge 336$ & 16.112 & $\ge 1251.24$ & $\ge 14.56$ \\
4U3S & 13 & 151.919 & 14.261 & 639.15 & 215.40 \\
4U3S & 14 & 236.787 & 15.235 & 932.56 & 527.35 \\
4U3S & 15 & $\ge 336$ & 14.559 & $\ge 1384.74$ & $\ge 265.51$ \\
4WEM & 13 & 2.746 & 5.452 & 30.21 & 5.50 \\
4WEM & 14 & 5.540 & 6.354 & 52.32 & 9.19 \\
4WEM & 15 & 70.224 & 12.104 & 348.10 & 83.04 \\
4WEM & 17 & $\ge 336$ & 15.178 & $\ge 1328.26$ & $\ge 197.67$ \\
4WWI & 9 & 2.841 & 14.213 & 11.99 & 4.27 \\
4WWI & 11 & 9.951 & 17.792 & 33.56 & 13.86 \\
4WWI & 12 & $\ge 336$ & 26.671 & $\ge 755.86$ & $\ge 102.30$ \\
4WYQ & 10 & 5.091 & 9.612 & 31.78 & 4.56 \\
4WYQ & 11 & 3.792 & 10.681 & 21.30 & 6.52 \\
4WYQ & 12 & 19.621 & 12.607 & 93.38 & 14.62 \\
4WYQ & 13 & 54.146 & 27.355 & 118.76 & 53.99 \\
4WYQ & 14 & 88.905 & 31.665 & 168.46 & 77.62 \\
4WYU & 11 & 2.867 & 9.729 & 17.68 & 5.67 \\
4WYU & 12 & 37.060 & 17.583 & 126.46 & 40.00 \\
4WYU & 13 & 246.657 & 22.341 & 662.44 & 102.78 \\
4Z80 & 13 & 1.718 & 19.743 & 5.22 & 3.20 \\
4Z80 & 14 & 22.296 & 23.537 & 56.84 & 14.67 \\
4Z80 & 15 & 81.441 & 39.306 & 124.32 & 45.06 \\
4Z80 & 16 & 203.441 & 40.461 & 301.68 & 88.90 \\
4ZNC & 12 & 1.637 & 8.710 & 11.28 & 4.96 \\
4ZNC & 13 & 2.446 & 9.916 & 14.80 & 6.41 \\
4ZNC & 14 & 11.764 & 14.325 & 49.27 & 23.19 \\
4ZNC & 15 & 14.827 & 14.656 & 60.70 & 29.72 \\
5A6Y & 16 & 18.197 & 17.560 & 62.18 & 20.54 \\
5A6Y & 17 & 87.508 & 18.837 & 278.73 & 69.04 \\
5A6Y & 18 & 307.305 & 24.773 & 744.30 & 195.51 \\
5D68 & 14 & 35.803 & 9.526 & 225.51 & 64.94 \\
5D68 & 15 & 118.462 & 12.220 & 581.65 & 134.81 \\
5D68 & 16 & $\ge 336$ & 15.295 & $\ge 1318.04$ & $\ge 8.92$ \\
5DC0 & 10 & 4.380 & 8.417 & 31.22 & 17.03 \\
5DC0 & 13 & 36.296 & 10.508 & 207.24 & 96.17 \\
5DC0 & 14 & 74.313 & 12.993 & 343.18 & 169.93 \\
5DC0 & 16 & $\ge 336$ & 15.744 & $\ge 1280.46$ & $\ge 197.81$ \\
5DC4 & 11 & 1.161 & 7.404 & 9.41 & 3.74 \\
5DC4 & 13 & 12.359 & 12.357 & 60.01 & 13.31 \\
5DC4 & 15 & 69.793 & 15.219 & 275.15 & 55.04 \\
5DC4 & 17 & $\ge 336$ & 21.767 & $\ge 926.17$ & $\ge 48.21$ \\
5EM2 & 12 & 1.513 & 32.415 & 2.80 & 0.60 \\
5EM2 & 13 & 3.590 & 38.646 & 5.57 & 1.18 \\
5EM2 & 14 & 6.755 & 48.944 & 8.28 & 2.02 \\
5IT3 & 13 & 2.287 & 5.231 & 26.24 & 18.95 \\
5IT3 & 14 & 17.267 & 6.880 & 150.58 & 54.03 \\
5IT3 & 15 & 77.187 & 8.288 & 558.76 & 230.24 \\
5IT3 & 16 & 148.697 & 8.195 & 1088.69 & 351.35 \\
5IT3 & 17 & $\ge 336$ & 8.834 & $\ge 2282.09$ & $\ge 582.94 $\\
\end{longtable}

\endgroup

\subsection{Generation and analysis of SHV-1--BLIP conformational ensembles}
\label{sec:mi-analysis}

\subsubsection{System preparation and ensemble generation.}
We compared wild-type (WT) BLIP and BLIP E73M in complex with SHV-1 at 298.15 K,
using the same scaffold derived from chains A and B of PDB ID 2G2U.
Chain A represents SHV-1 and chain B represents BLIP. The 13 flexible
positions were SHV-1 residues 103, 104, 105, 106, 132, and 167, and
BLIP residues 53, 72, 73, 74, 75, 112, and 143. Only BLIP position 73
changed amino acid identity; SHV-1 position 104 remained Asp in both
sequences. Residue labels use the deposited biological numbering.

The scaffold was generated using Reduce hydrogen placement, pdb4amber,
and local OpenMM minimization with ff14SB and OBC1 implicit solvent.
PACK* energy evaluation used the OSPREY AMBER/EEF1 energy function.
Backbone coordinates and all nonflexible residues were held unchanged
during ensemble generation. 
The mutant backbone was not substituted into the calculations.

Adaptive PACK* learning used 500 initial training draws, two-fold
cross-validation over 1-body/2-body shrinkage values 0/0, 2/5, 5/10, and
10/20, up to eight refits, 100--400 pilot draws, and 400 independent
validation draws. We used $\varepsilon=0.1$, $\delta=0.05$, $\tau=1$
kcal/mol, a conditional mean excess bound of 20, and hypothesis-test
significance level 0.05. Final sample counts were selected before final
sampling, with a minimum of 10000 and maximum of 20000. A sparse
proposal interaction graph used the dual $4$~\AA\ distance criterion;
target energies and importance weights retained all residue interactions.
Both backgrounds used seed 42 for the detailed analyses below. Training,
pilot, and validation draws were excluded from the structural analysis.

Both calculations returned partition function estimates after independent
validation. Final draw counts were 17678 for WT and 10000 for E73M;
effective sample sizes (ESS) were 12470.7 and 9235.0. The largest normalized
weights were 0.0000831 and 0.0002260. Direct full-energy evaluations of
32 final assignments per sequence agreed within $10^{-6}$ kcal/mol.
Saved angles, distances, assignments, and importance weights were aligned
draw by draw, including repeated assignments.

\begin{table}[htbp]
\centering
\caption{Importance-weighted SHV-1--BLIP ensembles from the independent
estimation samples, seed 42. WT denotes wild type; E73M is the BLIP mutant.
ESS denotes effective sample size. Contact probabilities use the minimum
distance from K74 $\mathrm{N}_{\zeta}$ to either D104 carboxylate oxygen.}
\label{tab:shv-ensembles}
\begin{tabular}{@{}lrrrr@{}}
\hline
Background & Draws & ESS & Mean distance (\AA) & Contact (\%) \\
\hline
WT & 17678 & 12470.7 & 4.2036 & 0.000464 \\
E73M & 10000 & 9235.0 & 3.4498 & 52.235514 \\
\hline
\end{tabular}
\end{table}

\subsubsection{Residue ranking by conformational distribution shift.}
We ranked mutation-induced changes in the side-chain conformational
distributions using joint-angle Jensen--Shannon divergence (JSD) between
the importance-weighted WT and E73M ensembles, using periodic kernel
density estimation (KDE) with the primary $5^{\circ}$ bandwidth and seed 42.
Excluding the mutated BLIP position 73 from the 13-position model left
12 unmutated residue positions for comparison.
D104, K74, and W112 had the largest JSD values
(Table~\ref{tab:shv-jsd-ranking}). Each residue used its complete
heavy-atom side-chain angle vector, retaining within-residue correlations
and identifying chemically equivalent terminal atoms.

\begin{table}[htbp]
\centering
\caption{Joint-angle Jensen--Shannon divergence (JSD) ranking for the
13-position SHV-1--BLIP model, using periodic kernel density estimation
(KDE) with the primary $5^{\circ}$ bandwidth and seed 42.
The 12 unchanged residue positions are ranked in descending order;
the mutated BLIP position 73 is excluded. JSD values are in nats.}
\label{tab:shv-jsd-ranking}
\begin{tabular}{@{}rlr@{}}
\hline
Rank & Residue & JSD (nats) \\
\hline
1 & SHV-1 D104 & 0.227777 \\
2 & BLIP K74 & 0.215150 \\
3 & BLIP W112 & 0.145454 \\
4 & SHV-1 T167 & 0.045533 \\
5 & SHV-1 N132 & 0.018544 \\
6 & SHV-1 Y105 & 0.009644 \\
7 & BLIP Q72 & 0.009296 \\
8 & BLIP Y53 & 0.007543 \\
9 & BLIP Y143 & 0.005080 \\
10 & SHV-1 V103 & 0.001019 \\
11 & SHV-1 S106 & 0.000424 \\
12 & BLIP L75 & 0.000002 \\
\hline
\end{tabular}
\end{table}

\subsubsection{Angular distributions and basin definitions.}
\label{sec:distribution-shift}
All angular distributions were computed from importance-weighted
estimation samples across all residue conformations.
Each draw contributes its importance weight divided by the sum of weights
in that sequence's final ensemble. D104 has two heavy-atom side-chain
angles, defined by
$\mathrm{N}$--$\mathrm{C}_{\alpha}$--$\mathrm{C}_{\beta}$--$\mathrm{C}_{\gamma}$
and $\mathrm{C}_{\alpha}$--$\mathrm{C}_{\beta}$--$\mathrm{C}_{\gamma}$--$\mathrm{O}_{\delta 1}$;
K74 has four, ending with
$\mathrm{C}_{\gamma}$--$\mathrm{C}_{\delta}$--$\mathrm{C}_{\epsilon}$--$\mathrm{N}_{\zeta}$.
D104 $\sidechi{2}$ is an axial angle with a
$180^{\circ}$ period, identifying equivalent $\mathrm{O}_{\delta 1}/\mathrm{O}_{\delta 2}$ orientations.
Other reporter angles have a $360^{\circ}$ period. Circular summaries
use these respective periods (Table~\ref{tab:shv-angles}).

Angle distributions were computed using importance-weighted periodic
von Mises KDE~\cite{dimarzio2011torus}, with primary bandwidth $5^{\circ}$.
For a period of L degrees and bandwidth h degrees, the kernel
concentration is the reciprocal square of 2 pi h/L; thus h specifies
the same local angular smoothing scale for both periods. Each density
integrates to one over its full period. Smoothing describes the saved
minimized conformations and does not represent thermal sampling inside
an RC voxel.

D104 $\sidechi{1}$ basins use the minimized angle modulo $360^{\circ}$:
gauche+ (g$^+$) spans $[0^{\circ},120^{\circ})$, trans spans
$[120^{\circ},240^{\circ})$, and gauche$-$ (g$^-$) spans
$[240^{\circ},360^{\circ})$.
Populations sum the unsmoothed normalized weights in each sector.
These angular basins combine RC labels and differ from individual RC
populations. For D104, this three-sector definition applies to $\sidechi{1}$,
not the symmetric terminal $\sidechi{2}$.
Table~\ref{tab:shv-basins} reports all three populations and joint
contact probabilities, including negligible populations.

\begin{table}[htbp]
\centering
\caption{D104 $\sidechi{1}$ basin populations and joint contact probabilities
from importance-weighted samples without kernel density estimation (KDE)
smoothing. WT denotes wild type; E73M is the BLIP mutant.
Basin labels: g$^+$, gauche+; g$^-$, gauche$-$.
Values are percentages of the full ensemble.}
\label{tab:shv-basins}
\begin{tabular}{@{}llrr@{}}
\hline
Background & Basin & Population (\%) & Basin and contact (\%) \\
\hline
WT & g$^+$ & 99.999501 & 0.000000 \\
WT & trans & 0.000499 & 0.000464 \\
WT & g$^-$ & 0.000000 & 0.000000 \\
E73M & g$^+$ & 47.764476 & 0.000000 \\
E73M & trans & 52.235524 & 52.235514 \\
E73M & g$^-$ & 0.000000 & 0.000000 \\
\hline
\end{tabular}
\end{table}

\begin{table}[htbp]
\centering
\caption{Circular means and circular standard deviations (SD) of
unsmoothed minimized angles, in degrees. WT denotes wild type; E73M is
the BLIP mutant. D104 $\sidechi{2}$ summaries use a 180-degree period;
other angles use 360 degrees.}
\label{tab:shv-angles}
\begin{tabular}{@{}ll lrr@{}}
\hline
Residue & Angle & Background & Mean & SD \\
\hline
D104 & $\sidechi{1}$ & WT & 59.364 & 0.388 \\
D104 & $\sidechi{2}$ & WT & -51.501 & 0.138 \\
K74 & $\sidechi{1}$ & WT & -65.992 & 0.702 \\
K74 & $\sidechi{2}$ & WT & 174.674 & 2.692 \\
K74 & $\sidechi{3}$ & WT & 176.559 & 2.779 \\
K74 & $\sidechi{4}$ & WT & 73.466 & 4.309 \\
D104 & $\sidechi{1}$ & E73M & 132.116 & 76.970 \\
D104 & $\sidechi{2}$ & E73M & -83.097 & 34.871 \\
K74 & $\sidechi{1}$ & E73M & -68.597 & 2.507 \\
K74 & $\sidechi{2}$ & E73M & 175.202 & 2.390 \\
K74 & $\sidechi{3}$ & E73M & 171.560 & 4.437 \\
K74 & $\sidechi{4}$ & E73M & 67.932 & 5.918 \\
\hline
\end{tabular}
\end{table}

\begin{figure}[htbp]
\centering
\includegraphics[width=\textwidth]{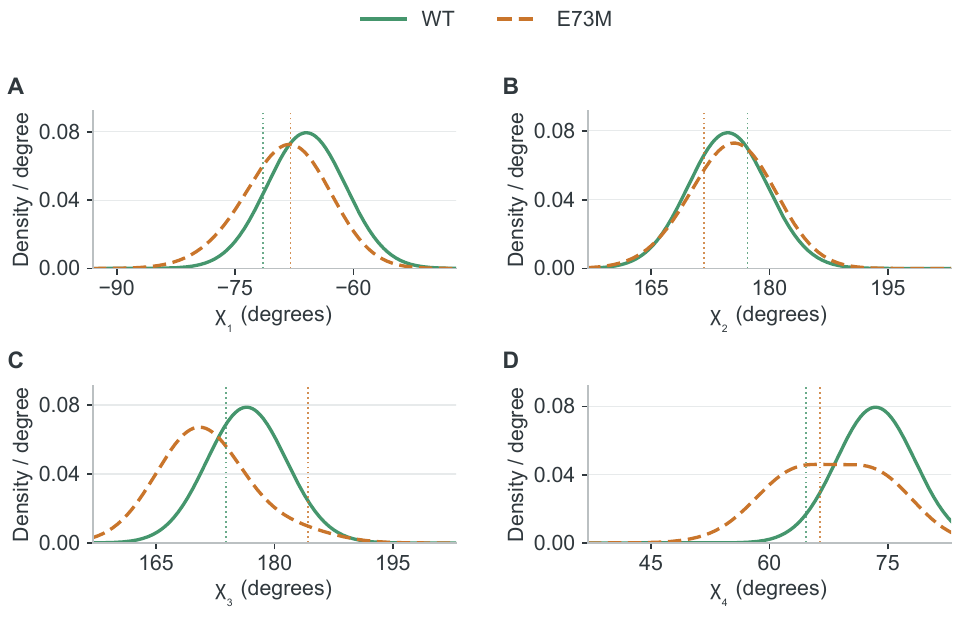}
\caption{K74 side-chain dihedral-angle distributions after constrained
local minimization across all residue conformations (RCs).
Panels A--D show $\sidechi{1}$--$\sidechi{4}$. Distributions were computed
using importance-weighted periodic kernel density estimation (KDE)
with $5^{\circ}$ bandwidth.
Green solid curves: wild type (WT); orange dashed curves: BLIP E73M.
Green vertical dotted lines: WT crystal angles; orange vertical dotted
lines: E73M crystal angles. The displayed ranges highlight angular
changes within the dominant rotamer basins. Densities remain normalized
over the complete $360^{\circ}$ period; samples outside the displayed
ranges are included in all numerical summaries. Angles near trans are
shown on a continuous branch around $180^{\circ}$. Both ensembles use
the 13-position model and seed 42.}
\label{fig:within-rc-redistribution}
\end{figure}

\subsubsection{Contact geometry and experimental comparison.}
We defined the minimum oxygen--nitrogen distance in each minimized
conformation as
\[
 d=\min\bigl\{
 \operatorname{dist}(\mathrm{D104}\,\mathrm{O}_{\delta 1},\mathrm{K74}\,\mathrm{N}_{\zeta}),
 \operatorname{dist}(\mathrm{D104}\,\mathrm{O}_{\delta 2},\mathrm{K74}\,\mathrm{N}_{\zeta})
 \bigr\},
\]
where $\operatorname{dist}$ denotes the Euclidean distance between the
named atoms. Conformations with $d\le 3.5$~\AA\ counted as satisfying
the salt-bridge contact proxy~\cite{lenguyen2025saltbridges}.
The empirical cumulative distance distributions and contact probabilities
use unsmoothed normalized weights. The proxy does not impose an additional
donor-angle criterion. Weighted contact populations were 0.000464\%
for WT and 52.2355\% for E73M; mean distances were 4.2036 and
3.4498~\AA. At thresholds of 3.0 and 4.0~\AA, E73M contact populations
were 52.2084\% and 52.2355\%, versus 0.000464\% and 0.001003\% in WT.

The deposited crystal distances were 4.4826~\AA\ in WT and
2.7366~\AA\ in E73M. Reynolds et al.~\cite{reynolds2008shv}
identified salt-bridge formation as a structural response to E73M and
demonstrated sensitivity to local side-chain and backbone geometry.
Our increase in short-distance population agreed qualitatively with
this observation. The experimental structures supply individual
conformations, not measurements of solution basin populations; the
calculated populations provide an ensemble prediction.

D104 crystal $\sidechi{1}$ changed from
$52.4072^{\circ}$ to $-166.3079^{\circ}$ and $\sidechi{2}$ from
$-46.5679^{\circ}$ to $20.4613^{\circ}$.
In the E73M ensemble, 52.2355\% lay within $9^{\circ}$ of the mutant
crystal $\sidechi{1}$, but only 2.7974\% lay within $9^{\circ}$ of its
$\sidechi{2}$, after identifying equivalent oxygens. Applying the same three-sector convention to each of
K74's four angles, the dominant joint basin (g$^-$, trans, trans, g$^+$)
accounted for 99.8264\% of WT and 99.9574\% of E73M importance weight.
Thus K74's changes occurred primarily within existing rotamer basins.

Main-text Fig.~\ref{MAIN-fig:mi-analysis}A shows WT and E73M crystal
structures and calculated example conformations. The calculated examples
are the lowest-energy sampled WT conformation and the lowest-energy
sampled E73M conformation satisfying $d\le 3.5$~\AA. These examples
illustrate geometry; population estimates use all importance-weighted
estimation samples.

We repeated both calculations with ten random seeds (42--51), keeping
the model and algorithm settings unchanged. Across seeds, the contact
population remained near zero for WT and approximately one-half for E73M,
while recovery of the mutant crystal D104 $\sidechi{2}$ remained limited
(Table~\ref{tab:shv-seed-variability}).

\begin{table}[htbp]
\centering
\caption{Variability across random seeds in the SHV-1--BLIP ensembles.
Seeds were weighted equally. Ranges are minima and maxima, not confidence
intervals. All final batches were included descriptively.}
\label{tab:shv-seed-variability}
\begin{tabular}{@{}p{0.56\textwidth}rr@{}}
\hline
Quantity & Mean (\%) & Range across seeds (\%) \\
\hline
WT contact population & 0.00165 & 0--0.0144 \\
E73M contact population & 50.65 & 48.29--52.24 \\
E73M D104 $\sidechi{2}$ within $9^{\circ}$ of the mutant crystal & 2.53 & 2.09--2.95 \\
\hline
\end{tabular}
\end{table}

\subsubsection{Bandwidth sensitivity.}
\label{sec:shv-bandwidth-sensitivity}
We evaluated joint-angle Jensen--Shannon divergence
(JSD)~\cite{lin1991divergence} between equally weighted background
distributions at $1^{\circ}$, $2^{\circ}$, and $5^{\circ}$ bandwidths.
Each residue used its complete heavy-atom side-chain angle vector,
retaining within-residue correlations and all RCs. Chemically equivalent
terminal atoms were identified for Asp and Tyr. Excluding the mutated
position B73 left 12 unchanged flexible positions.
Four independent quadrature replicates each drew 1024 queries per
background by weighted conformation selection followed by periodic kernel
noise. JSD uses natural logarithms and ranges from zero to log(2) nats.
Numerical quadrature standard errors measure integration variability,
rather than molecular-seed or experimental uncertainty.
Table~\ref{tab:distribution-shift-all-sites} lists all positions and
bandwidths. Rankings and magnitudes depend on bandwidth; the primary
plots and interpretation use $5^{\circ}$.
\begin{table}[htbp]
\centering
\caption{Joint-angle Jensen--Shannon divergence in nats for all 12 unchanged flexible residue positions, ordered by the primary 5-degree result. Chemically equivalent terminal atoms are identified. Numerical quadrature errors are supplied in the accompanying data.}
\label{tab:distribution-shift-all-sites}
\begin{tabular}{@{}lrrr@{}}
\hline
Residue & 1 degree & 2 degrees & 5 degrees \\
\hline
SHV-1 D104 & 0.277247 & 0.239861 & 0.227777 \\
BLIP K74 & 0.675788 & 0.527063 & 0.215150 \\
BLIP W112 & 0.551307 & 0.445843 & 0.145454 \\
SHV-1 T167 & 0.075524 & 0.050711 & 0.045533 \\
SHV-1 N132 & 0.041727 & 0.040192 & 0.018544 \\
SHV-1 Y105 & 0.212081 & 0.091980 & 0.009644 \\
BLIP Q72 & 0.190362 & 0.054972 & 0.009296 \\
BLIP Y53 & 0.156847 & 0.045791 & 0.007543 \\
BLIP Y143 & 0.117388 & 0.032251 & 0.005080 \\
SHV-1 V103 & 0.029103 & 0.009530 & 0.001019 \\
SHV-1 S106 & 0.010245 & 0.002506 & 0.000424 \\
BLIP L75 & 0.000052 & 0.000013 & 0.000002 \\
\hline
\end{tabular}
\end{table}


\bibliographystyle{splncs04}
\bibliography{refs}

\end{document}